\documentclass[sigconf,10pt]{acmart}
\newcommand{\ignore}[1]{}
\usepackage{fancyhdr}
\usepackage[normalem]{ulem}

\usepackage{subfig}
\usepackage{graphicx}
\usepackage{caption}
\usepackage{dblfloatfix}
\usepackage{multirow}
\usepackage{fixltx2e}
\usepackage{xspace}
\usepackage{pifont}
\usepackage{booktabs}
\usepackage{setspace}
\usepackage{color}
\usepackage{xcolor}
\usepackage{balance}

\usepackage[]{algorithm2e}
\usepackage{xfrac}

\usepackage{gensymb}
\let\oldcelsius\celsius
\renewcommand{\celsius}{~\oldcelsius~$ $}

\newcommand{\squishlist} {
	\begin{list}{$\bullet$} {
		\setlength{\itemsep}{-2pt}
		\setlength{\parsep}{2pt}
		\setlength{\topsep}{0pt}
		\setlength{\partopsep}{0pt}
		\setlength{\leftmargin}{1.0em}
		\setlength{\labelwidth}{1em}
		\setlength{\labelsep}{0.5em}
	}
}
\newcommand{\squishend} {
	\end{list}
}

\newcommand{\boxbegin} {
	\begin{tcolorbox}[enhanced, frame hidden, colback=gray!50, breakable]
}

\newcommand{\boxend} {
	\end{tcolorbox}
}

\newcommand{\yboxbegin} {
	\begin{tcolorbox}[enhanced, frame hidden, colback=yellow!50, breakable]
}

\newcommand{\yboxend} {
	\end{tcolorbox}
}

\newcommand{\dimms}[1]{96\xspace}

\newcommand{\navdimms}[0]{24\xspace}

\newcommand{\vdd}{{{$V_{\text{DD}}$}}\xspace}

\newcommand{\hvdd}{{{$V_{\text{DD}}/2$}}\xspace}

\newcommand{\trcdN}{\texttt{{tRCD}}}
\newcommand{\trasN}{\texttt{{tRAS}}}
\newcommand{\trpN}{\texttt{{tRP}}}
\newcommand{\twrN}{\texttt{{tWR}}}

\newcommand{\trcd}{\texttt{{tRCD}}\xspace}
\newcommand{\tras}{\texttt{{tRAS}}\xspace}
\newcommand{\trp}{\texttt{{tRP}}\xspace}
\newcommand{\twr}{\texttt{{tWR}}\xspace}

\newcommand{\cmdact}{\texttt{{ACTIVATION}}\xspace}
\newcommand{\cmdread}{\texttt{{READ}}\xspace}

\newcommand{\cmdprech}{\texttt{{PRECHARGE}}\xspace}

\newcommand{\myprofiling}{{DIVA Profiling}\xspace}

\newcommand{\myshuffling}{{DIVA Shuffling}\xspace}

\newcommand{\mydram}{{DIVA-DRAM}\xspace}

\newcommand{\changes}[1]{\textcolor{black}{#1}}

\newcommand{\dhlii}[1]{\textcolor{black}{#1}}
\newcommand{\dhliii}[1]{\textcolor{black}{#1}}
\newcommand{\dhliv}[1]{\textcolor{black}{#1}}
\newcommand{\sgI}[1]{\textcolor{black}{#1}}
\newcommand{\sgII}[1]{\textcolor{black}{#1}}
\newcommand{\sgIII}[1]{\textcolor{black}{#1}}

\newcommand{\module}[3]{{{\textit #1}$_{\mathrm{#2}}^{\mathrm{#3}}$}\xspace}

\usepackage{amsmath}    % \begin{align*} (asterisk = no equation numbers)

\usepackage[text-rm]{siunitx}
\mathchardef\mhyphen="2D

\usepackage{soul}
\usepackage{tikz}
\usepackage{indentfirst}
\usepackage[most]{tcolorbox}
\tcbuselibrary{skins}
\usepackage{lipsum}
\usepackage{siunitx}

\usepackage{booktabs}
\renewcommand{\textrightarrow}{$\rightarrow$}

\usepackage{booktabs}

\renewcommand{\textrightarrow}{$\rightarrow$}

\setcopyright{none}
\renewcommand\footnotetextcopyrightpermission[1]{}

\begin{document}

\title{Understanding and Exploiting\\%
Design-Induced Latency Variation in Modern DRAM Chips}

\newcommand{\authspace}[0]{\hspace{18pt}}
\newcommand{\affilspace}[0]{\hspace{15pt}}

\author{\vspace{-3pt}
	Donghyuk Lee$^\dag$$^\ddag$ \authspace
	Samira Khan$^\aleph$ \authspace
	Lavanya Subramanian$^\dag$ \authspace
	Saugata Ghose$^\dag$\vspace{-9pt}\\
	Rachata Ausavarungnirun$^\dag$ \authspace
	Gennady Pekhimenko$^\dag$$^\P$ \authspace
	Vivek Seshadri$^\dag$$^\P$ \authspace
    Onur Mutlu$^\dag$$^\S$}
\affiliation{\vspace{-2pt}
	$^\dag$Carnegie Mellon University \affilspace
	$^\ddag$NVIDIA \affilspace
	$^\aleph$University of Virginia \affilspace
	$^\P$Microsoft Research \affilspace
    $^\S$ETH Z{\"u}rich\vspace{18pt}}

	\begin{abstract}

Variation has been shown to exist across the cells within a modern DRAM chip.
Prior work has studied and exploited several forms of variation, such as
manufacturing-process- or temperature-induced variation. We empirically
demonstrate a new form of variation that exists within a real DRAM chip, {\em
induced by the design and placement} of different components in the DRAM
chip\dhlii{:} different regions in DRAM, based on their relative
\dhlii{distances} from the peripheral structures, require different minimum
access latencies for reliable operation. In particular, we show that in most
real DRAM chips, cells closer to the peripheral structures can be accessed much
faster than cells that are farther. We call this phenomenon {\em design-induced
variation in DRAM}. Our goals are to {\em i)} understand design-induced
variation that exists in real\dhlii{,} state-of-the-art DRAM chips\dhlii{,} {\em
ii)} exploit it to develop low-cost mechanisms that can dynamically find and use
the {\em lowest latency at which to operate a DRAM chip reliably}\dhlii{,} and,
thus\dhlii{,} \dhlii{\em iii)} improve overall system performance while ensuring
reliable system operation.

To this end, we first experimentally demonstrate and analyze designed-induced
variation in modern DRAM devices by testing and characterizing \dimms~DIMMs (768
DRAM chips). Our characterization identifies DRAM regions that are {\em
vulnerable} to errors, if operated at lower latency, and finds consistency in
their locations across a given DRAM chip generation, due to design-induced
variation. Based on our \dhlii{extensive} experimental analysis, we develop two
mechanisms that reliably reduce DRAM latency. First, \myprofiling uses runtime
profiling to \dhlii{\emph{dynamically}} identify the lowest DRAM latency that
does not introduce failures. \myprofiling exploits design-induced variation and
periodically profiles {\em only} the \dhlii{\emph{vulnerable regions}} to
determine the lowest DRAM latency at low cost. It is the first mechanism to
\dhlii{\emph{dynamically}} determine the lowest latency that can be used to
operate DRAM \dhlii{\emph{reliably}}. \myprofiling reduces the latency of
read/write requests by 35.1\%/57.8\%, respectively, at 55\celsius. Our second
mechanism, \myshuffling, shuffles data such that values stored in vulnerable
regions are mapped to multiple error-correcting \dhlii{code} (ECC)
\dhlii{codewords}. As a result, \myshuffling can correct 26\% more multi-bit
errors than conventional ECC. \dhlii{Combined together, our two mechanisms
reduce read/write latency by} 40.0\%/60.5\%, which translates to an overall
system performance improvement of 14.7\%/13.7\%/13.8\% (in 2-/4-/8-core systems)
\dhliii{across} a variety of workloads, while ensuring reliable
operation.\\\vspace{-0.1in}

\end{abstract}

	\maketitle

    \vspace{-10pt}
	\section{Introduction} 
\label{sec:intro}

In modern systems, DRAM-based main memory is significantly slower than the
processor. Consequently, processors spend a long time waiting to access data
from main memory~\cite{mutlu-hpca2003, ailamaki-vldb1999}, making the long main
memory access latency one of the most critical bottlenecks in achieving high
performance~\cite{mutlu-imw2013, lee-hpca2015, mutlu-superfri2014}.
Unfortunately, the latency of DRAM has remained almost constant in the past
decade~\cite{lee-hpca2013, jung-2005, borkar-cacm2011, 
% samsung-spec,
patterson-commun2004, chang-sigmetrics2016, chang-thesis2017, lee-thesis2016}.
The main reason for this is that DRAM is optimized for cost-per-bit
\dhliii{(i.e., storage density)}, rather than access latency. Manufacturers
leverage technology scaling to pack more DRAM cells in the same area, thereby
enabling high DRAM density, as opposed to improving latency.

As the DRAM cell size scales to smaller technology nodes, the variation among
DRAM cells increases~\cite{kang-memforum2014}. This variation can take several
forms, such as manufacturing-process- or temperature-induced variation, and can
widen the gap between the access latencies of the fastest and the slowest
cells\sgII{~\cite{lee-hpca2015, chandrasekar-date2014, chang-sigmetrics2016,
kim-thesis2015}}. DRAM vendors do {\em not} currently exploit this variation:
instead, they use a fixed standard latency. In order to increase yield and
reduce cost, instead of discarding chips with slow cells to improve the standard
latency, vendors use a {\em pessimistic} standard latency that guarantees
correct operation for the {\em slowest} cell in {\em any} acceptable chip.

In this work, we experimentally demonstrate, analyze and take advantage of a
unique, previously-unexplored form of variation in cell latencies in real DRAM
chips. We observe that there is variation in DRAM cells' access latencies based
on their {\em physical location in the DRAM chip}. Some cells can be accessed
faster than others because they happen to be closer to peripheral structures,
e.g., sense amplifiers or wordline drivers~\cite{vogelsang-micro2010,
lee-hpca2013, keeth-book}. This phenomenon is unique: in contrast to other
commonly-known and experimentally demonstrated forms of variation, such as
manufacturing-process- or temperature-induced variation in DRAM
cells~\cite{lee-hpca2015, chandrasekar-date2014, chang-sigmetrics2016}, it is
{\em induced by the design and placement} of different components, hence
physical organization, in a real DRAM chip. Hence, we refer to this phenomenon
as {\em design-induced variation}.\footnote{Note that other
works~\cite{vogelsang-micro2010, son-isca2013, lee-hpca2013} observe that the
access latency of a cell depends on its \dhlii{distance from} the peripheral
structures, but none of these works \dhlii{characterize} or \dhliv{exploit} this
phenomenon in real DRAM chips.}

\begin{figure}[h]
	\vspace{-0.10in}
	\centering
	\includegraphics[height=1.3in]{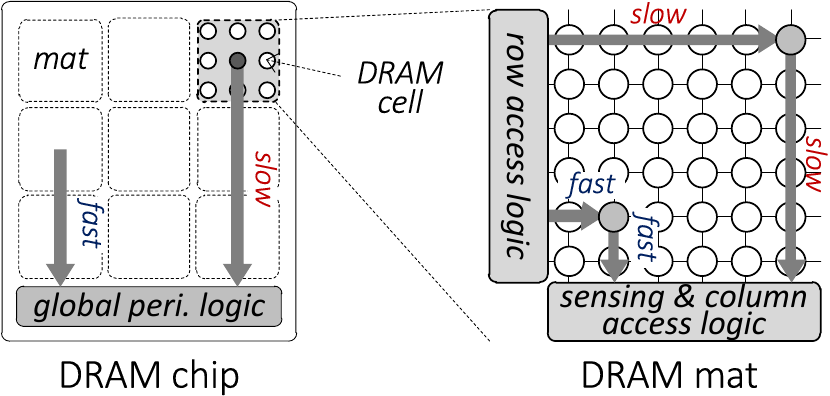}
	\vspace{-0.10in}
	\caption{Design-Induced Variation in a DRAM Chip}
	\label{fig:intro}
	\vspace{-0.10in}
\end{figure}

\sgI{Design-induced variation occurs because different cells in DRAM have
different distances between the cell and the peripheral logic used to access the
cell, as shown in Figure~\ref{fig:intro}.}
% Design-induced variation arises from the \dhlii{differences} in the
% \dhliii{distances of} \dhlii{different} cells and the peripheral logic that is
% used to access \dhlii{those} \dhliii{cells} (Figure~\ref{fig:intro}). 
The wires connecting the cells to peripheral logic exhibit large resistance and
large capacitance~\cite{lee-hpca2013, lee-hpca2015}. Consequently, cells
experience different RC delays based on their \dhliii{relative} distances from
the peripheral logic. Cells located closer to the peripheral logic experience
smaller delay and can be accessed faster than the cells located farther from the
peripheral logic.

Design-induced variation in latency is present in both vertical and horizontal
directions in a 2D DRAM cell array (called a mat): {\em i)} Each vertical {\em
column of cells} is connected to a {\em sense amplifier} and {\em ii)} each
horizontal {\em row of cells} of a mat is connected to a {\em wordline driver}.
Variations in the vertical and horizontal dimensions, together, divide the cell
array into heterogeneous latency regions, where cells in some regions require
larger access latencies for reliable operation. This variation in latency has
direct impact on the reliability of the cells. Reducing the latency {\em
uniformly across all regions} in DRAM would improve performance, but can
introduce failures in the {\em inherently slower} regions that require long
access latencies for correct operation. We refer to these inherently slower
regions of DRAM as design-induced {\em vulnerable regions}.

{\bf Our goals} are to {\em i)} experimentally demonstrate, characterize and
understand design-induced variation in modern DRAM chips, and {\em ii)} develop
new, low-cost mechanisms that leverage design-induced variation to \dhlii{\em
dynamically} find and use the lowest latency at which to operate DRAM reliably,
and thus improve overall system performance while ensuring reliable system
operation.

We first identify the {\em design-induced vulnerable regions} of \dhliii{real}
\dhlii{DRAM chips}. Doing so is not an easy task due to two major challenges.
First, {\em identifying design-induced vulnerable regions requires a detailed
knowledge of DRAM internals.} Modern DRAM cells are organized in a hierarchical
manner, where cells are subdivided into multiple mats and these mats are
organized as a matrix (Figure~\ref{fig:intro}). Due to this hierarchical
organization, the vulnerability of cells does {\em not} necessarily increase
linearly with increasing row and column addresses, but depends on {\em i)} the
location of the cell in \dhlii{the} mat and {\em ii)} the location of the mat in
the chip.

Second, {\em identifying design-induced vulnerable regions is difficult due to
the current DRAM interface that does not expose how data corresponding to an
address is mapped inside of DRAM}. Even though certain regions in DRAM might be
more vulnerable due to the design and placement of cells, internal scrambling
\dhlii{of addresses}~\cite{khan-dsn2016} and remapping of rows and
columns~\cite{liu-isca2013} scatters and distributes that region \dhlii{across}
the address space. In this work, we provide a detailed analysis on how to
identify such vulnerable regions despite the limitations posed by the modern
DRAM interface. 

To understand design-induced variation in modern DRAM chips, we build an
FPGA-based \dhlii{DRAM} testing infrastructure, similar to that used by prior
works\sgII{~\cite{lee-hpca2015, kim-isca2014, liu-isca2013, chandrasekar-date2014,
chang-sigmetrics2016, hassan-hpca2017, khan-sigmetrics2014, khan-cal2016,
chang-sigmetrics2017, khan-dsn2016, chang-thesis2017, lee-thesis2016,
kim-thesis2015}}. Our \dhlii{extensive} experimental study of \dimms~real DIMMs
(768 DRAM chips) \dhlii{using this infrastructure} shows that {\em i)} modern
DRAM chips exhibit design-induced latency variation in both row and column
directions, {\em ii)} design-induced vulnerability gradually increases in the
row direction within a mat and this pattern repeats in every mat, and {\em iii)}
some columns are more vulnerable than others due to the internal hierarchical
design of the DRAM chip.

We develop two new mechanisms that exploit design-induced variation to enable
low DRAM latency \dhliii{at high reliability and low cost}. First, we propose to
reduce the DRAM latency at runtime, by \dhlii{\em dynamically} identifying the
lowest DRAM latency that ensures reliable operation. To this end, we develop an
online DRAM testing mechanism, called {\em \myprofiling}. The key idea is to
periodically test {\em only} the regions vulnerable to design-induced variation
in order to find the minimum possible DRAM latency (for reliable operation), as
these regions would exhibit failures earlier than others when the access latency
is reduced and, therefore, would indicate the latency boundary where further
reduction in latency would hurt reliability. \myprofiling achieves this with
much lower overhead than conventional DRAM profiling mechanisms that must test
{\em all} of the DRAM cells~\cite{nair-isca2013, liu-isca2012,
venkatesan-hpca2006, khan-sigmetrics2014}. For example, for a 4GB DDR3-1600
DIMM, \myprofiling takes 1.22ms, while conventional profiling takes 625ms.

Second, to avoid uncorrectable failures (due to lower latency) in systems with
ECC, we propose \myshuffling, a mechanism to \dhlii{\em reduce multi-bit
failures} while operating at a lower latency. The key idea is to leverage the
understanding of the error characteristics of regions vulnerable to
design-induced variation in order to remap or shuffle data such that the failing
bits get spread over multiple ECC \dhliii{codewords} and \dhlii{thereby} become
correctable by ECC.

We make the following {\bf contributions}:

\squishlist

	\item To our knowledge, this is the first work to experimentally demonstrate,
	characterize and analyze the phenomenon of design-induced variation that
	exists in real\dhlii{,} state-of-the-art DRAM chips. Due to this phenomenon,
	when DRAM latency is reduced, we find that certain regions of DRAM are more
	vulnerable to failures than others, based on their relative distances from the
	peripheral logic.

	\item We identify the regions in DRAM that are most vulnerable to
	design-induced variation based on the internal hierarchical organization of
	DRAM bitlines and wordline drivers. We experimentally demonstrate the
	existence of design-induced vulnerable regions in DRAM by testing and
	characterizing \dimms~real DIMMs (768 DRAM chips).

	\item We develop two new mechanisms, called \myprofiling and \myshuffling,
	which exploit design-induced variation to improve both latency and reliability
	of DRAM at low cost. \myprofiling is the first mechanism to dynamically
	determine the lowest latency at which to operate DRAM reliably: it \dhlii{\em
	dynamically} reduces the latencies of read/write operations by 35.1\%/57.8\%
	at 55\celsius, while ensuring reliable operation. \myshuffling is the first
	mechanism that takes advantage of design-induced variation to improve
	reliability by making ECC more effective: on average, it corrects 26\% of
	total errors that are {\em not} correctable by conventional ECC\dhliii{, while
	operating at lower latency}. We show that the combination of our \dhlii{two}
	techniques, \mydram, leads to a raw DRAM latency reduction of 40.0\%/60.5\%
	(read/write) and an overall system performance improvement of
	14.7\%/13.7\%/13.8\% (2-/4-/8-core) over a variety of workloads in our
	evaluated systems, while ensuring reliable system operation. We also show that
	\mydram outperforms Adaptive-Latency DRAM (AL-DRAM)~\cite{lee-hpca2015}, a
	state-of-the-art technique that lowers DRAM latency by exploiting temperature
	and process variation (but \dhlii{\em not} designed-induced
	variation).\footnote{A second important benefit \dhlii{of DIVA-DRAM} over
	AL-DRAM is that \mydram is {\em not vulnerable} to changes in \dhliii{DRAM}
	latency characteristics over time due to issues such as aging and wearout,
	since \mydram determines latency {\em dynamically} based on runtime profiling
	of latency characteristics. As AL-DRAM does {\em not} determine latency
	dynamically and instead relies on \dhlii{\em static} latency parameters, it is
	vulnerable to dynamic changes in latency characteristics, which leads to
	either \dhlii{potential} reliability problems or large latency margins to
	prevent \dhlii{potential} failures. See Section~\ref{sec:mech_lowlatency} for
	a more detailed discussion of this.}

\squishend

	\section{Modern DRAM Architecture} 
\label{sec:background}

We first provide background on DRAM organization and operation that is useful to
understand the cause, characteristics and implications of {\em design-induced
variation}.

\subsection{DRAM Organization} 
\label{sec:dram}

\begin{figure}[h]
	\vspace{-0.2in}
	\centering
	\subfloat[Chip (8 banks)] {
		\includegraphics[height=1.1in]{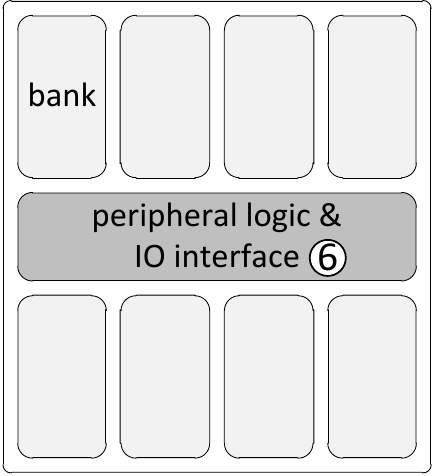}
		\label{fig:dram_chip}
	}
%	\hspace{0.01in}
	\subfloat[Bank] {
		\includegraphics[height=1.1in]{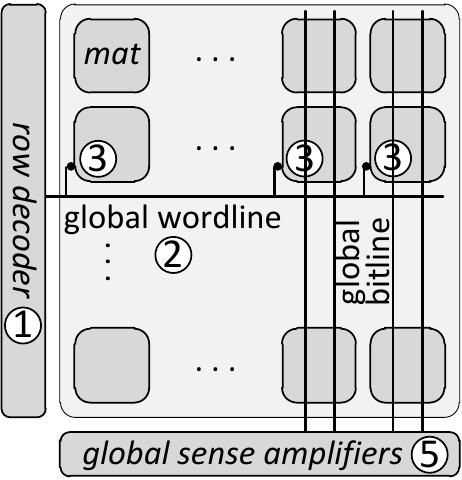}
		\label{fig:dram_bank}
	}
%	\hspace{0.01in}
	\subfloat[Mat (cell array)] {
		\includegraphics[height=1.1in]{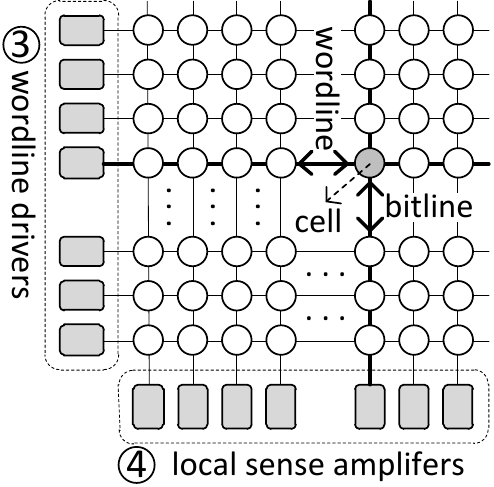}
		\label{fig:dram_mat}
	}
	\vspace{-0.1in}
	\caption{Hierarchical Organization of a DRAM System}
	\label{fig:dram_org}
	\vspace{-0.15in}
\end{figure}

DRAM is organized in a hierarchical manner where each DIMM consists of multiple
chips, banks, and mats\dhlii{,} as shown in Figure~\ref{fig:dram_org}. A DRAM
{\em chip} (shown in Figure~\ref{fig:dram_chip}) consists of {\em i)} multiple
banks and {\em ii)} peripheral logic that is used to transfer data to the memory
channel through the IO interface. Each {\em bank} (shown in
Figure~\ref{fig:dram_bank}) is subdivided into multiple {\em mats}. In a bank,
there are two global components that are used to access the mats: {\em i)} a
{\em row decoder} that selects a row of cells {\em across} multiple mats and
{\em ii)} {\em global sense amplifiers} that transfer a fraction of data from
the row through the global bitlines, based on the column address.
Figure~\ref{fig:dram_mat} shows the organization of a {\em mat} that consists of
three components: {\em i)} a 2-D cell array in which the cells in each row are
connected horizontally by a wordline, and the cells in each column are connected
vertically by a bitline, {\em ii)} a column of wordline drivers that drive each
wordline to appropriate voltage levels in order to activate a row during an
access and {\em iii)} a row of {\em local sense amplifiers} that sense and latch
data from the activated row.

\subsection{DRAM Operation} 
\label{sec:dram_access}

On a memory request (e.g., to read a cache line), there are two major steps
involved in accessing the requested data. First, to access a row, the memory
controller issues an \cmdact command along with the row address to select a row
in a bank. On receiving this command, DRAM transfers all the data in the row to
the corresponding local sense amplifiers. Second, in order to access a specific
cache line from the activated row, the memory controller issues a \cmdread
command with the column address of the request. DRAM then transfers the selected
data from the local sense amplifiers to the memory controller, over the memory
channel.

While this is a high-level description of the two major DRAM operations, these
operations, in reality, consist of two levels of accesses through: {\em i)}
global structures across mats within a bank (global sense amplifiers, global
wordlines, and global bitlines) and {\em ii)} local structures within a mat
(local sense amplifiers, local wordlines, and local bitlines). A row-column
access goes through multiple steps in the global-local hierarchy\dhlii{, as
annotated in Figure}~\ref{fig:dram_org}: \ding{172} When the row decoder in a
bank receives a row address, it first activates the corresponding global
wordline in the bank. \ding{173} The global wordline, in turn, activates the
corresponding wordline driver in each mat that it is connected to. \ding{174}
The wordline driver in each mat activates the corresponding local wordline
connecting the row to the local sense amplifiers. \ding{175} These local
amplifiers sense and latch the entire row through the local bitlines in each mat
across the bank. \ding{176} When DRAM receives the column address, a fraction of
data from each mat is transferred from the local sense amplifiers to the global
sense amplifiers, through the global bitlines. \ding{177} Data from the global
sense amplifiers is then sent to the memory channel through the IO interfaces of
the DRAM chip.

Both DRAM row and column accesses are managed by issuing row and column access
commands to DRAM. The minimum time between these commands is determined by
\dhlii{internal DRAM} operation considerations\dhlii{,} such as how long it
takes to sense data from cells in a selected wordline, how long it takes to
transfer data from the local to the global sense amplifiers~\cite{micron-dram,
lee-hpca2015, lee-hpca2013, kim-isca2012}. There are four major \dhlii{\em
timing parameters} for managing row and column accesses. \tras (\trp) is the
minimum time needed to select (deselect) a row in a bank for activation. \trcd
is the minimum time needed to access a column of a row after activating the
row\dhlii{.} \twr is the minimum time needed to update the data in a column of a
row after activating the row. \dhlii{More detailed information on these timing
parameters and DRAM operation can be found in}~\cite{kim-isca2012, lee-hpca2015,
lee-hpca2013, chang-sigmetrics2016}.

	\section{Design-Induced Variation} 
\label{sec:arch}

In this work, we show that DRAM access latency varies based on the location of
the cells in the DRAM hierarchy. Intuitively, transferring data from the cells
near the IO interfaces \dhlii{(and sensing structures)} incurs less time than
transferring data from the cells farther away from the IO interfaces \dhlii{(and
sensing structures)}. We refer to this variability in cell latency caused by the
physical organization and design of DRAM as {\em design-induced variation}.
\dhlii{Since} DRAM is organized as a multi-level hierarchy (in the form of
chips, banks and mats), design-induced variation exists at multiple levels.
Design-induced variation has \dhlii{several} specific characteristics that
clearly distinguish it from other known types of variation observed in
DRAM\dhliii{,} e.g., process variation and temperature
dependency~\cite{lee-hpca2015, chandrasekar-date2014}\dhlii{:}

\squishlist

	\item {\bf Predetermined at design time.} Design-induced variation depends on
	the internal DRAM design, predetermined at {\em design time}. This is unlike
	other types of variation, (e.g., process variation and temperature induced
	variation~\cite{lee-hpca2015, chandrasekar-date2014}), which depend on the
	manufacturing process \dhlii{and operating conditions} after design.

	\item {\bf Static distribution.} The distribution of design-induced variation
	is static\dhlii{, determined by the location of cells.} For example, a cell
	closer to the sense amplifier is {\em always} faster than a cell farther away
	from the sense amplifier, assuming there are no other sources of variation
	(e.g., process variation). On the other hand, prior works show that
	variability due to process variation follows a \dhlii{\em random}
	distribution~\cite{lee-hpca2015, chandrasekar-date2014}, \dhlii{independent of
	the location of cells}.

	\item {\bf Constant.} Design-induced variation depends on the physical
	organization, which remains constant over time. Therefore, it is different
	from other types of variation that change over time (e.g., variable retention
    time\sgII{~\cite{liu-isca2013, khan-sigmetrics2014, kim-edl2009, qureshi-dsn2015,
    patel-isca2017, yaney-iedm1987, restle-iedm1992, mori-iedm2005}}, wearout due
	to aging~\cite{sridharan-sc2012, sridharan-sc2013, meza-dsn2015,
	schroeder-tdsc2010, li-atc2007, hwang-asplos2012, wee-jssc2000, min-vlsi2001,
	tanabe-jssc1992}).

	\item {\bf Similarity in DRAMs with the same design.} DRAMs that share the
	same internal design exhibit similar design-induced variation
	(Section~\ref{sec:profile_rowint}). \dhlii{This is} unlike process variation
	that manifests \dhlii{itself} significantly differently in different DRAM
	chips with the same design.

\squishend

{\bf The goals} of this work are to {\em i)} experimentally demonstrate,
characterize, and understand the design-induced variation in real DRAM chips,
especially within and across mats, and {\em ii)} leverage this variation and our
understanding of it to reduce DRAM latency at low cost in a reliable way.
Unfortunately, detecting the design-induced vulnerable regions is not trivial
and depends on two factors: {\em i)} how bitline and wordline drivers are
organized internally, {\em ii)} how data from a cell is accessed through the
DRAM interface. In order to define and understand the design-induced variation
in modern DRAM, we investigate three major research questions related to the
impact of DRAM \dhliii{{\em organization}, {\em interface}, and {\em operating
conditions}} on design-induced variation in the following sections.

\subsection{Impact of DRAM Organization} 
\label{sec:overview_existence}

The first question we answer is: {\em how does the DRAM organization affect the
design-induced vulnerable regions?} To answer this, we present \dhlii{\em i)}
the expected characteristics of design-induced variation and \dhlii{\em ii)}
systematic methodologies to identify these characteristics in DRAM chips.

\begin{figure}[h]
	\vspace{-0.10in}
	\centering
	\subfloat[Conceptual Bitline] {
		\includegraphics[width=.48\linewidth]{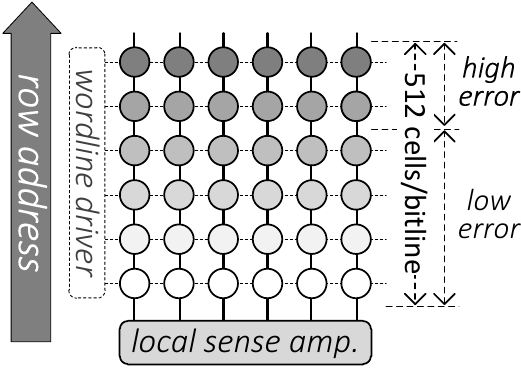}
		\label{fig:arch_row_concept}
	}
	\subfloat[Open Bitline Scheme] {
		\includegraphics[width=.48\linewidth]{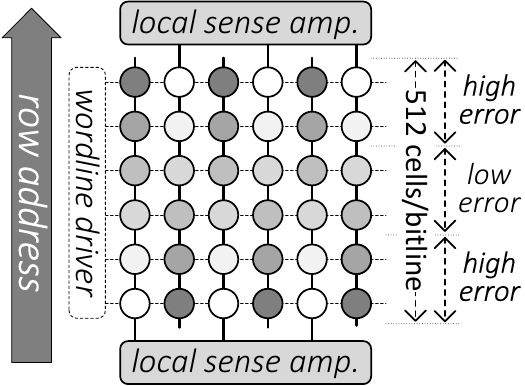}
		\label{fig:arch_row_real}
	}
	\vspace{-0.05in}
	\caption{Design-Induced Variation Due to Row Organization}
	\label{fig:arch_row}
	\vspace{-0.10in}
\end{figure}

{\bf Effect of Row Organization.} As discussed in Section~\ref{sec:dram}, a mat
consists of a 2D array of DRAM cells along with peripheral logic needed to
access this data. In the vertical direction, DRAM cells (typically, 512
cells~\cite{vogelsang-micro2010, kim-isca2012}), connected through a bitline,
share a local sense amplifier. As a result, access latency gradually increases
as the distance of a row from the local sense amplifier increases (due to the
longer propagation delay through the bitline). This variation can be exposed by
\dhlii{reading data from DRAM faster} by using smaller values for DRAM timing
parameters. Cells in the rows closer to the local sense amplifiers can be
accessed faster \dhlii{in a reliable manner}. Hence, they exhibit no failures
due to \dhlii{shorter timing parameters}. On the contrary, cells located farther
away from the sense \dhliii{amplifiers} take longer to access \dhlii{in a
reliable manner}, and might start failing when smaller values are used for the
timing parameters. As a result, accessing rows in ascending order starting from
the row closest to the sense amplifiers should exhibit a gradual increase in
failures due to design-induced variation, as shown in
Figure~\ref{fig:arch_row_concept}. In this figure, the \dhliii{\em darker}
color indicates \dhliii{\em slower} cells, which are \dhliii{\em more
vulnerable} to failures when we reduce the access latency.

In the open-bitline scheme~\cite{inoue-jssc1988}, alternate bitlines within a
mat are connected to two different rows of sense amplifiers (\dhlii{at} the top
and \dhlii{at the} bottom of the mat), as shown in
Figure~\ref{fig:arch_row_real}. In this scheme, even cells and odd cells in a
row located at the edge of the mat exhibit very different distances from their
corresponding sense amplifiers, leading to different access latencies. On the
other hand, cells in the middle of a mat have \dhlii{a} similar distance from
both the top and bottom sense amplifiers, exhibiting similar latencies. Due to
this organization, we observe that there are more failures in rows located on
both ends of a mat, but there is a gradual decrease in failures in rows in the
middle of the mat.

Based on these \dhlii{observations about row organization}, we define two
\dhlii{expected} characteristics of vulnerable regions across the rows when we
reduce DRAM latency uniformly. First, {\bf the number of failures would
gradually increase with increased distance from the sense amplifiers}. Second,
{\bf this gradual increase in failures would periodically repeat in every mat
(every 512 rows)}. We experimentally demonstrate these characteristics in
Section~\ref{sec:profile_bitline}.

\begin{figure}[h]
	%\vspace{0.05in}
	\centering
	\includegraphics[width=\linewidth]{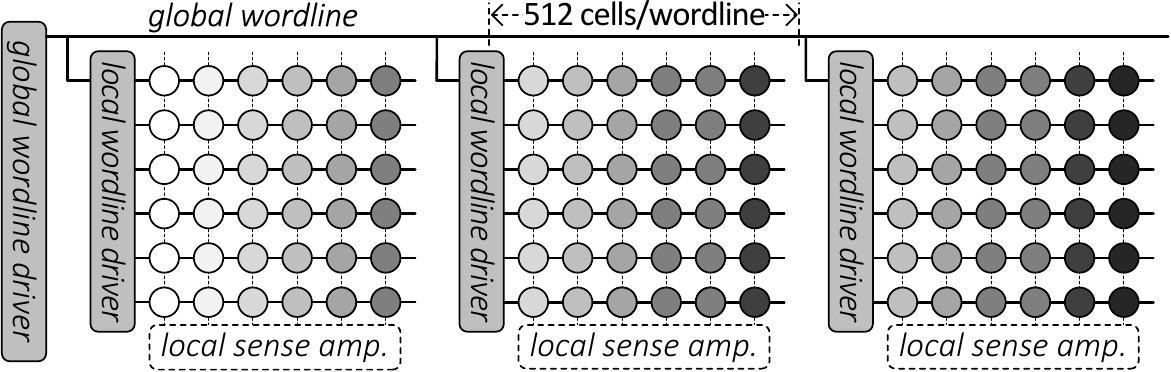}
	\vspace{-0.15in}
	\caption{Design-Induced Variation in Column Organization}
	\label{fig:arch_col}
	\vspace{-0.10in}
\end{figure}

{\bf Effect of Column Organization.} As we discussed in
Section~\ref{sec:dram_access}, the wordline drivers in DRAM are organized in a
hierarchical manner\dhlii{:} a \dhlii{strong} global wordline \dhlii{driver} is
connected to {\em all} mats over which a row is distributed and a local wordline
driver activates a row within a mat. This {\em hierarchical wordline
organization} leads to latency variation at two levels. First, a \dhlii{local}
wordline in a mat located closer to the global wordline driver starts activating
the row earlier than \dhlii{that in} a mat located farther away from the global
wordline driver ({\em design-induced variation due to the global wordline}).
Second, within a mat, a cell closer to the local wordline driver gets activated
faster than a cell farther away from the local wordline driver ({\em
design-induced variation due to the local wordline}). Therefore, columns that
have the same distance from the local wordline driver, but are located in two
different mats, have different latency characteristics (\dhlii{see}
Figure~\ref{fig:arch_col}, where a darker color indicates slower cells, which
are more vulnerable to failures if/when we reduce \dhlii{the} access latency).
\dhlii{However, exact latency characteristics of different columns in different
mats depend on the strength of the global versus local wordline drivers and the
location of the respective mats and columns.}

We define two \dhlii{expected} characteristics of vulnerable regions across
columns when we reduce DRAM latency uniformly. First, {\bf although some columns
are \dhlii{clearly} more vulnerable than others, the number of failures
\dhlii{likely} would {\em not} gradually increase with ascending column
numbers}. Second, {\bf the failure characteristics observed with ascending
column numbers would be similar for all rows}. We experimentally demonstrate
these characteristics in Section~\ref{sec:profile_wordline}.

\subsection{Impact of the Row/Column Interface} 
\label{sec:overview_interface}

Our second question is: {\em how does the row/column interface affect the
ability to identify the design-induced vulnerable regions in DRAM?}
Unfortunately, identifying design-induced vulnerable regions becomes challenging
due to a limited understanding of how data corresponding to an address is mapped
inside DRAM. While it is possible to identify vulnerable regions based on
location, exposing and exploiting such information through the row/column DRAM
addressing interface is challenging due to two reasons.

{\bf Row Interface (Row Address Mapping).} DRAM vendors internally \dhliii{\em
scramble} the row addresses in DRAM\dhlii{.} \dhlii{This causes} the address
known to the system \dhlii{to be} different from the actual physical
address~\cite{vandegoor-delta2002, khan-dsn2016, liu-isca2013}. As a result,
consecutive row addresses issued by the memory controller can be mapped to
entirely different regions of DRAM. Unfortunately, the \dhlii{internal} mapping
of the row addresses is not {\em exposed} to the system and varies across
products from different generations and manufacturers. \dhliii{In
Section}~\ref{sec:overview_existence}, we showed that if the access latency is
reduced, accessing rows in \dhlii{a mat} in ascending row number order would
exhibit a gradual increase in failures. Unfortunately, due to row remapping,
accessing rows in ascending order of {\em addresses known to the memory
controller} will \dhlii{likely} exhibit irregular and scattered failure
characteristics.

\begin{figure}[h]
	\vspace{-0.07in}
	\centering
	\subfloat[Data Mapping] {
		\includegraphics[height=1.3in]{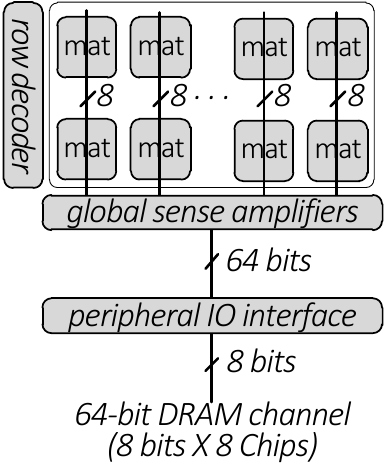}
		\label{fig:arch_col_data_mapping}
	}
	\subfloat[Data Burst (Data Out)] {
		\includegraphics[height=1.3in]{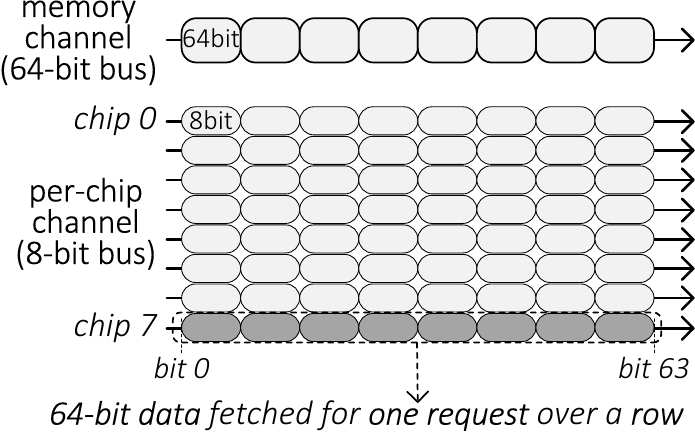}
		\label{fig:arch_col_burst}
	}
	\vspace{-0.10in}
	\caption{Accessing Multiple Mats in a Data Burst}
	\label{fig:arch_col_interface}
	\vspace{-0.05in}
\end{figure}

{\bf Column Interface (Column Address Mapping).} In the current interface, the
bits accessed by a column command are \dhlii{\em not} mapped to consecutive
columns in a mat. This makes it challenging to identify the vulnerable regions
in a wordline. When a column address is issued, 64 bits of data from a row are
transferred over the global bitlines (typically, 64-bit
\dhlii{wide}~\cite{vogelsang-micro2010}). This data is transferred in eight
8-bit bursts over the IO channel\dhlii{,} as shown in
Figure~\ref{fig:arch_col_interface}. 
\sgI{However, the data transferred with each column address comes from cells
that \dhliv{are} in different mats, and have different distances from their
global and local wordline drivers. This makes it impossible to determine the
physical column organization by simply sweeping the column address in
\emph{ascending} order.}
% \dhliii{However, data transferred with each column address comes from cells that
% are different mats and have different distances from their global wordline
% drivers and local wordline drivers, making it impossible to determine the
% physical column organization by simply sweeping column accesses with {\em
% increasing} the column address.}

% However, data transferred with each column address comes from different mats,
% making it impossible to always access {\em consecutive} physical columns in a
% mat by simply {\em increasing} the column address.

In this work, we provide alternate ways to identify design-induced vulnerable
regions using the current row/column interface in DRAM. We describe the key
ideas of our methods.

\squishlist

	\item {\em Inferring vulnerable rows from per-row failure count.} In order to
	identify the gradual increase in design-induced variability with increasing
	row addresses in mats (in terms of internal DRAM physical address), we try to
	reverse engineer the row mapping in DRAM. We hypothesize the mapping for one
	mat and then verify that mapping in other DRAM mats in different chips that
	share the same design. The key idea is to correlate the number of failures to
	the physical location of the row. For example, the most vulnerable row would
	be the one with the most failures and hence should be located at the edge of
	the mat. Section~\ref{sec:profile_rowint} provides experimental
	\dhlii{analysis and} validation of our method.

	\item {\em Inferring vulnerable columns from per-bit failure count in the IO
	channel.} A column access transfers 64 bits of data from a DRAM chip over the
	IO channel. These 64 bits come from 64 bitlines that are distributed over
	different mats across the entire row. Our key idea to identify the vulnerable
	bitlines in the column direction is to examine each bit in a 64-bit burst. We
	expect that due to design-induced variation, some bits in a 64-bit burst that
	are mapped to \dhlii{relatively slow} bitlines are more vulnerable than other
	bits. In Section~\ref{sec:profile_colint}, we experimentally identify the
	location of bits in bursts that consistently exhibit more failures, validating
	the existence of design-induced variation in columns.

\squishend

\subsection{Impact of Operating Conditions} 
\label{sec:overview_sensitivity}

The third question we answer is: {\em Does design-induced variation in latency
show similar characteristics at different operating conditions?} DRAM cells get
affected by temperature and the refresh interval~\cite{lee-hpca2015,
liu-isca2013, khan-sigmetrics2014, patel-isca2017}. Increasing the temperature
within the normal system operating range (45\celsius~to 85\celsius) or
increasing the refresh interval increases the leakage in cells, making them more
vulnerable to failure. However, as cells get similarly affected by changes in
operating conditions, we observe that the {\em trends} due to design-induced
variation remain similar at different temperatures and refresh intervals, even
though the absolute number of failures may change. We provide detailed
experimental analysis of design-induced variation at different operating
conditions, in Section~\ref{sec:profile_sensitivity}.

	\section{DRAM Testing Methodology}
\label{sec:pmethod}

In this section, we describe our FPGA-based DRAM testing infrastructure and the
testing methodology we use for our experimental studies in
Section~\ref{sec:profile}.

{\bf FPGA-Based DRAM Testing Infrastructure.} We build \dhlii{an} infrastructure
similar to that used in previous works\sgII{~\cite{lee-hpca2015, kim-isca2014,
liu-isca2013, chandrasekar-date2014, chang-sigmetrics2016, hassan-hpca2017,
khan-dsn2016, khan-sigmetrics2014, chang-sigmetrics2017, khan-cal2016,
kim-thesis2015, lee-thesis2016, chang-thesis2017}}. Our infrastructure provides
the ability to: {\em i)} generate test patterns with flexible DRAM timing
parameters, {\em ii)} provide an interface from a host machine to the FPGA test
infrastructure, and {\em iii)} maintain a stable DRAM operating temperature
during experiments. We use a Xilinx ML605 board~\cite{ml605} that includes an
FPGA-based memory controller connected to a DDR3 SODIMM socket. We designed the
memory controller~\cite{mig} with the flexibility to change DRAM parameters. We
connect this FPGA board to the host machine through \dhliii{the} \sgI{PCIe}
interface~\cite{pcie}. We manage the FPGA board from the host machine and
preserve the test results in the host machine's storage. In order to maintain a
stable operating temperature for the DIMMs, during our experiments, we place the
FPGA board in a heat chamber that consists of a temperature controller, a
temperature sensor, and a heater which enables us to test at different
temperatures.

{\bf Profiling Methodology.} The major purpose of our experiments is to
characterize design-induced variation in \dhlii{DRAM} latency. We would like to
{\em i)} determine the characteristics of failures when we reduce timing
parameters beyond the error-free operation region, and {\em ii)} observe any
correlation between the error characteristics and the internal design of the
tested DRAMs. To this end, we analyze the error characteristics of DRAM by
lowering DRAM timing parameters below the values specified for error-free
operation.

An experiment consists of three steps: {\em i) writing background data}, {\em
ii) changing timing parameters}, and {\em iii) verifying cell content}. In Step
1, we write a certain data pattern to the entire DIMM with standard DRAM timing
parameters, ensuring that correct (\dhlii{i.e.,} the intended) data is written
into all cells. In Step 2, we change the timing parameters. In Step 3, we verify
the content of the DRAM cells after the timing parameters are changed. To pass
verification, a DRAM cell must maintain its data value until the next refresh
operation. To complete the verification step, \dhlii{we let DRAM cells remain
idle and leak charge for the {\em refresh interval} and read and verify the
data.} If the data read in Step 3 does not match the data written in Step 1, we
log the addresses corresponding to the \dhlii{failures} and \dhlii{the failed}
bits in the failed \dhlii{addresses}.

% An experiment consists of three steps: {\em i) writing background data}, {\em
% ii) changing timing parameters}, and {\em iii) verifying cell content}. In
% Step 1, we write a certain data pattern to the entire DIMM with standard DRAM
% timing parameters, ensuring that correct (\dhlii{i.e.,} the intended) data is
% written into all cells. In Step 2, we change the timing parameters. In Step 3,
% we verify the content of the DRAM cells after the timing parameters are
% changed. To pass verification, a DRAM cell must maintain its data value until
% the next refresh operation. To complete the verification step, we let DRAM
% cells remain idle and leak charge for the {\em refresh interval} after writing
% background data and only after that read and verify the data. If the data read
% in Step 3 does not match the data written in Step 1, we log the addresses
% corresponding to the \dhlii{failures} and \dhlii{the failed} bits in the
% failed \dhlii{addresses}.

{\bf Data Patterns.} In order to exercise worst-case latency behavior, we use a
row stripe pattern, wherein a test pattern is written in odd rows and an
inverted test pattern is written in even rows~\cite{vandegoor-delta2002,
kim-isca2014}. This pattern drives the bitlines in opposite directions when
accessing adjacent rows. The patterns we have used in our tests are {\tt 0000},
{\tt 0101}, {\tt 0011}, and {\tt 1001}. We perform the test twice per pattern,
once with the test data pattern and once with the inverted version of the test
data pattern, in order to test every cell in charged (e.g., data 1) and
non-charged states (e.g., data 0). We report the sum of failures from these two
cases for each test. We perform 10 iterations of the same test to make sure the
errors are consistent.

We evaluate four DRAM timing parameters: \trcd, \tras, \trp, and \twr. For each
timing parameter, our evaluations start from the \dhlii{standard values}
(13.75/35.0/13.75/15.0ns for \trcd/\tras/\trp/\twr,
respectively)~\cite{micron-dram} and reduce the timing parameters to the lowest
values that our DRAM infrastructure allows (5ns for \trcd/\tras/\twr, and \trcd
+ 10ns for \tras). We use \dimms~DIMMs, comprising 768 DRAM chips, from three
DRAM vendors for our experiments. Appendix~\ref{sec:appendix_summary} lists
evaluated DIMMs and their major characteristics. We \dhlii{provide} detailed
results for each DIMM online~\cite{safari-git}.

	\section{Characterization of Design-Induced Variation in DRAM} 
\label{sec:profile}

In this section, we present the results of our profiling studies that
demonstrate the presence of design-induced variation in both the vertical
(bitline) and horizontal (wordline) directions. We {\em i)}~show the existence
of design-induced variation in Sections~\ref{sec:profile_bitline}
and~\ref{sec:profile_wordline}, {\em ii)}~analyze the impact of the row and
column interface in Sections~\ref{sec:profile_rowint}
and~\ref{sec:profile_colint}, and {\em iii)}~characterize the impact of
operating conditions on design-induced variation in
Section~\ref{sec:profile_sensitivity}. We then provide a summary of our analysis
on design-induced variation across 96 DIMMs (768 DRAM chips) in
Section~\ref{sec:profile_dimms}. In Appendix~B, we present the results of our
\dhlii{supporting} circuit-level SPICE simulation studies that \dhlii{validate
our hypotheses on} design-induced variation in a mat.

\subsection{Design-Induced Variation in Bitlines} 
\label{sec:profile_bitline}

As we explain in Section~\ref{sec:overview_existence}, we expect different error
characteristics for \dhlii{different} cells connected to a bitline, depending on
the \dhliii{relative distances} \dhlii{of the cells} from the local sense
amplifiers. To demonstrate the existence of design-induced variation in a
bitline, we design a test pattern that sweeps the row address.

\begin{figure}[h]
	\vspace{-0.15in}
	\centering
	\subfloat[\trp 12.5ns] {
		\includegraphics[height=1.0in]{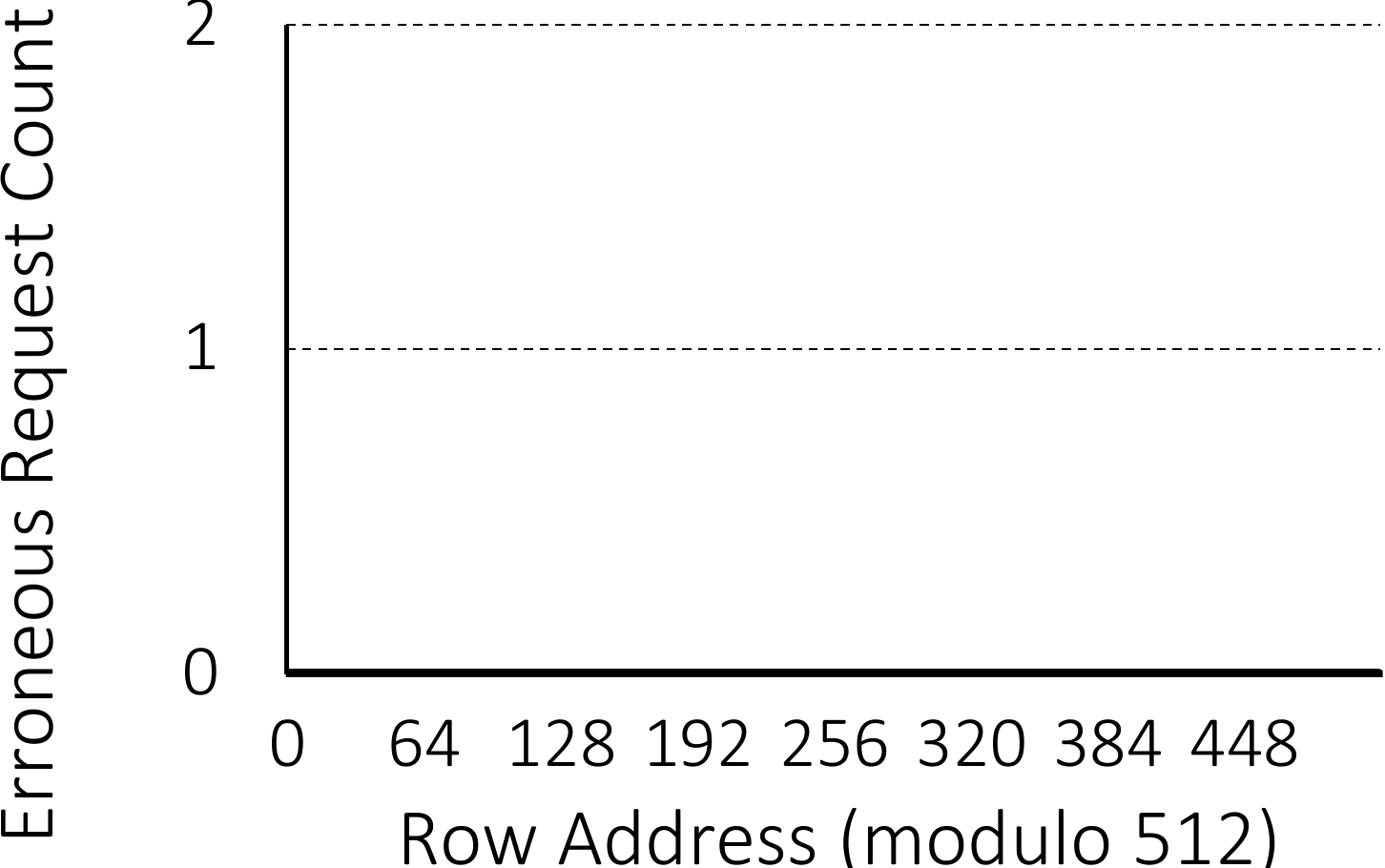}
		\label{fig:profile_row_125ns}
	}
	\subfloat[\trp 10.0ns] {
		\includegraphics[height=1.0in]{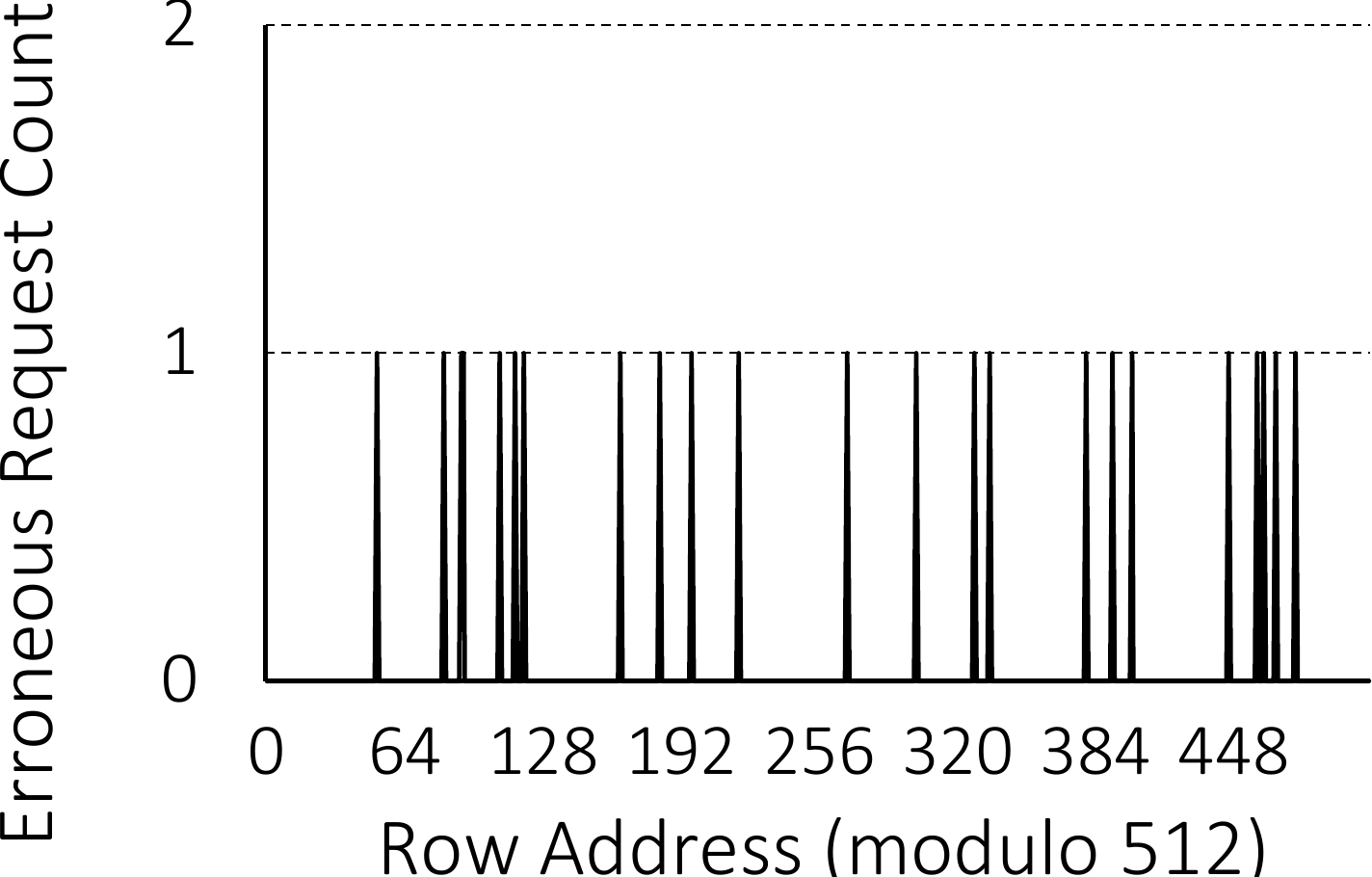}
		\label{fig:profile_row_10ns}
	}

	\vspace{-0.05in}
	\subfloat[\trp 7.5ns] {
		\includegraphics[height=1.0in]{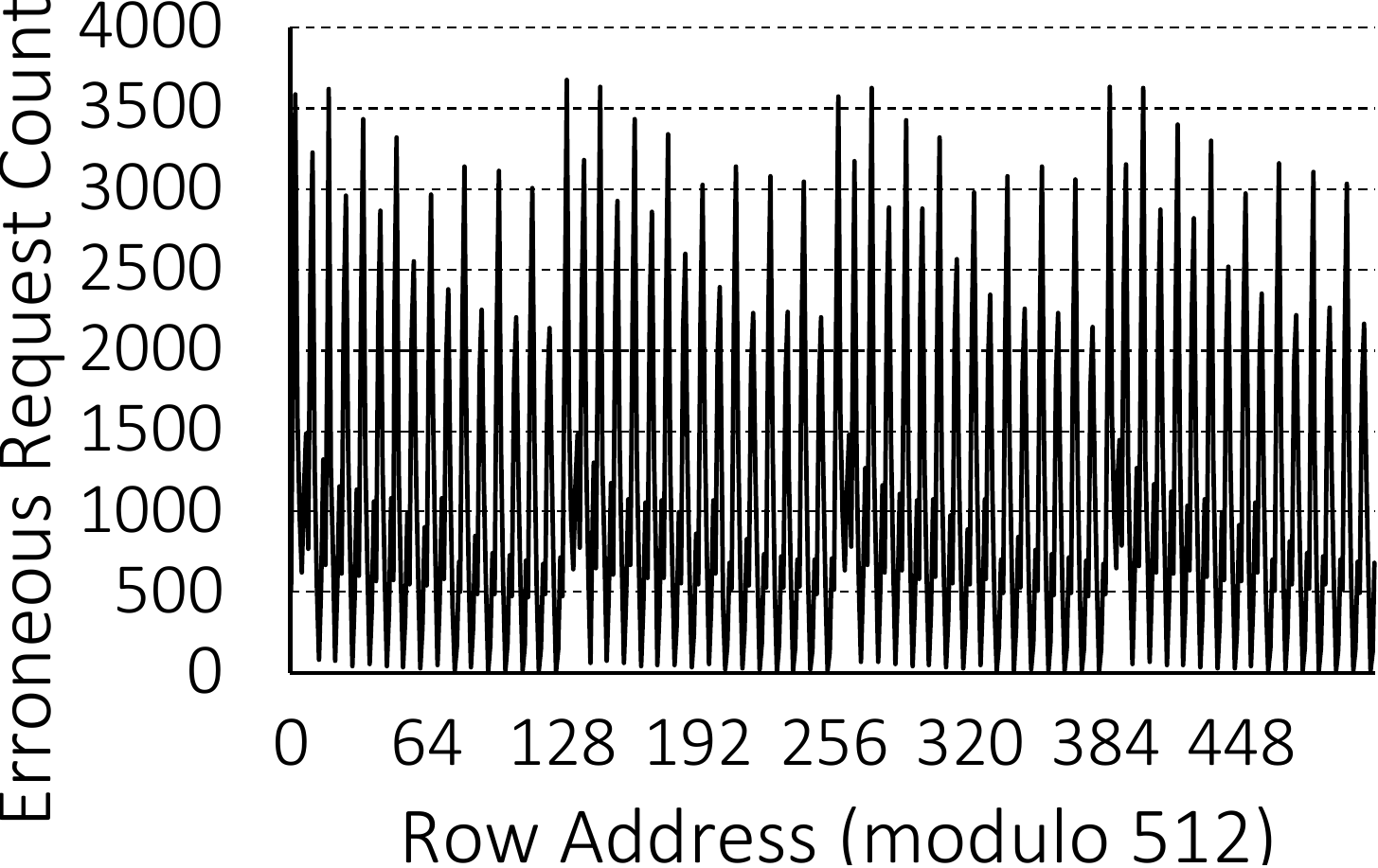}
		\label{fig:profile_row_75ns}
	}
	\subfloat[\trp 5.0ns] {
		\includegraphics[height=1.0in]{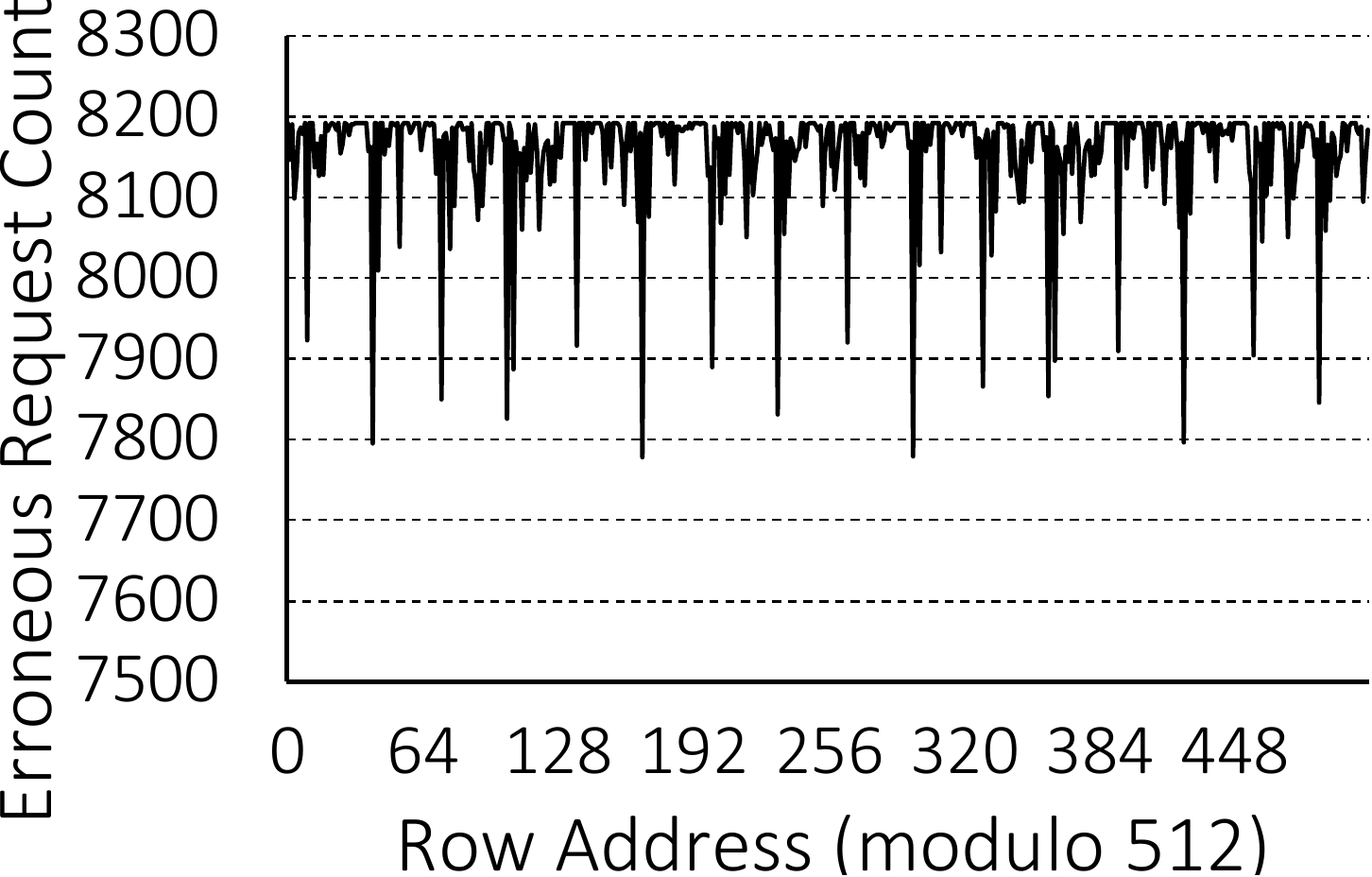}
		\label{fig:profile_row_5ns}
	}
	\vspace{-0.05in}
	\caption{Erroneous Request Count When Sweeping Row Addresses with \dhlii{a}
	Reduced \trp Timing Parameter}
	\label{fig:profile_row_trp}
	\vspace{-0.10in}
\end{figure}

{\bf Per-Row Error Count with Row Address Sweeping.}
Figure~\ref{fig:profile_row_trp} plots the error count for four values of a DRAM
timing parameter, \trp (whose standard value is 13.75ns), with a refresh
interval of 256 ms (greater than the normal 64 ms refresh interval to emphasize
the effects of access latency~\cite{lee-hpca2015}) and an ambient temperature of
85\celsius. We tested all rows (and 16 columns) in a DIMM and plot the number of
erroneous accesses for \dhliii{each} \dhlii{set of {\em row address modulo 512}
rows}.\footnote{Even though there are redundant cells (rows), DRAM does
\dhlii{\em not} allow direct access to redundant cells. Therefore, we can only
access a 512$\times$512 cell mat ($2^n$ data chunk).
Figure~\ref{fig:profile_row_trp} plots the number of erroneous requests in every
\dhlii{512-cell} chunk.} We \dhlii{aggregate the error count across} errors
\dhlii{every set of {\em row address modulo 512} rows} because each bitline is
connected to 512 cells. Hence, our expectation is that the design-induced
variation pattern will repeat every 512 cells.\footnote{Note that Figure~6 shows
the {\em sum} of all error counts for {\em all} rows with the same \dhlii{\em
row number modulo 512}. In other words, each value on the x-axis of Figure~6c
represents a modulo value {\em i}, where \dhlii{the corresponding y-axis value}
shows the \dhlii{aggregated} number of errors for \dhlii{the set of rows --}
Row~\emph{i}, Row 512+\emph{i}, Row 1024+\emph{i}, etc. We provide each
\dhlii{individual} row's error count in Figure 7b to substantiate this further.}
We \dhlii{make} two key observations. First, reducing a timing parameter
\dhlii{enough} below its standard value induces errors, and reducing it further
induces more errors. \dhlii{At a \trp of 12.5ns, there are no errors, due to the
latency margin that exists in DRAM cells, as shown in previous
works}~\cite{lee-hpca2015, chang-sigmetrics2016}. At a \trp of 10.0ns (3.75ns
reduction from the standard value), the number of errors is small, as shown in
Figure~\ref{fig:profile_row_10ns} while at a \trp of 5.0ns, we observe a large
number of errors, as shown in Figure~\ref{fig:profile_row_5ns}. Second, we
observe \dhliii{significant} error count variation across 512 row chunks only at
7.5ns (with error counts ranging from 0 to more than 3500 in
Figure~\ref{fig:profile_row_75ns}), while most errors are {\em randomly}
distributed at 10.0ns (Figure~\ref{fig:profile_row_10ns}) and most rows show
very high error counts at 5.0ns (Figure~\ref{fig:profile_row_5ns}).

\begin{figure}[h]
	\vspace{0.05in}
	\centering
	\subfloat[\sgI{Erroneous Requests Aggregated Across All Row Chunks, with the Rows in a Row Chunk Sorted by Ascending Error Count}] {
		\includegraphics[height=1.05in]{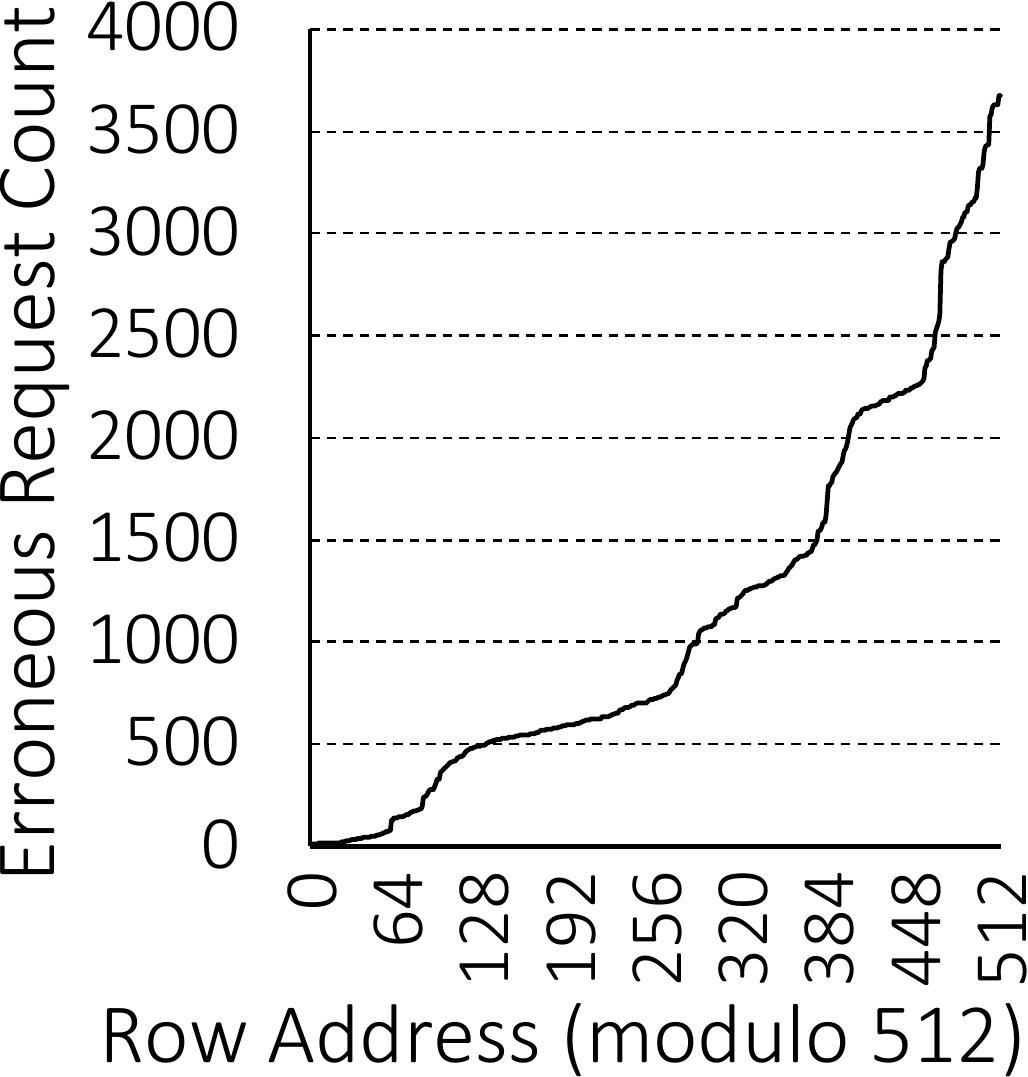}
		\label{fig:profile_row_sort}
	}\hspace{0.05in}
	\subfloat[\dhlii{Erroneous Requests of Individual \dhliii{Rows}, \changes{Sorted Using the Row Ordering from} Figure}~\ref{fig:profile_row_sort}]{
		\includegraphics[height=1.05in]{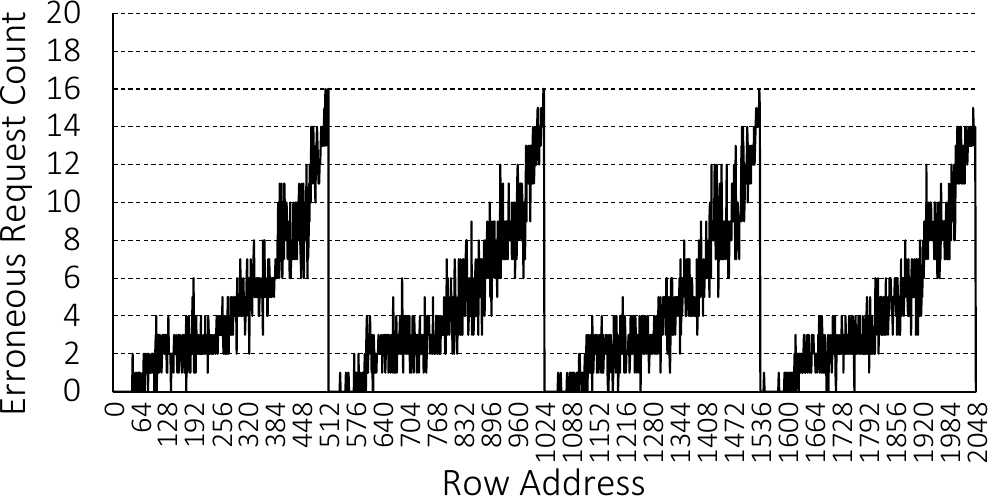}
		\label{fig:profile_row_order}
	}
	\vspace{-0.10in}
	\caption{Periodicity in \dhliii{Erroneous} Request Count (\trp 7.5ns)}
	\label{fig:profile_row}
	\vspace{-0.20in}
\end{figure}

{\bf Periodicity in Per-Row Error Count.} To understand these trends better, we
break down the error counts further for a \trp of 7.5ns. As we expect the
variation pattern to repeat every 512~rows, we use the value of \dhlii{\em row
address modulo 512} (which we refer to as a {\em row chunk}) to tally all of the
number of errors observed in the DIMM, as shown in
Figure~\ref{fig:profile_row_75ns}. We then sort the row chunks based on the
number of errors, shown in Figure~\ref{fig:profile_row_sort}. To see whether
periodicity exists, \dhlii{we then reorder the erroneous request counts of
\dhliii{each} {\em individual row} within {\em every set of 512 rows} by using
the sorted order in Figure~\mbox{\ref{fig:profile_row_sort}}, which we show in
Figure~\mbox{\ref{fig:profile_row_order}.}}
% we then use the sorting of the modulo (i.e., row chunk) values {\em within
% every set of 512 rows} to show the number of errors observed {\em in each
% row}, which we show in Figure~\ref{fig:profile_row_order}. 
We reorder the per-row data in this manner as\dhliii{,} without the sorting, it
is difficult to observe the periodicity that exists in the error count.

As expected, there is periodicity in error counts across 512 row chunks.
Therefore, we conclude that {\em error count shows periodicity with row
address}, confirming our expectation that there is predictable design-induced
variation in the latency of cells across a bitline. We will \dhliii{understand}
the reason why this periodicity does \dhliii{\em not} show up with increasing
row addresses in Section~\ref{sec:profile_rowint}.

\subsection{Design-Induced Variation in Wordlines} 
\label{sec:profile_wordline}

As we explained in Section~\ref{sec:overview_existence}, we expect
design-induced variation across cells in a wordline, depending on the distance
from the wordline driver. To confirm the existence of design-induced variation
across a wordline, we use a similar evaluation methodology as the one used in
Section~\ref{sec:profile_bitline}, except that {\em i)} we sweep the column
address instead of the row address, {\em ii)} \dhlii{aggregate} errors in the
same column across multiple rows (128 columns per row). In order to minimize the
impact of variation across a bitline and focus on variation across a wordline,
we test all columns in only 16 rows.

\begin{figure}[h]
	%\vspace{-0.05in}
	\centering
	\subfloat[\trp 10ns \& Aggregated] {
		\includegraphics[height=1.0in]{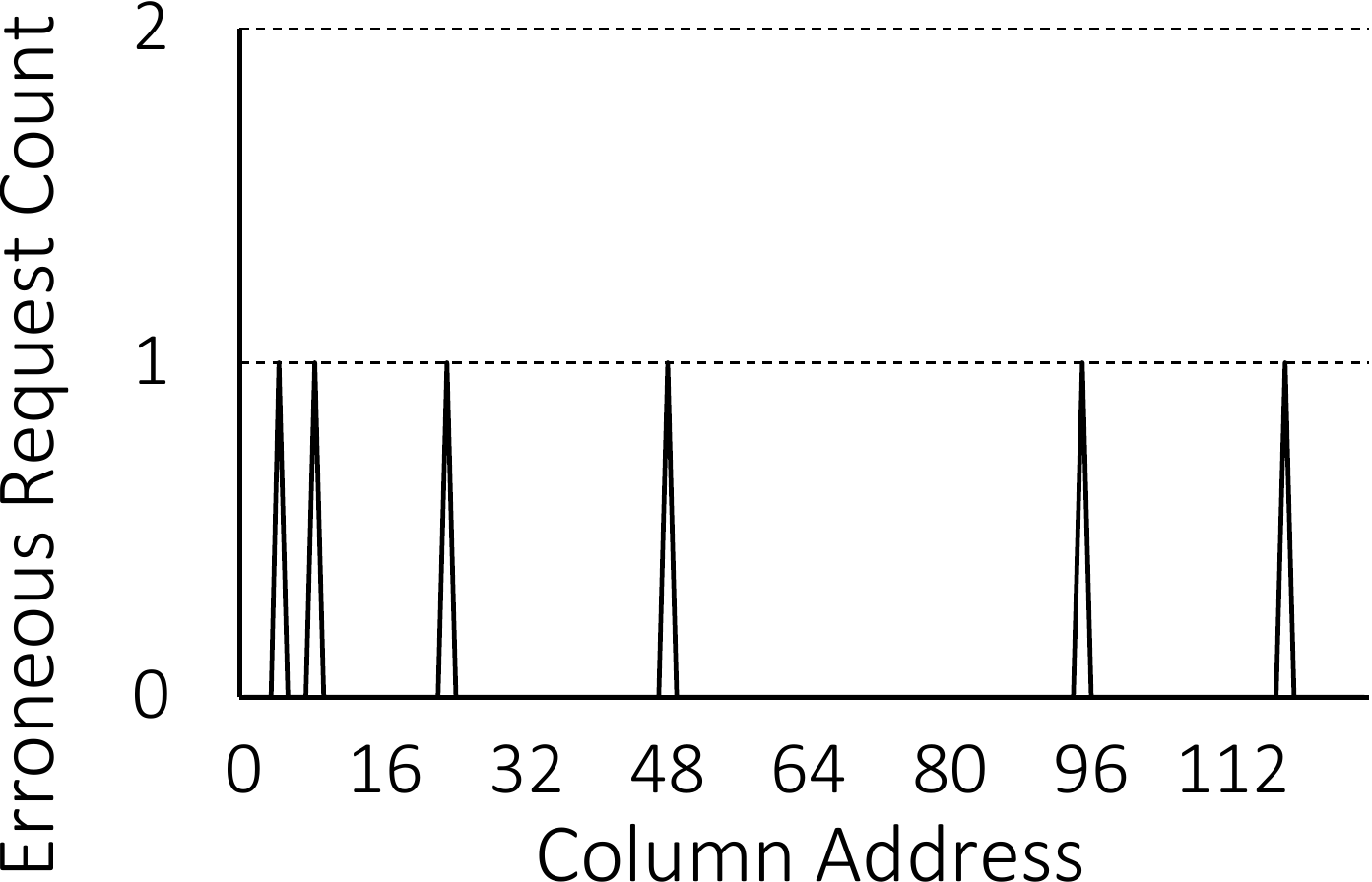}
		\label{fig:profile_col_addr_std}
	}
	\subfloat[\trp 7.5ns \& Aggregated] {
		\includegraphics[height=1.0in]{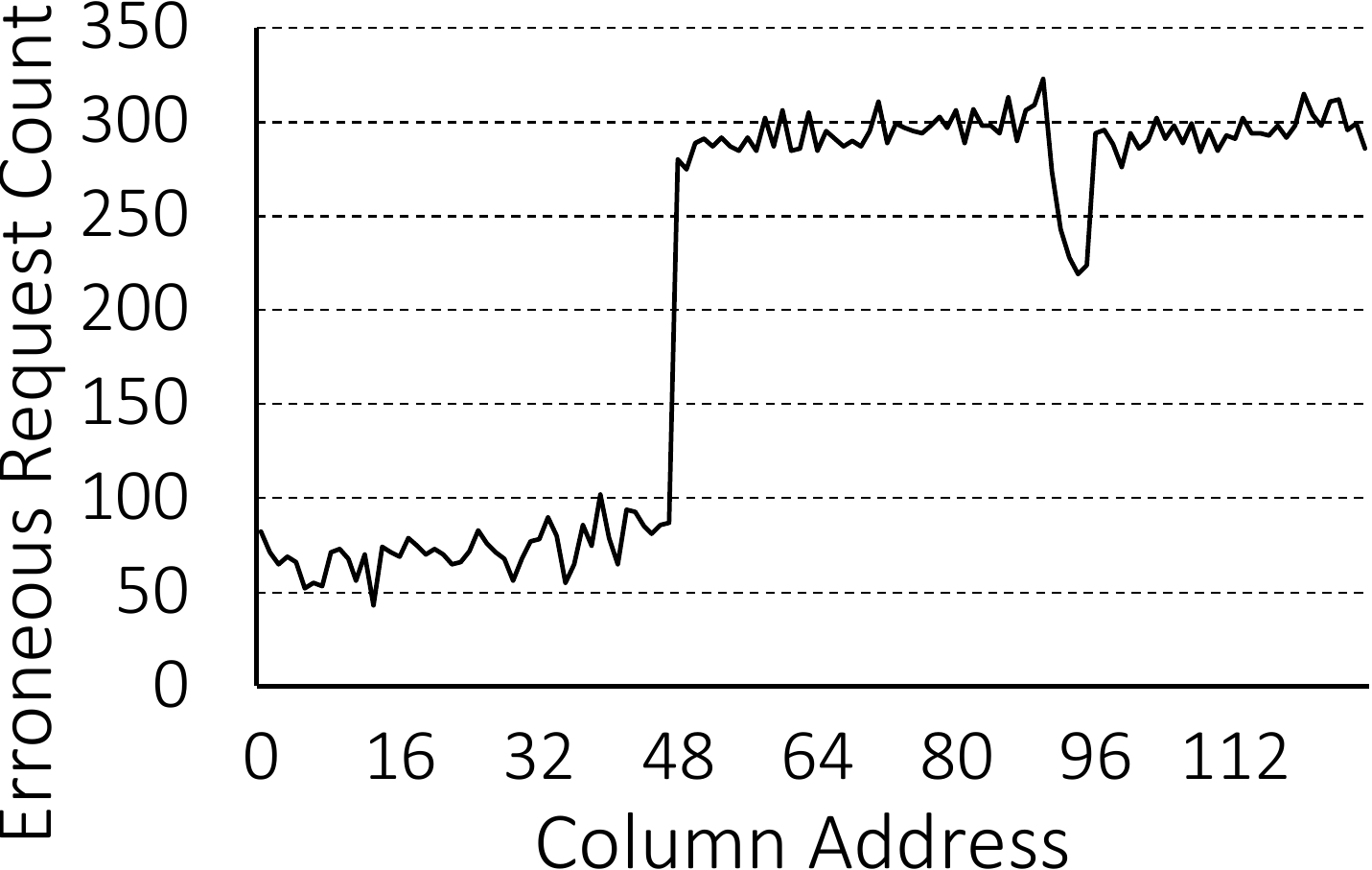}
		\label{fig:profile_col_addr_low}
	}

	\vspace{-0.05in}
	\subfloat[\trp 7.5ns \& Case 1] {
		\includegraphics[height=1.0in]{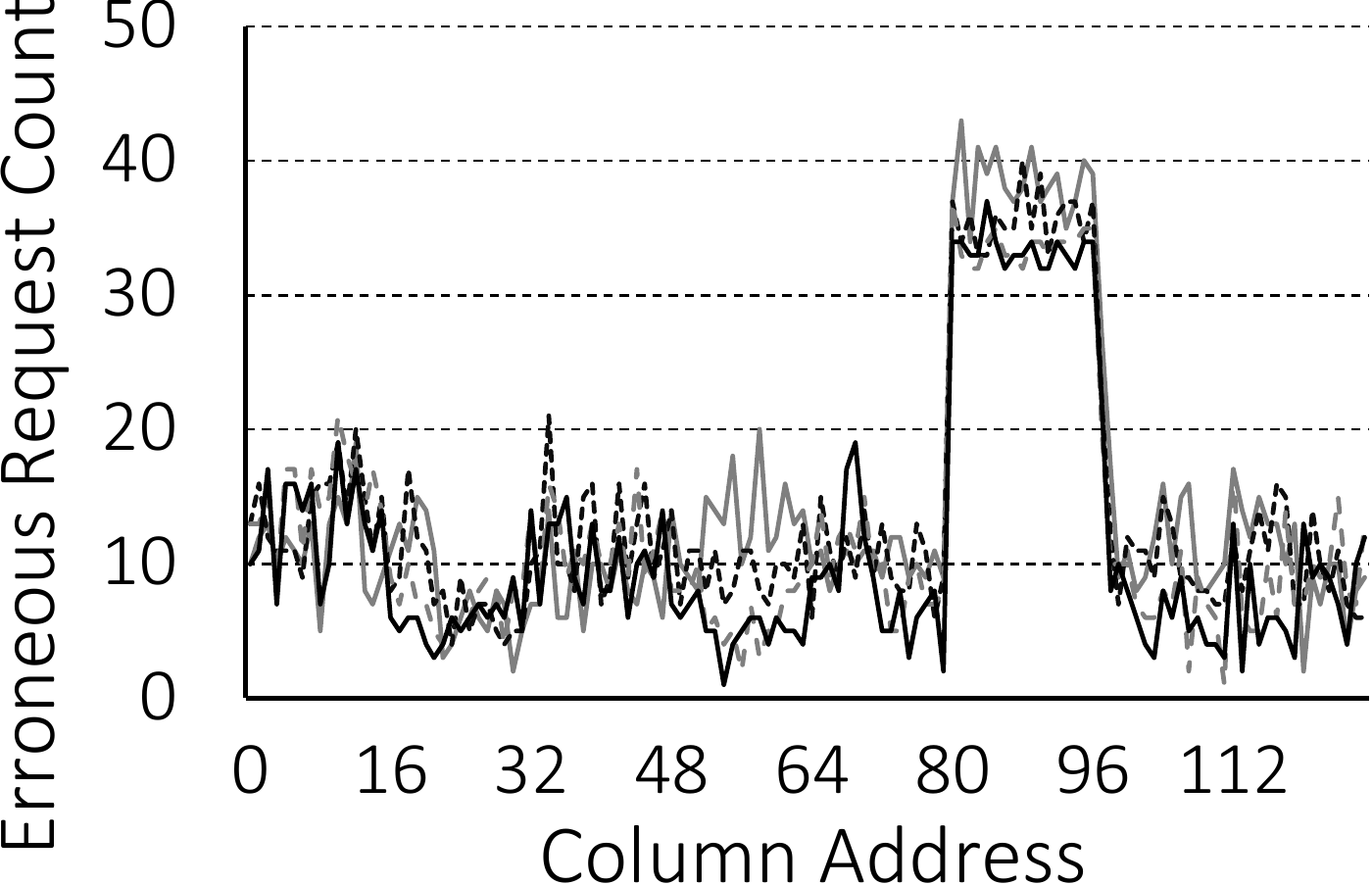}
		\label{fig:profile_col_case1}
	}
	\subfloat[\trp 7.5ns \& Case 2] {
		\includegraphics[height=1.0in]{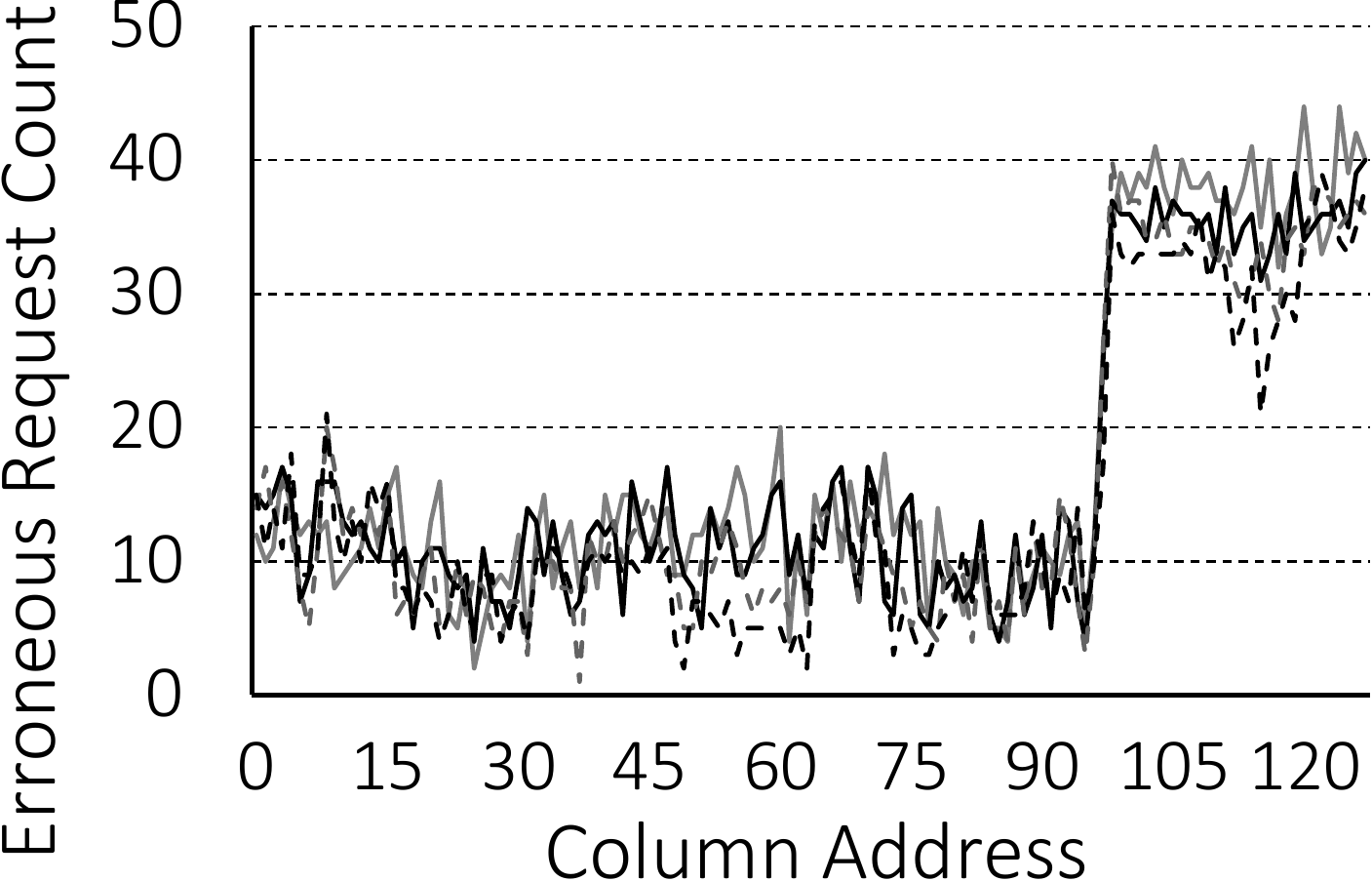}
		\label{fig:profile_col_case2}
	}
	\vspace{-0.10in}
	\caption{Erroneous Request Count When Sweeping Column Addresses with \dhlii{a}
	Reduced \trp Timing Parameter}
	\label{fig:profile_col_addr}
	\vspace{-0.10in}
\end{figure}

{\bf Per-Column Error Count with Column Address Sweeping.}
Figure~\ref{fig:profile_col_addr} provides results with two \trp values (10ns
and 7.5ns). Similar to the evaluation with sweeping row addresses, we see that
the number of errors is small and the distribution is random when \trp is
reduced by a small amount, as shown in Figure~\ref{fig:profile_col_addr_std}.
However, the number of errors is large when \trp is reduced significantly, as
shown in Figure~\ref{fig:profile_col_addr_low}. We observe variations in error
counts across different column addresses at a \trp of 7.5ns. Besides other
variations, there is a large jump near the 48th column and a dip in error count
near the 96th column, as shown in Figure~\ref{fig:profile_col_addr_low}.

To understand these, we separately plot each row's error count, which displays
different patterns. We provide two such types of patterns (\dhlii{obtained} from
multiple rows) in Figures~\ref{fig:profile_col_case1}
and~\ref{fig:profile_col_case2}. In one such type, shown in
Figure~\ref{fig:profile_col_case1}, the error count drastically increases at
around the 80th column and drops at around the 96th column (There are other
types of patterns with similar shapes but with the jumps/drops happening at
different locations). In the type of pattern shown in
Figure~\ref{fig:profile_col_case2}, the error count drastically increases at the
96th column and stays high. We attempt to correlate such behavior with the
internal organization of DRAM.

\begin{figure}[h]
	\vspace{-0.10in}
	\centering
	\includegraphics[width=\linewidth]{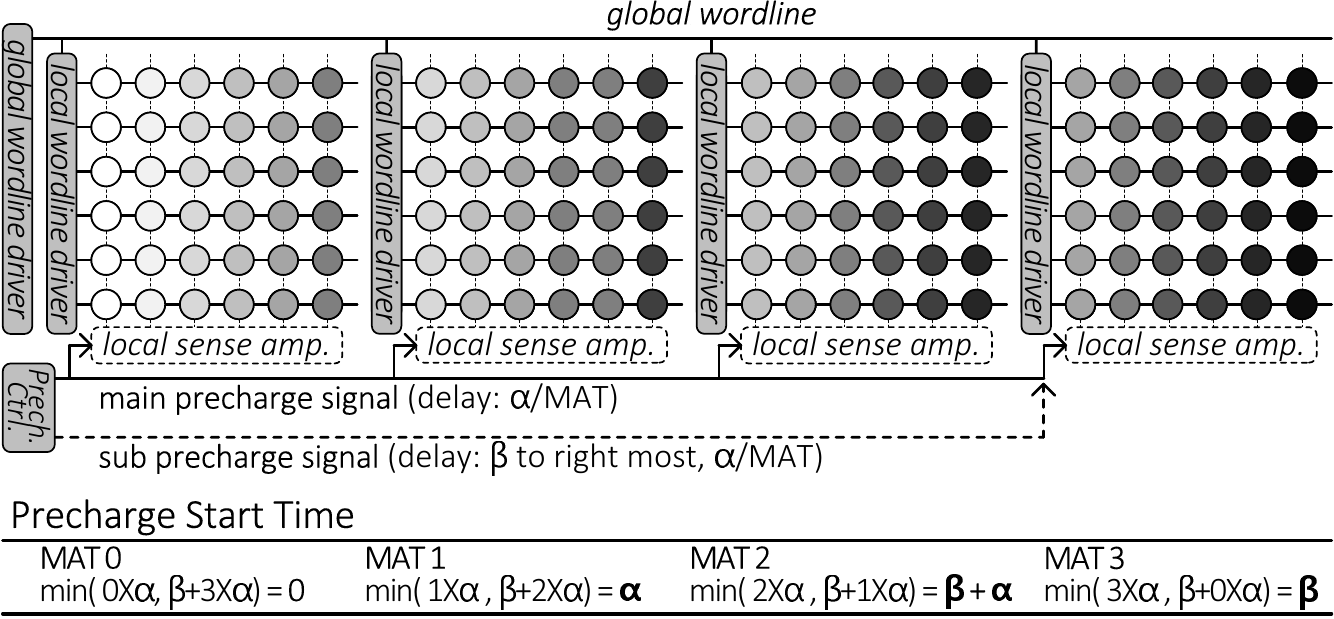}
	\vspace{-0.15in}
	\caption{Design-Induced Variation Due to Precharge Control}
	\label{fig:prech_signal}
	%\vspace{-0.10in}
\end{figure}

Figure~\ref{fig:prech_signal} shows an illustration of how the precharge control
signal flows across mats. The timing parameter \trp dictates how long the memory
controller should wait after it issues a precharge command before it issues the
next command. When a precharge command is issued, the precharge signal
propagates to the local sense amplifiers in each mat, leading to propagation
delay (higher for sense amplifiers that are farther away). To mitigate this
variation in the delay of the precharge control signal, DRAM uses two signals,
{\em i)} \dhlii{a} main precharge signal -- propagating from left to right, and
{\em ii)} \dhlii{a} sub precharge signal -- that directly reaches the right and
propagates from right to left.

The main and sub precharge signals arrive at different times at the different
mats due to parasitic capacitance on the propagation path. The main precharge
signal is delayed by $\alpha$ per mat going from left to right, while the sub
precharge signal is delayed by $\beta$ when it reaches the rightmost mat where
$\alpha > \beta$, since the sub precharge signal does not have any load going
from left to right. However, after that, the sub precharge signal exhibits a
delay of $\alpha$ per mat when propagating through mats from right to left. The
sense amplifiers in a mat respond to the faster one of the two precharge
signals. For instance, in Figure~\ref{fig:prech_signal}, mat 3 receives the
precharge signal the last. Hence, accesses to it would exhibit more errors than
accesses to other mats if \trp is reduced. Such control signal delays result in
the kind of jumps in errors at particular column addresses we see in real DRAM
chips (e.g., Figures~\ref{fig:profile_col_addr_low},
\ref{fig:profile_col_case1}, \ref{fig:profile_col_case2}). We conclude that
error count varies across columns, based on the column's distance from the
wordline and control signal drivers. While such control signal delays explain
why such jumps occur, knowledge of the exact location of mats and how they are
connected to the control signals is necessary to \dhlii{understand and explain
why jumps occur at \dhliii{\em particular} column addresses.}
% tie back the control signal propagation to the specific column addresses at
% which the jumps occur.

\subsection{Effect of the Row Interface} 
\label{sec:profile_rowint}

As shown in Figure~\ref{fig:profile_row_75ns}, the error count across a bitline
does not linearly increase with increasing {\em DRAM-external row address}
\dhlii{(i.e., the address issued by the memory controller over the memory
channel). However,} we observe periodicity when rows are sorted by error count,
\dhliii{as shown in} Figure~\ref{fig:profile_row}. This behavior could occur
because the DRAM-external row address is {\em not} directly mapped to the
internal row address in a DRAM mat~\cite{liu-isca2013}. Without information on
this mapping, it is difficult to tie the error count periodicity to specific
external row addresses. In this subsection, we estimate the 
\dhliv{\emph{most-likely mapping}} between the DRAM-external row address and the DRAM-internal
row address ({\em estimated row mapping}) \dhliii{\em based on the observed
error count}. We then analyze the similarity of the estimated row address
mapping across multiple DIMMs manufactured by the same DRAM company (in the same
time frame).

\begin{figure}[h]
	\vspace{-0.10in}
	\centering
	\hspace{-0.05in}
	\subfloat[Internal\,Address] {
		\includegraphics[width=1.0in]{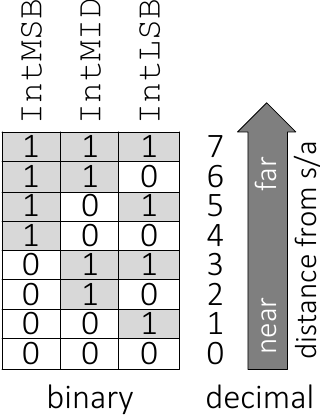}
		\label{fig:addr_physical_row}
	}
	\hspace{0.05in}
	\subfloat[External\,Address] {
		\includegraphics[width=1.0in]{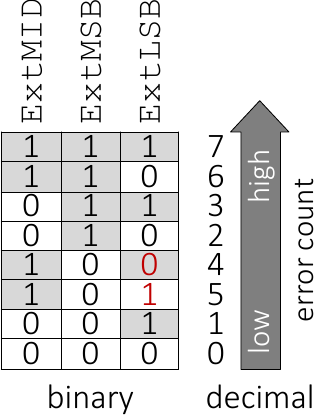}
		\label{fig:addr_logical_row}
	}
%	\hspace{0.02in}
	\subfloat[Estimation] {
		\includegraphics[width=1.0in]{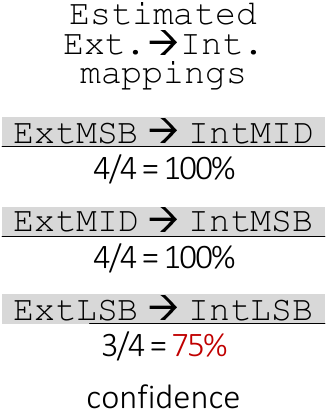}
		\label{fig:addr_mapping}
	}
	\vspace{-0.12in}
	\caption{DRAM-Internal vs. DRAM-External Row Addressing \dhliv{and Estimated
	Mapping Based on Observed Error Counts for the External Addresses}}
	\label{fig:arch_tran_row}
	\vspace{-0.18in}
\end{figure}

{\bf Methodology for Estimating Row Address Mapping.} We explain our estimation
methodology using a simple example shown in Figure~\ref{fig:arch_tran_row},
which has a 3-bit row address (eight rows per mat).
Figure~\ref{fig:addr_physical_row} shows the DRAM-internal row address in both
decimal and binary, increasing in the order of distance between the row and the
sense amplifier.

Figure~\ref{fig:addr_logical_row} shows DRAM-external row addresses \dhlii{that}
are {\em ranked based on the error counts}. As observed, the order is not the
same as the DRAM-internal address order in Figure~\ref{fig:addr_physical_row}.
To determine the estimated \sgI{external-to-internal} row mapping \dhliv{based
on the observed error counts for the external addresses}, \sgI{we explore all
possible permutations that rearrange the three bits in the row address. For each
of the eight rows in the mat, we have the error count. Our goal is to find an
ordering of the three bits, which we call the {\em internal row address}, for
which the error count monotonically increases with the number represented by the
three bits. For example, after rearranging, the row with an internal row address
of ``001'' should have a higher error count than the row with an internal row
address of ``000''. We find that by mapping the \dhliv{MSB of the internal} row
address (\dhliv{\tt IntMSB}) to the middle bit of the external row address
(\dhliv{\tt ExtMID}), and by mapping the middle bit of the \dhliv{internal} row
address (\dhliv{\tt IntMID}) to the MSB of the \dhliv{external} row address
(\dhliv{\tt ExtMSB}), as shown in Figure~\ref{fig:addr_mapping}, the row error
count increases monotonically with the internal row address.} The estimated
mapping (in the logical address) is indicated by dark boxes when the expected
bit is ``1'' and light boxes when the expected bit is ``0''. There are cases
when this mapping does \sgIII{\emph{not}} match with the actual external address (indicated in
red). \dhlii{Figure~\mbox{\ref{fig:addr_mapping}} shows that, in this example,
external to internal mapping can be estimated with high confidence.} \dhliv{For
example, we can say with 100\% confidence that the external address bit {\tt
ExtMID} maps to the internal address bit {\tt IntMSB} since the observed error
counts for the {\tt ExtMID} \sgIII{bit} match the expected error counts from \sgIII{the {\tt IntMSB} bit}.}

\begin{figure}[h]
	\vspace{0.05in}
	\centering
	\includegraphics[width=\linewidth]{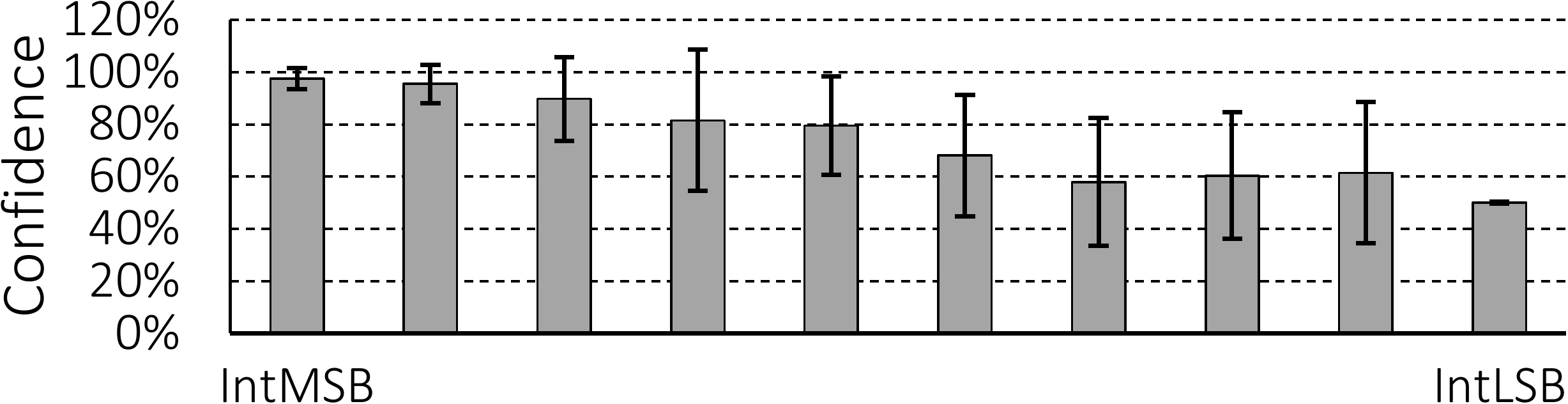}
	\vspace{-0.20in}
	\caption{Confidence in Estimated Row Mapping} 
	\label{fig:profile_addr_tran}
	\vspace{-0.15in}
\end{figure}

{\bf Estimated Row Address Mapping in Real DIMMs.} We perform such an external
to internal address mapping comparison and mapping exercise on eight DIMMs
manufactured by the same company in a similar time frame.
Figure~\ref{fig:profile_addr_tran} shows the average confidence level over all
rows in a DIMM, for the estimated row mapping. We use error bars to show the
standard deviation of the confidence over eight DIMMs. We make three
observations. First, all DIMMs show the \dhliii{\em same} estimated row mapping
(with fairly high confidence) for at least the five most significant bits. This
result shows that DIMMs manufactured by the same company at the same time have
similar design-induced variation. Second, the confidence level is almost always
less than 100\%. This is because process variation \dhlii{and row repair
mechanisms introduce} perturbations \dhliii{in addition to} design-induced
variation, which can change the ranking of rows (determined based on error
counts \dhliii{as we explained earlier}). Third, the confidence level drops
gradually from \dhliv{\tt IntMSB} to \dhliv{\tt IntLSB}. This is also due to the
impact of process variation and row repair \dhlii{mechanisms}. The noise from
process variation and row repair can change row ranking and grouping by error
count. Address bits closer to \dhliv{\tt IntMSB} tend to divide rows into groups
at a larger granularity than address bits closer to \dhliv{\tt IntLSB}.
Therefore, the higher order bits show higher confidence. Based on these
observations, we conclude that DRAMs that have the same design display similar
error characteristics due to design-induced latency variation.

In summary, we observe predictable row address mapping (similar to Figure 11)
when testing DIMMs from the same vendor that were manufactured around the same
time \dhlii{frame} (i.e., they likely have the same internal circuit design).

\subsection{Effect of the Column Interface} 
\label{sec:profile_colint}

\begin{figure}[h]
	\vspace{-0.10in}
	\centering
	\includegraphics[width=\linewidth]{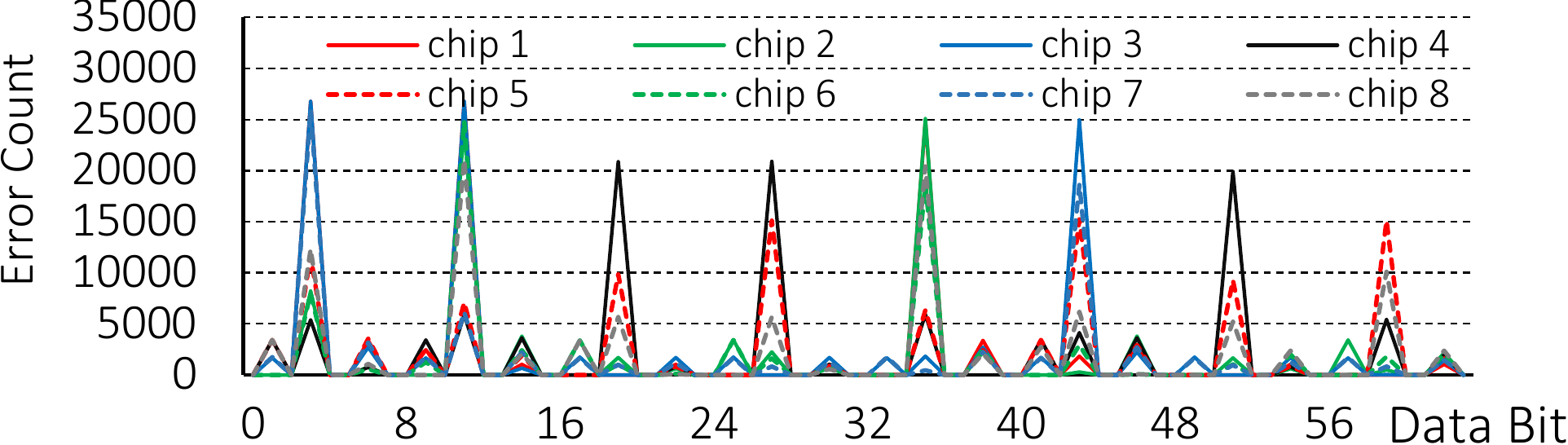}
	\vspace{-0.15in}
	\caption{Error Count in \dhlii{Data-Out} Bit Positions}
	\vspace{-0.15in}
	\label{fig:profile_col_io}
\end{figure}

Another way to observe the error characteristics in the wordline organization is
\dhliii{by} using the {\em mapping between the global sense amplifier and the IO
channel}. As we explained, global sense amplifiers in a DRAM chip concurrently
read 64-bit data from different locations of a row, leading to variation in
errors. Figure~\ref{fig:profile_col_io} plots errors in 64-bit data-out (as
shown in Figure~\ref{fig:arch_col_interface}) in the IO channel (For example,
first eight bits (\dhlii{bits} 0 -- 7) are the first burst of data transfer). We
draw three conclusions. First, there is large variation in the amount of errors
in the IO channel. For example, more than 26K errors happen in the third bit
while no errors \dhlii{are observed} in the first bit of the IO channel. Second,
the error characteristics of eight DRAM chips show similar trends. Third, while
we observed regular error distribution at different bit positions from DIMMs
that show design-induced variation, we also observed that the \dhlii{error}
patterns from different DIMMs (e.g., DIMMs from different vendors) were
different. Section~\ref{sec:mech_shuffling} uses these observations to develop
a new error correction mechanism, called \myshuffling.

\subsection{Effect of Operating Conditions} 
\label{sec:profile_sensitivity}

\begin{figure}[h]
	\vspace{-0.05in}
	\centering
	\subfloat[Varying Retention Time] {
		\includegraphics[width=0.48\linewidth]{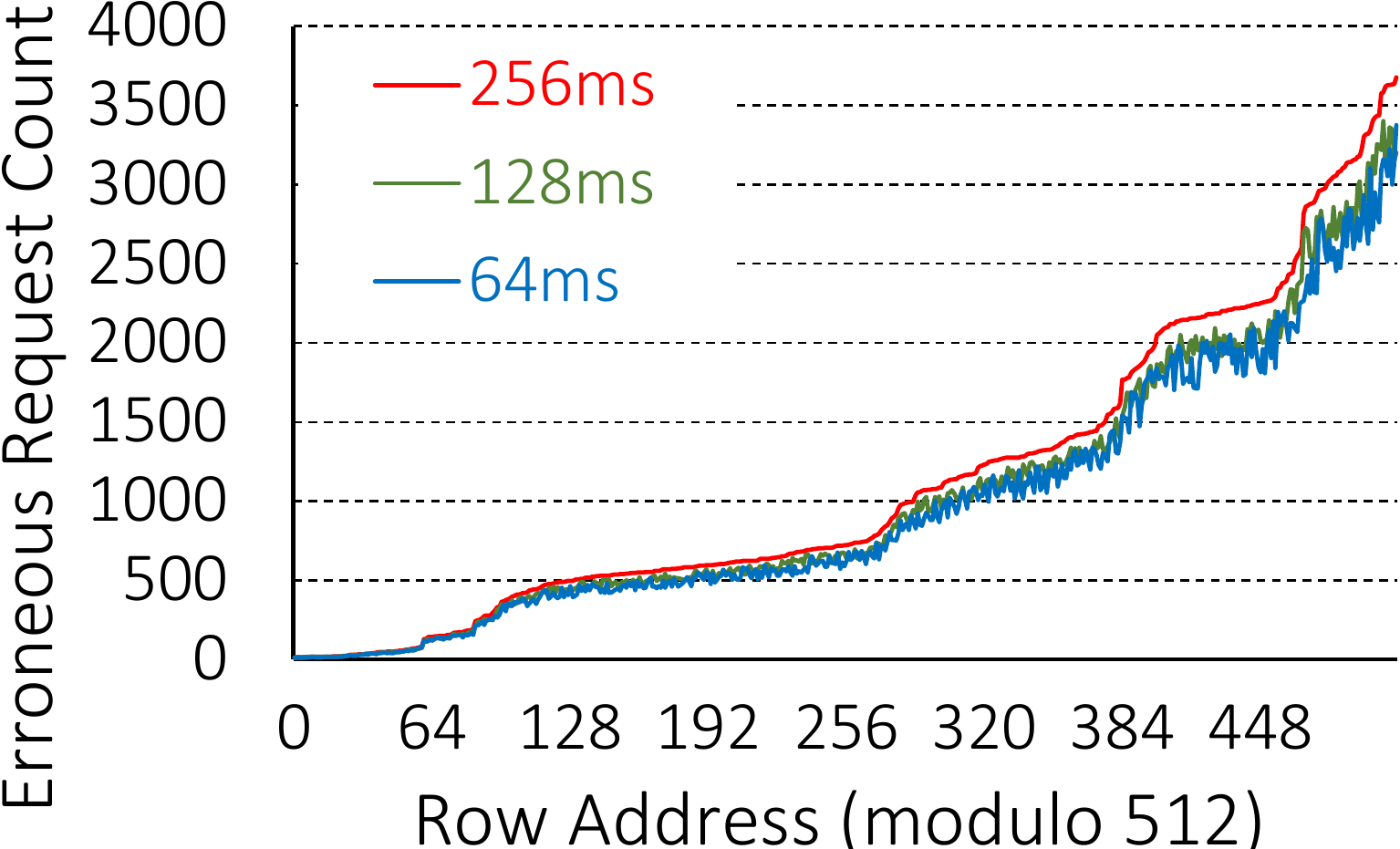}
		\label{fig:profile_ret}
	}
	\subfloat[Varying Temperature] {
		\includegraphics[width=0.48\linewidth]{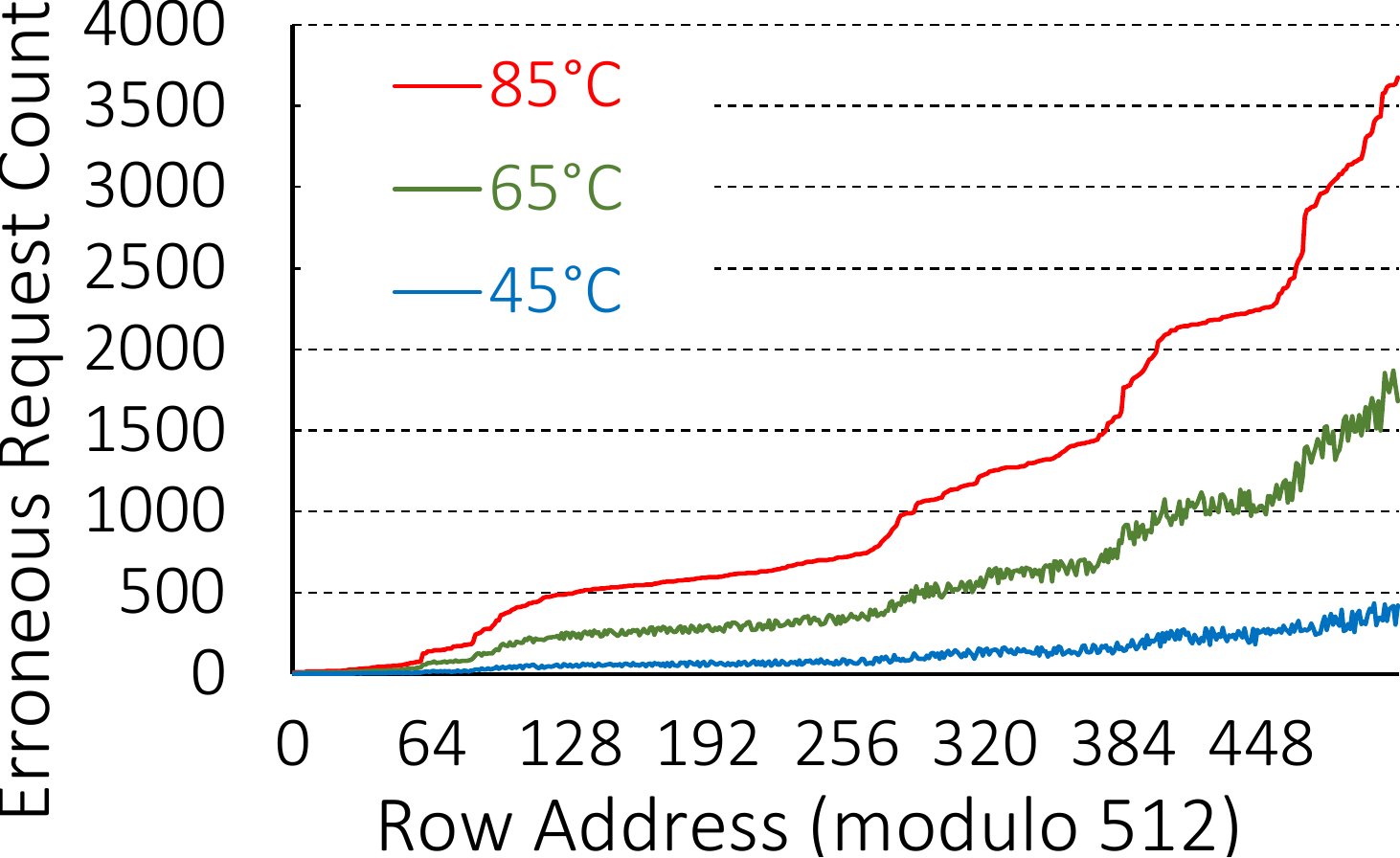}
		\label{fig:profile_temp}
	}
	\vspace{-0.10in}
 	\caption{Design-Induced Variation vs. Operating Conditions}
	\label{fig:profile_sensitivity}
	\vspace{-0.15in}
\end{figure}

Figure~\ref{fig:profile_sensitivity} shows the error count sensitivity to the
refresh interval and the operating temperature by using the same method as row
sweeping (\dhlii{aggregating the error count across every set of row address
modulo 512 rows,} as done in Section~\ref{sec:profile_bitline}). We make three
observations. First, neither the refresh interval nor temperature changes the
overall trends of design-induced variation (\dhliii{i.e.}, \dhliv{the}
variability characteristics in different row addresses remain the same, though
the absolute number of errors changes). Second, reducing the refresh interval or
the ambient temperature within the normal system operating conditions (i.e.,
45\celsius~to 85\celsius) leads to fewer errors. Third, the variability in cells
is much more sensitive to the ambient temperature than the refresh interval.
When changing the refresh interval, the total error count does not change
drastically\dhliii{:} \dhliii{it} exhibits only \dhliii{a} 15\% decrease with
\sgI{a} 4X reduction in refresh interval. On the other hand, changing the
ambient temperature has a large impact on the total error count\dhliii{: error
count reduces by} 90\% with \dhliii{a 40}\celsius~change in temperature. This is
due to the fact that frequent refreshes make only the cells
faster\sgII{~\cite{lee-hpca2013, hassan-hpca2016, shin-hpca2014}}, whereas reducing
temperature makes not only the cells but also the peripheral circuits faster.
Based on these observations, we conclude that temperature or refresh interval do
not change the trends in \dhlii{design-induced variation}, \dhliii{but} they
impact the total number of failures in vulnerable regions at different rates. 

\subsection{Summary Results of \dimms~DIMMs} 
\label{sec:profile_dimms}

\begin{figure}[h]
	\vspace{-0.10in}
	\centering
	\subfloat[\trp~(7.5ns)] {
		\includegraphics[width=0.48\linewidth]{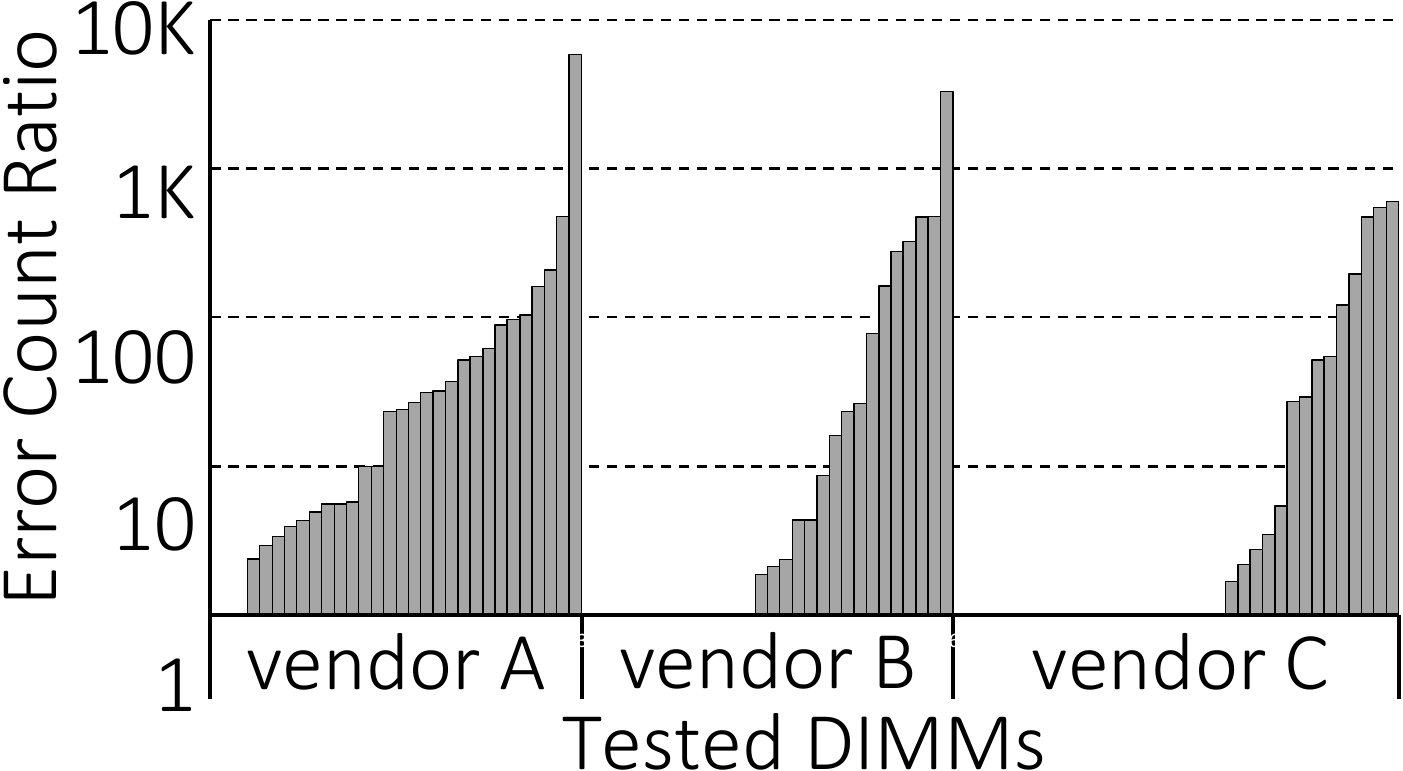}
		\label{fig:summary_trp}
	}
	\subfloat[\trcd~(7.5ns)] {
		\includegraphics[width=0.48\linewidth]{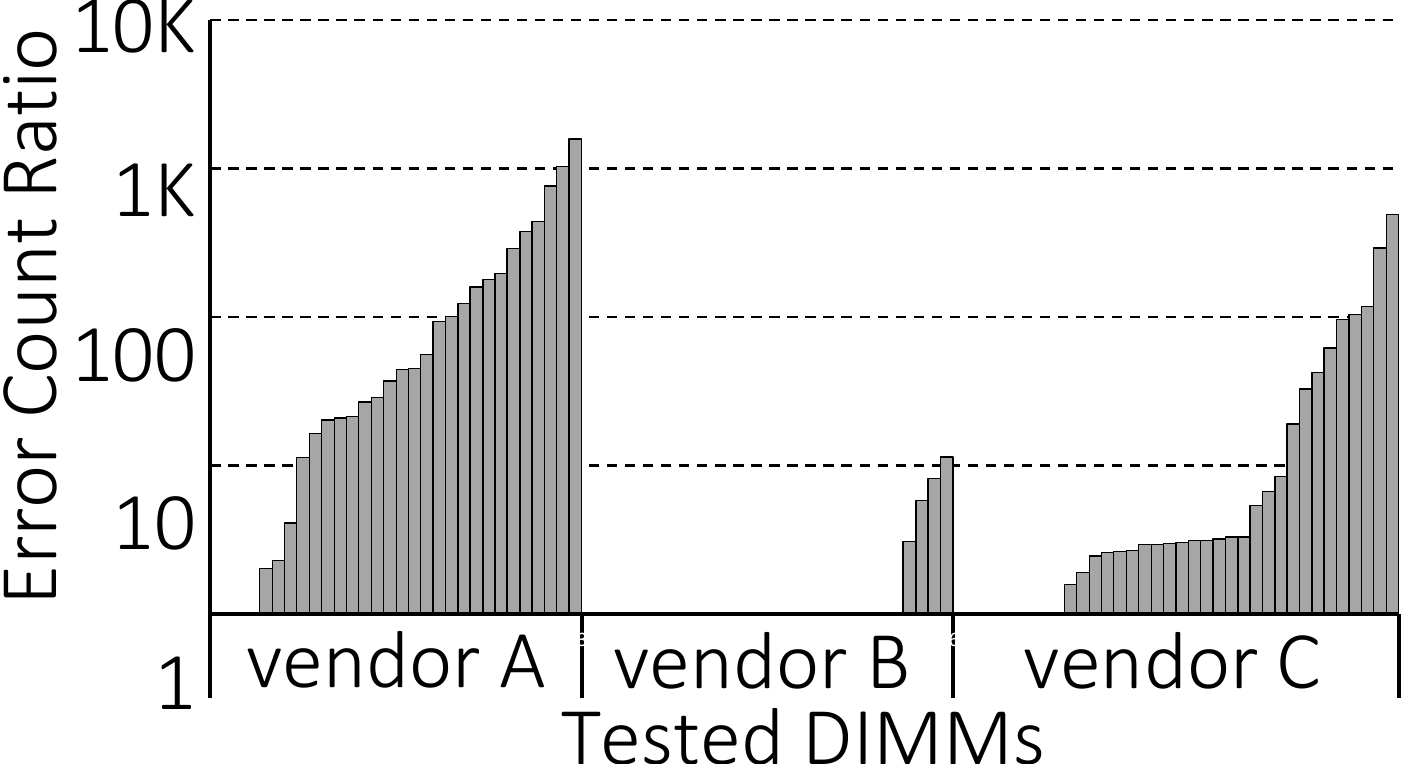}
		\label{fig:summary_trcd}
	}
	\vspace{-0.10in}

	\caption{Vulnerability Ratio: \dhlii{the error \dhliii{count} ratio
	\dhliii{between} the top 10\% most vulnerable and \dhliii{the top 10\%} least
	vulnerable rows}}

	\label{fig:profile_summary}
	\vspace{-0.15in}
\end{figure}

We profile \dimms~DIMMs with 768 chips from three vendors to characterize the
design-induced variation in DRAM chips. We observe similar trends and
characteristics in DIMMs from the same generation, though the absolute number of
failures are different. In Figure~\ref{fig:profile_summary}, we show the
\dhliii{error count} difference \dhliii{between} the most vulnerable region vs.
the least vulnerable region in each of the tested \dhlii{DIMMs}. We define the
difference as {\em vulnerability ratio} and calculate it using the error
\dhliii{count} ratio \dhliii{between the error count of} the top 10\% most
vulnerable \dhliii{rows and the error count of the top 10\%} least vulnerable
rows.\footnote{Note that the results show the {\em variation} of error
distribution, which does {\em not} represent either the performance or the
reliability of DIMMs from different vendors.}

We make two observations from this figure. First, most of the DIMMs exhibit
large design-induced variation in terms of vulnerability ratio (e.g., as high as
5800 times, notice the log scale). Second, we did not observe design-induced
variation in \navdimms~DIMMs. However, we believe that this is in part due to a
limitation of our infrastructure, where we can reduce timing parameters only at
a coarser granularity (i.e., at a step size of 2.5~ns) due to the limited FPGA
frequency, similar to the DRAM test infrastructures used in prior
works\sgII{~\cite{lee-hpca2015, kim-isca2014, chandrasekar-date2014, liu-isca2013,
hassan-hpca2017, chang-sigmetrics2016, khan-sigmetrics2014, khan-dsn2016,
chang-sigmetrics2017, khan-cal2016, kim-thesis2015, lee-thesis2016,
chang-thesis2017}}. As a result, it is \dhlii{sometimes} possible that reducing a
step of a timing parameter \dhlii{causes} the tested DIMM to \dhlii{transition}
from a \dhliii{no-error or very-low-error} \dhlii{state} to a \dhlii{state}
where latency is low enough to make all cells fail, missing the timing where
design-induced variation is clearly visible. In real machines where
state-of-the-art DRAM uses \dhlii{a} much lower clock period (e.g., DDR3-2133:
0.94ns), design-induced variation might be prevalent. Third, DRAMs from the same
vendor and from similar production time frames show similar characteristics to
each other, including whether or not they are susceptible to design-induced
\dhliii{variation related} errors. For example, DRAMs from Vendor B have
drastically high error counts \dhliii{across} most regions when \trcd is reduced
below a certain value. We include summary results for each DIMM that we tested
in \dhlii{Appendix}~\ref{sec:appendix_summary}. We \dhlii{provide} detailed
results for each DIMM online~\cite{safari-git}.

\dhliii{In summary}, we have experimentally demonstrated that \dhliii{\em i)}
design-induced variation is prevalent across a large number of DIMMs and
\dhliii{\em ii)} our observations hold true in most of the DIMMs. We validate
these observations on the existence of design-induced variation in DRAM using
circuit-level SPICE simulations in Appendix~B. We conclude that modern DRAMs are
amenable to reducing latency by exploiting design-induced variation.

	\section{Mechanisms to Exploit Design-Induced Variation} 
\label{sec:case}

In this section, we present two mechanisms that leverage design-induced
variation to reduce DRAM latency while maintaining reliability: {\em i)}
Design-Induced Variation Aware online DRAM Profiling (\myprofiling) to determine
\dhliii{by} how much DRAM latency can be safely reduced while still achieving
failure-free operation, and {\em ii)} Design-Induced Variation Aware data
Shuffling (\myshuffling) to avoid uncorrectable failures (due to lower latency)
in systems with ECC. We intentionally aim \dhlii{to design} intuitive and simple
mechanisms, such that they are practical and easy to integrate into real
systems.

\subsection{\myprofiling} 
\label{sec:mech_lowlatency}

Previous works observe that the standard DRAM timing parameter values are
determined based on the worst-case impact of \dhliii{{\em process variation} and
worst-case operating conditions,} and leverage this observation to reduce
overall DRAM latency \dhlii{under} \dhliii{common-case} operating
conditions~\cite{lee-hpca2015, chandrasekar-date2014}. We leverage \dhliii{\em
design-induced variation} in DRAM to develop a {\em dynamic} and {\em low-cost}
DRAM \dhliii{latency/error} profiling technique. We call this technique {\em
Design-Induced Variation Aware Online DRAM Profiling} ({\em \myprofiling}). The
key idea is to \dhliii{\em separate reduced-latency-induced errors into two
categories}, one caused by design-induced variation and the other caused by
process variation, and then employ \dhliii{\em different error mitigation
techniques} for these two error categories.

\myprofiling avoids two shortcomings faced by prior work on exploiting latency
variation to reduce overall DRAM latency~\cite{lee-hpca2015,
chandrasekar-date2014}. These prior works, which do not exploit {\em
design-induced latency variation}, are \dhliii{\em unable} to perform effective
\dhliii{\em online} profiling to dynamically determine DRAM latency, since
online profiling can incur high performance overhead~\cite{singh-mtdt2005,
elm-mtdt1994, rahman-prdc2011, patel-isca2017}. As a result, these prior works
rely on {\em static} profiling, which leads to two key shortcomings. First,
prior works do {\em not} present any concrete way to identify the lowest
possible values of timing parameters that \dhlii{guarantee} reliability. Second,
these works do {\em not} account for dynamic changes in minimum DRAM latency
that happen over time due to circuit aging and wearout. Therefore, implementable
mechanisms based on these works have to assume conservative margins to ensure
reliable operation in the presence of aging and wearout. This causes the
realistic latency reductions with such mechanisms to be lower than what we
optimistically show for these mechanisms~\cite{lee-hpca2015} in our evaluations
(Section~\ref{sec:mech_result}). By employing low-cost online profiling,
\myprofiling can attain much more aggressive latency reductions while
maintaining reliable operation.\footnote{Note that our evaluation of AL-DRAM
does \dhlii{\em not} factor in dynamic latency \dhlii{increases} due to {\em
aging and wearout}, giving AL-DRAM an unfair advantage in our results,
overestimating its latency benefit.}

\begin{figure}[h]
	\vspace{-0.10in}
	\centering
	\subfloat[Process] {
		\includegraphics[height=1.1in]{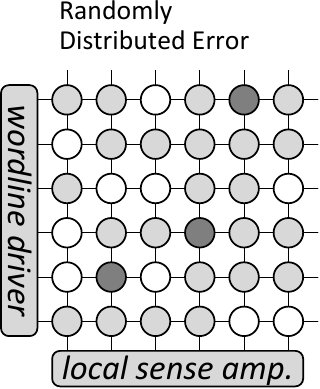}
		\label{fig:failure_rand}
	}
	\hspace{0.05in}
	\subfloat[Design] {
		\includegraphics[height=1.1in]{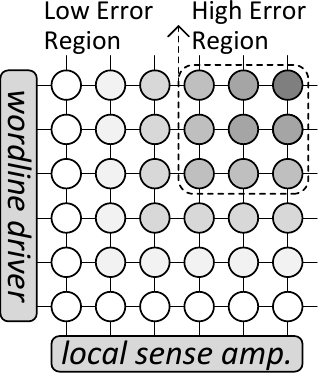}
		\label{fig:failure_arch}
	}
	\hspace{0.05in}
	\subfloat[Proc.$+$Design] {
		\includegraphics[height=1.1in]{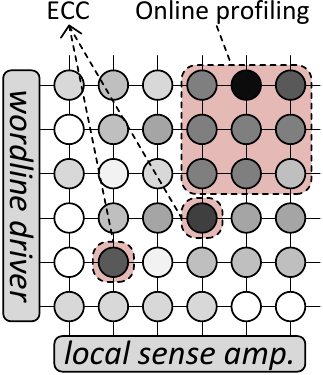}
		\label{fig:failure_both}
	}
	\vspace{-0.10in}
	\caption{Latency Variation in a Mat (Darker: Slower)}
	\label{fig:failure}
	\vspace{-0.10in}
\end{figure}

{\bf Design-Induced Variation vs. Process Variation.} The error characteristics
from process variation and design-induced variation are very different.
Figure~\ref{fig:failure} shows the error patterns from these two types of
variation (darker cells are more error prone). First, the errors caused by
process variation are usually randomly distributed over the entire DRAM
chip~\cite{lee-hpca2015, chandrasekar-date2014} (Figure~\ref{fig:failure_rand}).
Because these errors are random, existing ECC mechanisms (e.g.,
SECDED)~\cite{meza-dsn2015, luo-dsn2014} can detect and recover these random
errors. On the other hand, the errors caused by design-induced variation are
more systematic and are concentrated in specific regions in the DRAM chip
(Figure~\ref{fig:failure_arch}). For instance, when timing parameters are
aggressively reduced, cells that are farther away from both the row driver and
the local sense amplifiers are prone to more errors. As these high-error cells
are concentrated on a specific region of the mat, they typically result in
multi-bit errors that cannot be corrected by simple ECC (e.g., SECDED). To avoid
these undesirable multi-bit errors, we propose to periodically profile only the
\dhliii{high-error (i.e., vulnerable)} regions and track whether any of these
regions \dhlii{fail} under a specific set of timing parameters, which incurs
much less overhead than profiling the entire DRAM, and then tune the timing
parameters appropriately based on the failure information.

{\bf \myprofiling Mechanism.} \myprofiling combines SECDED \dhliii{ECC}, which
stores \dhliii{ECC codewords} in a separate chip on the DIMM (similar to
commodity DRAM), with online profiling in a synergistic manner to reduce DRAM
latency while maintaining high reliability. Due to design-induced variation,
there is a specific region within each subarray of the DRAM that requires the
highest access latency in the subarray. The DIVA-profiling-based memory system
uses this slowest region, which we call the {\em latency test region}, to
perform online latency profiling. To address the \dhliii{random} effect of
\dhlii{process} variation \dhlii{\em across} different subarrays in the entire
DRAM chip, our mechanism employs \emph{per-subarray} latency test
regions.\footnote{We further discuss the effect of process variation in
Appendix~C.}

Note that actual \dhliii{useful data (e.g., application or system data)} is
\dhliii{\em not} stored in \dhlii{these per-subarray latency test regions}. A
memory controller with DIVA Profiling support periodically accesses
\dhlii{these} latency test \dhlii{regions} and determines the smallest value of
DRAM timing parameters required for reliable operation in \dhlii{all of the
latency test regions} (without causing multi-bit errors). The system then adds a
small margin to the timing parameters obtained from this profiling (e.g., one
clock cycle increase) to determine the timing parameters for the other regions
({\em data region}), \dhliii{which store the actual useful data required by the
system and the programs.}

{\bf System Changes to Enable \myprofiling}. We require three changes to the
system. First, we need to account for the repair/remapping process employed by
DRAM vendors to increase yield. As we describe in
Section~\ref{sec:overview_interface}, when faulty cells are identified during
post-manufacturing test, the rows or columns corresponding to these faulty cells
are remapped to other rows or columns by blowing fuses after
manufacturing~\cite{arndt-iemt1999}. If a row from the latency test region is
remapped to a different row, this will affect the profiling phase of our
mechanism. In order to avoid such interactions with the repair/remapping process
(and potential inaccuracies in identification of the lowest latency at which to
operate a DRAM chip reliably), we propose an approach where rows from the
latency test regions are {\em not} remapped by DRAM vendors. Faulty cells in the
\dhliii{latency} test region are instead repaired using {\em column remapping},
another repair mechanism that is already implemented in commercial
DRAM~\cite{horiguchi-book}. Our mechanism finds a {\em uniform} latency for an
entire DIMM, at which all rows in \dhliii{all latency test regions of} the DIMM
operate reliably, by selecting the smallest latency that guarantees reliable
operation of \dhliii{all such test rows}. Therefore, the profiled latency can be
used to reliably operate all non-test rows (both normal rows and redundant
rows). This approach is straightforward to implement, since DRAM vendors are
likely to know the most vulnerable regions in the DRAM chip (based on their
design knowledge). Since rows in the latency test regions do \dhliv{\emph{not}}
store any \dhliii{useful} data, this approach maintains system reliability.

Second, systems with DIVA Profiling require the ability to change DRAM timing
parameters online. Since DIVA Profiling uses only one set of timing parameters
for the entire DIMM, the only required change is updating the timing parameters
in the memory controller with the smallest latency values that still ensure
reliable operation.

Third, DIVA Profiling requires a way of exposing the design-induced variation to
the \dhlii{memory controller}. The most intuitive approach is to expose either
the internal organization or the location of the slowest region as part of the
DRAM specification or the SPD (Serial Presence Detect) data in DIMMs
\dhliii{(e.g., as done in \sgII{\cite{kim-isca2012, chang-sigmetrics2016,
lee-hpca2013}})}. Address scrambling techniques in the \dhlii{memory controller}
need not impact DIVA Profiling since \dhlii{memory controller {\em i)}} knows
how the addresses are scrambled, and {\em ii)} can generate requests for
profiling without applying scrambling.

{\bf DIVA Profiling Overhead.} There are several overheads to consider when
implementing DIVA Profiling. First, in terms of {\em area overhead within the
DRAM array}, \myprofiling reduces the memory capacity slightly by reserving a
small region of the DRAM for latency testing. In a conventional DRAM, which
typically contains 512 rows per subarray, the area overhead is 0.2\% \dhlii{(one
row per subarray)}. Second, in terms of {\em \dhlii{latency} overhead},
\myprofiling requires additional memory accesses, which could potentially delay
demand memory requests. However, we expect the latency overhead of profiling to
be low, since \myprofiling reserves only the slowest rows as \dhlii{the} test
region (one row per subarray), and only these rows need to be profiled.
\myprofiling is much faster than conventional online profiling mechanisms that
must test {\em all} of the DRAM cells~\cite{nair-isca2013, liu-isca2012,
venkatesan-hpca2006, khan-sigmetrics2014}: \myprofiling takes 1.22ms \sgI{per
data pattern}\footnote{\dhliv{A DRAM manufacturer can select and provide the
worst-case data pattern(s) DIVA Profiling should use for each DRAM module. This
information can be conveyed via the Serial Presence Detect (SPD) circuitry
present in each DRAM module (as done in \cite{kim-isca2012,
chang-sigmetrics2016, lee-hpca2013}).}} to profile a 4GB DDR3-1600 DIMM,
\dhlii{whereas conventional profiling takes 625ms} (see
Appendix~\ref{sec:appendix_overhead} for \dhlii{the} detailed calculation). We
can employ intelligent and optimized profiling mechanisms that can further
reduce the impact of the overhead. For example, one simple and low overhead
mechanism can conduct online profiling as part of the DRAM refresh operation
\dhliii{(e.g., similar to methods that parallelize refresh operations and memory
accesses~\mbox{\cite{chang-hpca2014}})}, which would have minimal effect on
memory system performance. Third, in terms of {\em storage overhead within the
memory controller}, systems with DIVA Profiling require a very small amount of
additional storage (e.g., \dhliii{as low as} 16~bits for a 4GB DIMM) to
implement the profiling mechanism: {\em one bit} per DIMM to track if any rows
fail for the current timing parameters being tested, and {\em one row address
register} per DIMM, which points to the slowest region in the DIMM.

In summary, our mechanism profiles only the slowest region that is most
\dhliii{affected} by design-induced variation, thereby incurring low profiling
overhead, while achieving low DRAM latency {\em and} high reliability.

{\bf Energy Consumption.} DIVA Profiling consumes similar energy for a single
DRAM operation (e.g., activation, read, write, and precharge) compared to
conventional DRAM. The profiling overhead is low since only the test region
needs to be profiled. Furthermore, \dhlii{the DRAM latency reductions enabled
by} DIVA Profiling reduces system execution time, as we will see in
Section~\ref{sec:mech_result}, and can thereby reduce \dhliii{\em system} energy
consumption.

{\bf Other Sources of Latency Variation in DRAM.} DIVA Profiling has been
designed with careful consideration of other sources of DRAM latency variations,
e.g., voltage (due to supply grid) \& temperature variation and VRT (Variable
Retention Time\sgII{~\cite{liu-isca2013, khan-sigmetrics2014, kim-edl2009,
qureshi-dsn2015, patel-isca2017, yaney-iedm1987, restle-iedm1992,
mori-iedm2005}}). As explained, we divide DRAM failures into two categories: i)
localized, \dhliii{systematic} failures (caused by design-induced variation);
and ii) random failures (caused by process variation and VRT). We then exploit
\dhliii{different} error mitigation techniques to \dhliv{handle} these two
\dhliii{different} categories of failures: \dhliv{\em i)} online profiling for
localized \dhliii{systematic} failures, and \dhliv{\em ii)} ECC for random
failures. Since the physical \dhlii{size} of a mat is very small (e.g.,
1415.6~$\micro m^2$ in 30~nm technology), the effects of voltage and temperature
variation are similar across a mat. The \dhliii{negative} effects of process
variation and VRT can be \dhliii{handled} by ECC. Furthermore, we tackle the
impact of \dhliii{\em sense amplifier offset} \dhlii{(\dhliii{i.e., the
phenomenon that a sense amplifier shows} different sensitivities for detecting
``0'' and ``1'' due to process variation}~\cite{keeth-book}) by profiling {\em
all columns} of the \dhliii{rows in \dhliv{\em all} latency test regions}.
Hence, the variation from sense amplifier offset is accounted for in determining
the smallest possible values of timing parameters that ensure reliable
operation.

There can be several opportunities for applying \dhliv{\em different timing
parameters} to \dhlii{exploit} process variation (e.g., variation across
\dhliv{subarrays}, variation across \dhliv{banks}\dhliii{, or variation across
\dhliv{chips}}). \dhliii{\myprofiling, for example, can be used to determine
\dhliv{\em different} timing parameters for \dhliv{\em different} subarrays,
banks, or chips within a DIMM.} While \dhlii{exploiting} the \dhlii{latency}
variation induced by process variation \dhliii{in such a manner} is promising,
we leave this for future work.\footnote{\dhlii{A recent
work}~\cite{chang-sigmetrics2016, chang-thesis2017} \dhliv{characterizes and
exploits} this type of process variation, providing promising results.}
\dhliii{In DIVA-DRAM,} we focus \dhliii{\em solely} on \dhliii{exploiting}
design-induced variation, which remains consistent across DRAM chips. To this
end, DIVA Profiling uses the \dhliii{\em same} timing parameters \dhlii{\em
across all chips in a DIMM}.

\subsection{\myshuffling} 
\label{sec:mech_shuffling}

\begin{figure}[h]
	\vspace{-0.10in}
	\centering
	\subfloat[Conventional Mapping] {
		\includegraphics[height=1.30in]{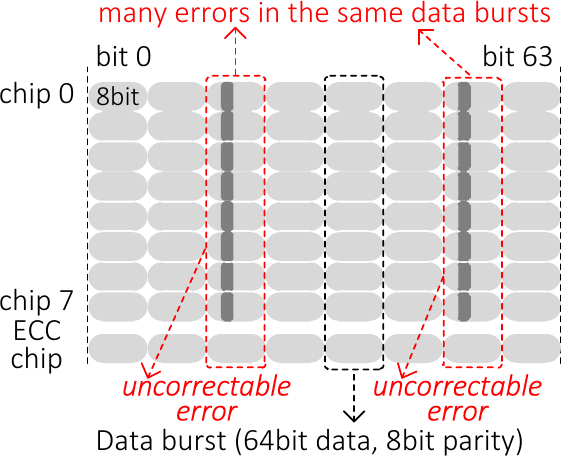}
		\label{fig:mech_map_ori}
	}
	\hspace{0.05in}
	\subfloat[Proposed Mapping] {
		\includegraphics[height=1.30in]{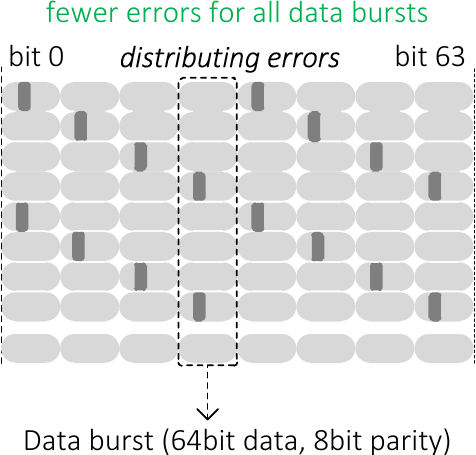}
		\label{fig:mech_map_new}
	}
	\vspace{-0.10in}
	\caption{Design-Induced Variation Aware Data Shuffling}
	\label{fig:mech_map}
	\vspace{-0.05in}
\end{figure}

Our second approach focuses on leveraging design-induced variation to mitigate
uncorrectable errors in memory systems with ECC \dhliii{(especially when DRAM is
operated at a lower latency than the standard latency)}. As we observed in
Section~\ref{sec:profile_colint}, when data is read out of a memory channel,
data in specific locations tends to fail more frequently. This happens because
data is delivered from locations \dhlii{that} are distributed across a wordline.
Due to design-induced variation in wordline and control signals, it takes longer
to access cells in specific locations compared to cells in other locations,
\dhlii{which could lead} to multi-bit errors in memory systems with ECC.
Figure~\ref{fig:mech_map_ori} shows the effect of design-induced variation in
systems with ECC. Data in the darker grey regions ({\em high-error
\dhlii{bits}}) tends to be more error-prone than data in the lighter grey
regions. These high-error bits are concentrated in a similar location across
different chips, \dhlii{and, as a result, are} part of the same data-transfer
burst. \dhlii{Since SECDED ECC can correct only one erroneous bit in a single
data burst~\mbox{\cite{luo-dsn2014}}, it is probable to \dhliii{observe}
uncorrectable errors for such data bursts.}\footnote{\dhliii{Note that
uncorrectable errors are reasonably common in the field, as reported by prior
work~\cite{meza-dsn2015}. While our \myshuffling mechanism can be used to
correct such errors as well, we leave the exploration of this to future work.}}

We tackle this problem and mitigate potential uncorrectable errors by leveraging
awareness of design-induced variation. Our key idea is {\em to distribute the
high-error bits across different ECC \dhliii{codewords}}. We call this mechanism
{\em design-induced-variation-aware data shuffling
(\myshuffling).}\footnote{While it is possible that different placement
algorithms for DIVA Shuffling could affect the latency and failure probability,
the search space of such algorithms is very large. We choose an intuitive
algorithm based on our observations of where errors and high-latency regions lie
within DRAM, and find that this algorithm results in high performance with
significant improvements in reliability.} 

There are \dhlii{two} potential ways in which such a shuffling mechanism can be
implemented. The first way is using DRAM chips that have different data-out
mappings\dhliii{,} by changing the DRAM chips internally during their
manufacturing. Since the data mapping is changed internally in the DRAM chips to
shuffle the high-error bits across different ECC \dhliii{codewords}, the address
decoding mechanism for reads and writes can remain identical across DRAM chips.
The second way is to shuffle the address mapping of DRAM chips within a DIMM. We
achieve this by connecting the address bus bits in a different order for
different DRAM chips in a DIMM, \dhlii{thereby enabling} different column
addresses \dhlii{to be} provided by different DRAM chips. Using these two
mechanisms, we can achieve data shuffling in the data output from DRAM
\dhliii{(as Figure~\mbox{\ref{fig:mech_map_new}} shows), which leads to fewer
errors in all data bursts.}

\begin{figure}[h]
	\vspace{-0.10in}
	\centering
	\includegraphics[width=\linewidth]{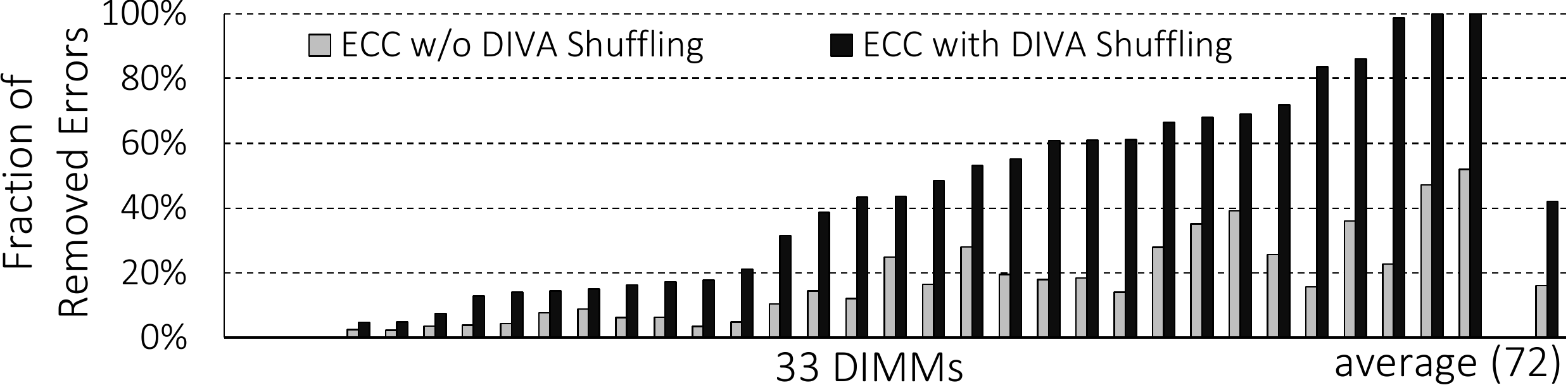}
	\vspace{-0.15in}
	\caption{Correctable Errors with/without DIVA Shuffling}
	\label{fig:mech_map_correction}
	\vspace{-0.10in}
\end{figure}

Figure~\ref{fig:mech_map_correction} shows the fraction of correctable errors
from a total of 72 DIMMs using SECDED ECC with \dhlii{and} without DIVA
Shuffling. We recorded the error locations and then filtered out correctable
errors assuming SECDED ECC. The Y-axis represents the total percentage of errors
with lower DRAM timing parameters, and the X-axis represents 33 (randomly
selected) DIMMs. The operating conditions \sgIII{(i.e., the reduced latencies)} were \dhlii{chosen} to make sure that
there are actually errors, so that ECC is useful.

Our \sgII{\myshuffling} mechanism corrects 26\% of the errors \dhlii{that} are
\dhliii{\em not} correctable by using {\em only} conventional ECC. In some
DIMMs, \sgII{\myshuffling corrects 100\% of the errors, while some other DIMMs}
still \dhlii{experience} errors even with \myshuffling. We believe that the
major cause for this is the malfunction of DRAM core operation, leading to
\sgII{excessively} high error \dhlii{rates}.
% \todo{add conditions here}} 
Overall, we conclude that using \myshuffling along with ECC can significantly
reduce the error rate than using conventional ECC alone.

\subsection{DRAM Latency \& Performance Analysis} 
\label{sec:mech_result}

\noindent{\bf DRAM Latency Profiling.} We profile \dimms~DIMMs, comprising 768
DRAM chips, for potential latency reduction. We use the same test methodology,
described in Section~\ref{sec:pmethod}, which is also similar to the methodology
of previous works~\cite{lee-hpca2015, chandrasekar-date2014}. We measure the
latency reduction of four timing parameters (\trcd, \tras, \trp, and \twr).

\begin{figure}[h]
	\vspace{-0.20in}
	\centering
	\subfloat[READ (\trasN$-$\trpN$-$\trcdN)] {
		\includegraphics[height=0.95in]{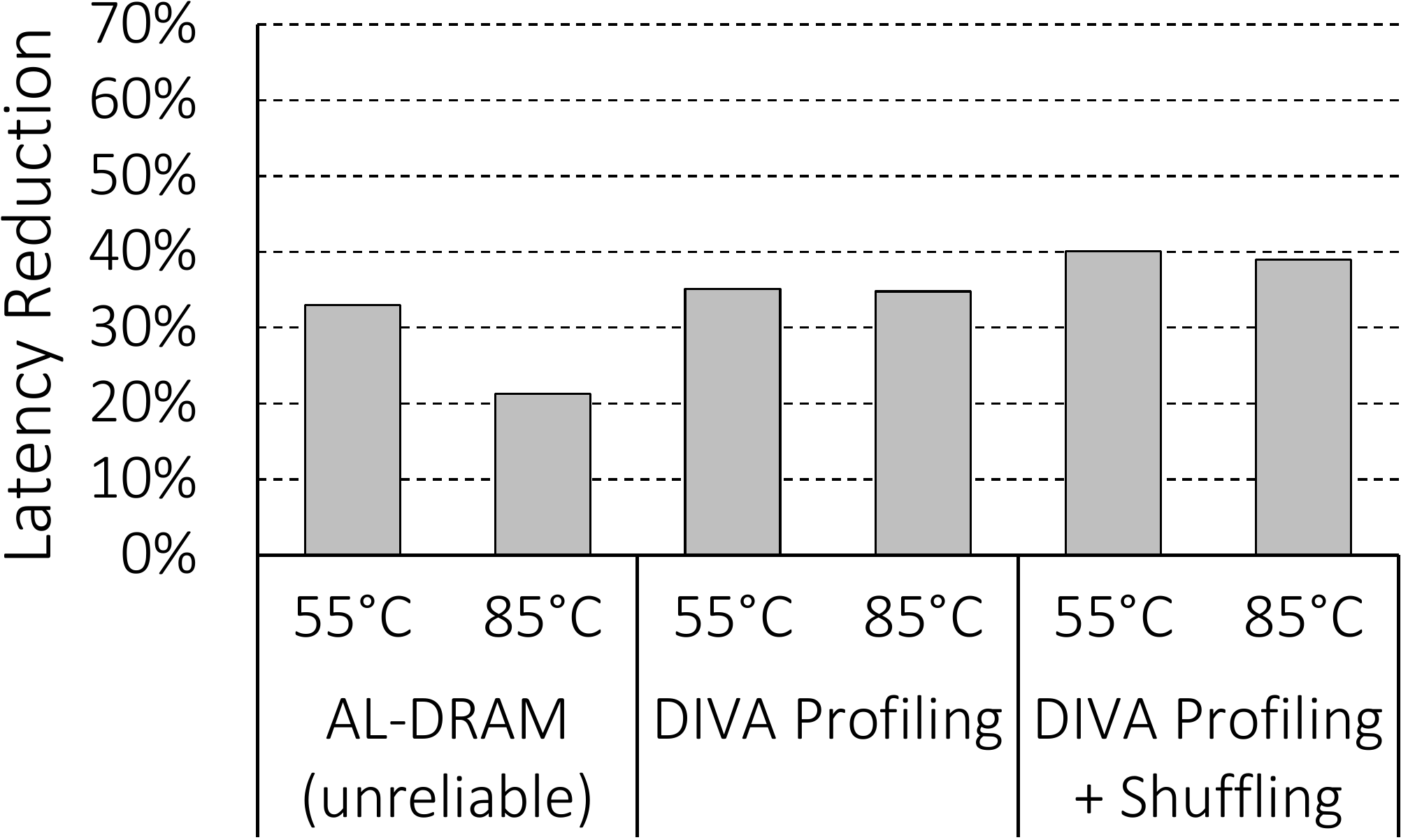}
		\label{fig:profile_timing_rd}
	}
	\subfloat[WRITE (\twrN$-$\trpN$-$\trcdN)] {
		\includegraphics[height=0.95in]{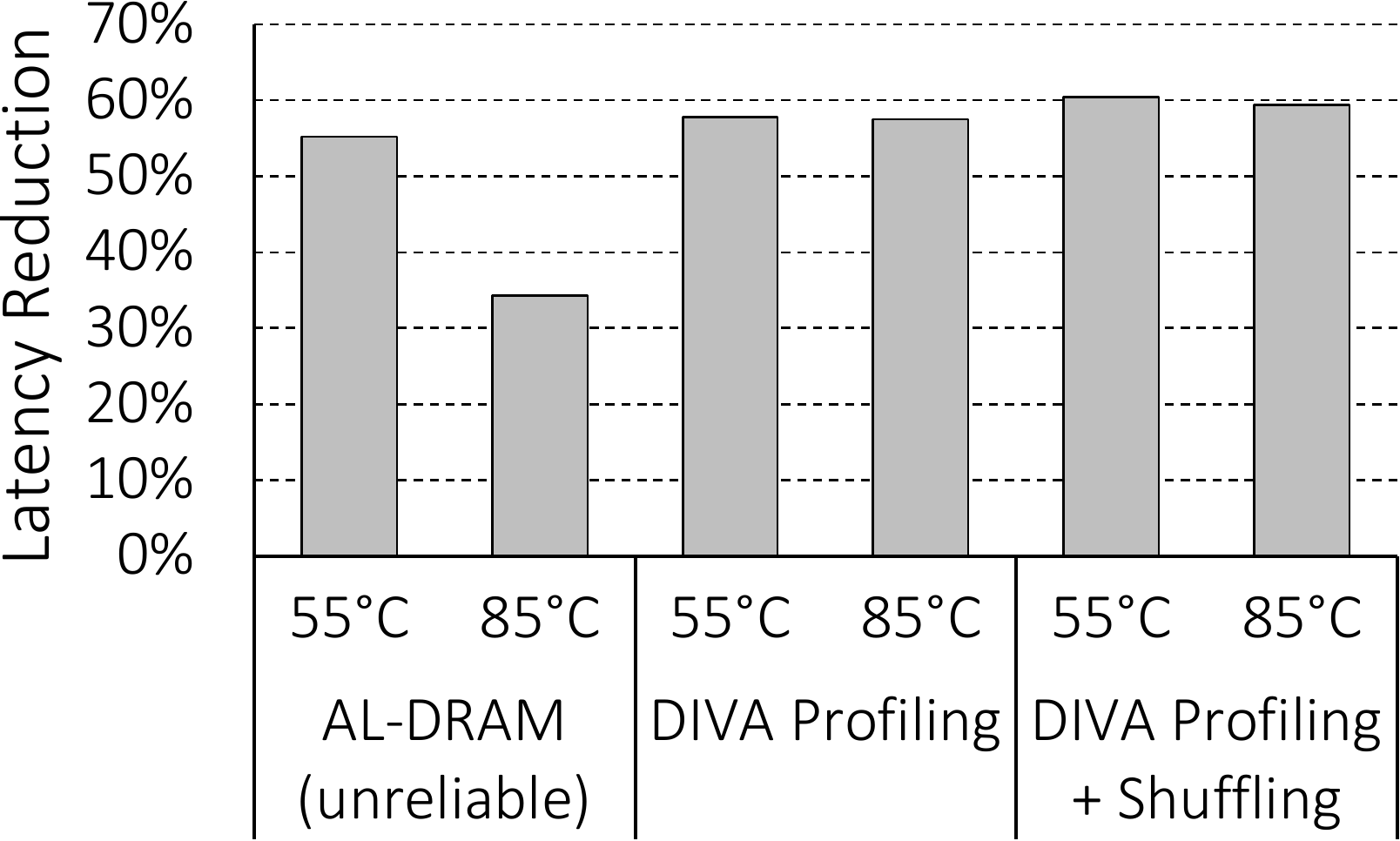}
		\label{fig:profile_timing_wr}
	}
	\vspace{-0.10in}
	\caption{Read and Write Latency Reduction}
	\label{fig:profile_rd_wr}
	\vspace{-0.10in}
\end{figure}

Figure~\ref{fig:profile_rd_wr} shows the average latency reduction for DRAM read
and write operations with three mechanisms --- AL-DRAM~\cite{lee-hpca2015},
\myprofiling, and \dhliii{the combination of \myprofiling and \myshuffling\ ---
normalized to} the sum of the corresponding \dhliii{baseline} timing parameters.
We compare these mechanisms at two operating temperatures, 55\celsius~and
85\celsius. We ignore the fact that AL-DRAM does \dhlii{\em not} account for
latency changes due to aging and wearout, and assume aggressive latency
reductions for it, giving AL-DRAM an unfair advantage.
AL-DRAM~\cite{lee-hpca2015} can reduce the latency for read/write operations by
33.0\% (18 cycles) and 55.2\% (18 cycles) at 55\celsius, and 21.3\% (12 cycles)
and 34.3\% (19 cycles) at 85\celsius, respectively. \myprofiling reduces the
corresponding latencies by 35.1\% (22 cycles) and 57.8\% (20 cycles) at
55\celsius, and 34.8\% (22 cycles) and 57.5\% (20 cycles) at 85\celsius,
respectively. Using \myshuffling on top of \myprofiling enables more latency
reduction (by 1.8\% on average). \dhlii{Thus, even though we give an unfair
advantage to AL-DRAM in our evaluation, our mechanisms achieve better latency
reduction compared to AL-DRAM, mainly because} ECC (and also ECC with
\myshuffling) can correct single-bit errors in an ECC \dhliii{codeword}.
Specifically, increasing \dhliii{the} temperature from 55\celsius~to 
\sgII{85\celsius}~with the same set of timing parameters mostly generates single-bit and
randomly distributed errors that can be corrected by ECC. \dhliii{Since AL-DRAM
does not employ ECC, its latency benefits degrade at high temperatures, whereas
our mechanism's latency benefits remain high at all temperatures.}

\noindent{\bf Performance Evaluation.} We simulate the performance \dhlii{of}
our DIVA Profiling mechanism using a modified version of
Ramulator~\cite{kim-cal2015}, a fast, cycle-accurate DRAM simulator that is
publicly available~\cite{ramulator}. We use Ramulator combined with a
cycle-level x86 multi-core simulator. Table~\ref{tbl:system} shows the system
configuration we model. We use PinPoints~\cite{luk-pldi2005, patil-micro2004} to
collect workload traces. We use 32 benchmarks from \dhlii{Stream}~\cite{stream,
moscibroda-usenix2007}, SPEC CPU2006~\cite{spec}, TPC~\cite{tpc} and
GUPS~\cite{gups}, each of which is used for a single-core workload. We construct
32 two-, four-, and eight-core workloads\dhliii{, for} a total of 96 multi-core
workloads (randomly selected from the 32 benchmarks). We measure single-core
performance using instructions per cycle (IPC) and multi-core performance using
the weighted speedup~\cite{snavely-asplos2000, eyerman-ieeemicro2008} metric. We
simulate 100 million instructions at 85\celsius~\dhliii{for each benchmark}. 

\begin{table}[h]
 %\vspace{-0.10in}
	\small{
	\centering
 	\begin{tabular}{lp{5.7cm}}
		\toprule
		\small{Component} & Parameters \\
		\midrule
		\multirow{2}{*}{Processor}				& 8 cores, 3.2GHz, 3-wide issue,\\
																			& 8 MSHRs/core, 128-entry inst. window\\
		\multirow{2}{*}{Last-level cache} & 64B cache-line, 16-way associative,\\
   															      & 512KB private cache-slice per core\\
		\multirow{1}{*}{Mem. Controller}  & 64/64-entry read/write queues,
																			FR-FCFS~\cite{rixner-isca2000, zuravleff-patent1997}\\
		\multirow{1}{*}{Memory system} 		& DDR3-1600~\cite{ddr3}, 2 channels,
																			2 ranks-per-channel\\
		\bottomrule
	\end{tabular}
	}
\caption{Configuration of Simulated Systems} \label{tbl:system}
\vspace{-0.10in}
\end{table}

\begin{figure}[h]
	\vspace{-0.10in}
	\centering
	\includegraphics[width=.95\linewidth]{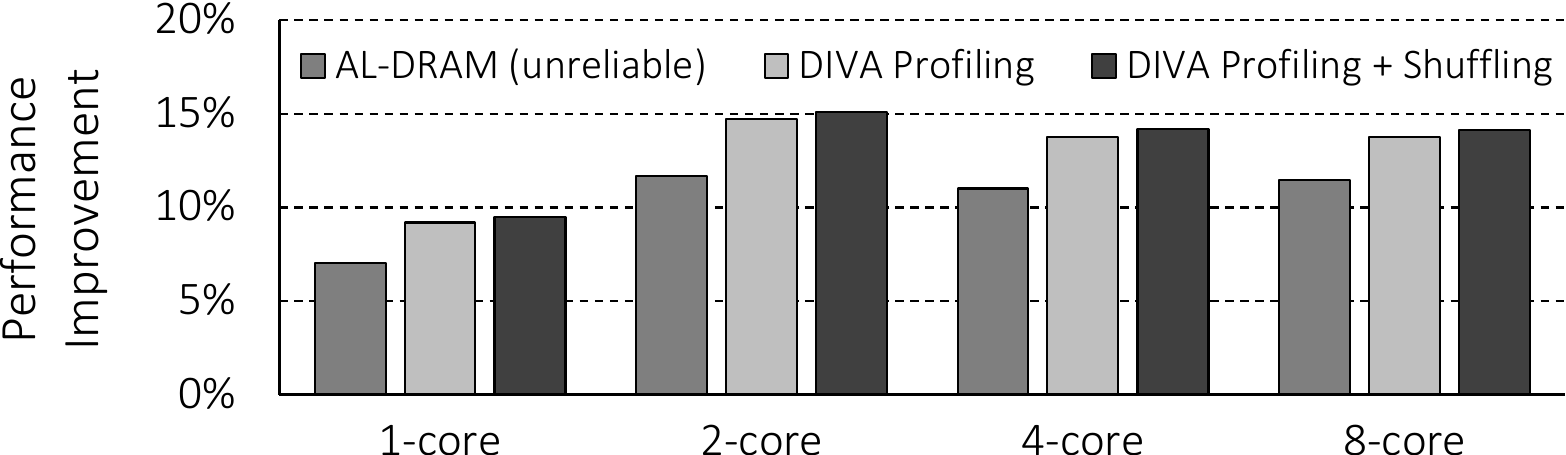}
	\vspace{-0.10in}
	\caption{Performance Improvement \dhliii{at 85}\celsius}
	\label{fig:mech_performance}
	\vspace{-0.10in}
\end{figure}

Figure~\ref{fig:mech_performance} shows the performance improvement with
\myprofiling and \myshuffling. We draw two major conclusions. First,
\myprofiling provides significant performance \dhlii{improvements} over the
baseline DRAM (9.2\%/14.7\%/13.7\%/13.8\% performance improvement in
single-/two-/four-/eight-core systems, respectively). This improvement is mainly
due to the reduction in DRAM latency. Second, using \myprofiling and
\myshuffling together provides even better performance improvements (by 0.5\% on
average) due to \dhlii{the} additional latency reductions \dhlii{enabled by}
\myshuffling.\footnote{Note that the \dhlii{main} reason we design DIVA
Shuffling is to improve reliability \sgII{(while using reduced \sgIII{latency} parameters)}, {\em not} performance.} \dhlii{Our
techniques} achieve \dhlii{these} performance \dhlii{improvements} by
dynamically monitoring and optimizing DRAM latency in a reliable manner (using
\myprofiling), while also improving DRAM reliability (using \dhliii{DIVA
Shuffling}). Third, DIVA-DRAM shows less performance sensitivity to temperature
when compared to AL-DRAM \dhliii{(as shown in
Figure~\mbox{\ref{fig:profile_rd_wr}})}. In general, increasing temperature
leads to more \sgII{randomly-distributed single-bit} errors, which limits the
performance benefits from AL-DRAM at high temperatures \dhliii{(as shown for
85\celsius~in Figure~\ref{fig:mech_performance})}. DIVA-DRAM incorporates ECC,
and, hence, is able to correct these single-bit errors, enabling latency
reductions (and performance improvement) similar to what we observe at lower
temperatures.

Figure~\ref{fig:mech_performance} also shows that our techniques outperform
AL-DRAM for all four configurations by 2.5\%/3.4\%/3.2\%/2.6\%, even though we
assume aggressive raw DRAM latency reductions for AL-DRAM
(Section~\ref{sec:mech_result}). We also ignore the fact that AL-DRAM is unable
to account for dynamic latency changes due to aging and wear-out, and is thus an
unrealistic mechanism (Section~\ref{sec:mech_lowlatency}). Considering that
aging or post-packaging failures affect a significant number of DRAM
parts~\cite{sridharan-sc2012, sridharan-sc2013, meza-dsn2015,
schroeder-tdsc2010, li-atc2007, hwang-asplos2012} and AL-DRAM \dhlii{\em cannot}
handle such failures, we conclude that our mechanisms would provide even higher
performance (and reliability) improvements over AL-DRAM in reality than we have
shown.

	\section{Related Work} 
\label{sec:related}

To our knowledge, this is the first work to {\em i)} experimentally demonstrate
and characterize {\em design-induced latency variation} across cells in {\em
real} DRAM chips, {\em ii)} develop mechanisms that take advantage of this
existing {\em design-induced variation} to reliably reduce DRAM latency as well
as \dhlii{to} mitigate errors, and {\em iii)} devise a \dhliii{practical}
mechanism to dynamically determine the lowest latency at which to operate DRAM
reliably.

{\bf Low Latency DRAM Organizations.} There are multiple proposals that aim to
reduce DRAM latency by changing DRAM internals. Our proposals can be combined
with these techniques to further reduce DRAM latency. Son et
al.~\cite{son-isca2013} enable low-latency access to banks near IO pads and
shorten bitlines to some subarrays, which reduces DRAM latency at the expense of
additional chip area~\cite{lee-hpca2013, kim-isca2012}. Our work, on the other
hand, performs a comprehensive experimental analysis of design-induced variation
across wordlines and bitlines at the mat level, and proposes new mechanisms to
take advantage of such mat-level latency variation. Lee et
al.~\cite{lee-hpca2013} propose TL-DRAM, a new subarray organization that
enables lower access latency to cells near local sense amplifiers. To achieve
this, TL-DRAM adds isolation transistors to separate a bitline into near and far
segments, thereby adding \dhlii{a small but} non-negligible area overhead to
DRAM. RL-DRAM reduces DRAM latency by using smaller subarrays~\cite{rldram}, but
this comes at a significant increase in chip area. In contrast to all these
works, \mydram reduces latency and mitigates DRAM errors with \dhlii{\em no
changes} to the DRAM mat design. Furthermore, while prior
works~\cite{lee-hpca2013, son-isca2013} are based on simulation results using a
circuit-level DRAM model, we profile {\em real} DIMMs and experimentally analyze
design-induced variation. Our \dhliii{new} method of finding the \dhlii{slowest}
regions \dhlii{in DRAM}, DIVA Profiling, is applicable to all these prior works.

{\bf Exploiting Process and Temperature Variations to Reduce DRAM Latency.} Lee
et al.'s AL-DRAM~\cite{lee-hpca2015} and Chandrasekar et
al.~\cite{chandrasekar-date2014} lower DRAM latency by leveraging latency
variation in DRAM due to the manufacturing process and temperature dependency.
In contrast to our work, these two works are different in two major ways. First,
they are not aware of and do not exploit {\em design-induced latency variation}
in DRAM, which is due to the design and placement of components in a DRAM chip
and is {\em independent of the manufacturing process and temperature}. Unlike
process variation, design-induced variation, as we have experimentally shown (in
Section~\ref{sec:profile}), {\em i)} is dependent on the internal design of
DRAM, {\em ii)} does not change over time, and {\em iii)} is similar across DRAM
chips that have the same design. Second, these two works do \dhlii{\em not}
provide an \dhlii{online} method for dynamically identifying the lowest latency
at which to operate DRAM reliably. Instead, they assume such latencies are
provided by the DRAM interface, which {\em i)} not only is difficult to achieve
due to increased cost on the DRAM manufacturer's end and \dhliii{the} difficulty
in changing the DRAM standard, {\em ii)} but also cannot adapt to
\dhlii{increases} in actual DRAM latency over time due to aging and wearout (and
therefore would lead to large margin in the provided latencies). Finally,
neither of these two works develop an online profiling or error correction
mechanism, which our work develops. We have already provided both extensive
qualitative (Section~\ref{sec:mech_lowlatency}) and quantitative
(Section~\ref{sec:mech_result}) comparisons to AL-DRAM and shown that our
mechanism significantly outperforms AL-DRAM, without requiring a priori
knowledge of the lowest latency at which to operate DRAM reliably (which AL-DRAM
does require), even when our simulations assume \dhlii{that} AL-DRAM provides
very aggressive latency reductions (ignoring the fact that AL-DRAM does not
account for aging and wearout).

{\bf Experimental Study of DRAM Failures.} Many works\sgII{~\cite{kim-isca2014,
khan-sigmetrics2014, lee-hpca2015, liu-isca2013, kim-edl2009,
chandrasekar-date2014, chang-sigmetrics2016, khan-cal2016, chang-sigmetrics2017,
patel-isca2017, khan-dsn2016, qureshi-dsn2015, kim-thesis2015, lee-thesis2016,
chang-thesis2017}} provide experimental studies and models for DRAM errors due to
different type of failures such as: {\em i)} retention time
failures~\cite{khan-sigmetrics2014, liu-isca2013, kim-edl2009, qureshi-dsn2015,
patel-isca2017, khan-dsn2016, khan-cal2016}, {\em ii)} wordline coupling
failures\sgII{~\cite{kim-isca2014, mutlu-date2017, kim-thesis2015}}, {\em iii)}
failures due to lower timing parameters\sgII{~\cite{lee-hpca2015,
chandrasekar-date2014, chang-sigmetrics2016, lee-thesis2016, chang-thesis2017}},
\dhliii{and {\em iv)} failures due to reduced-voltage
operation}~\cite{chang-sigmetrics2017, chang-thesis2017}. Specifically, Chang et
al.~\cite{chang-sigmetrics2016} observe the non-uniform distribution of DRAM
errors due to reduced latency, but \dhliii{do} not provide the fundamental
reasoning behind this non-uniformity. This work also proposes reducing DRAM
latency for some cells, but does not provide a mechanism for \dhliii{\em
finding} the lowest DRAM latency and instead assumes that the latency of each
cell is provided by the DRAM device. Our experiments and analyses focus on
understanding failures due to reducing latency in {\em design-induced
vulnerable} regions in DRAM, which has not been studied by any of these works.
Previous failure modes, e.g., Row Hammer\sgII{~\cite{kim-isca2014, mutlu-date2017,
kim-thesis2015}} or retention failures\sgII{~\cite{kim-edl2009, liu-isca2013,
patel-isca2017}}, do \dhliii{\em not} exhibit design-induced variation, i.e.,
they are not dependent on cell distance from peripheral DRAM structures\dhlii{,}
as shown in prior work~\cite{kim-isca2014, kim-edl2009}.

{\bf Study of DRAM Failures in Large Scale Systems.} Many previous
works~\cite{sridharan-sc2012, sridharan-sc2013, meza-dsn2015,
schroeder-tdsc2010, li-atc2007, hwang-asplos2012, schroeder-sigmetrics2009}
study DRAM errors in large scale systems (e.g., a server cluster \dhlii{or many
data centers}) and analyze the \dhlii{system-level impact on DRAM failures,
e.g.,} power fluctuation, operating temperature, wearout\dhlii{,} etc. Our
analyses are orthogonal to these studies and focus on the impact of internal
DRAM organization on latency and error characteristics.

{\bf DRAM Error Mitigation Techniques.} To increase system reliability and
efficiency, many error correction codes~\cite{alameldeen-isca2011, kim-hpca2015,
wilkerson-isca2010, luo-dsn2014} have been proposed specifically in the context
of DRAM error mitigation~\cite{khan-sigmetrics2014}.
VS-ECC~\cite{alameldeen-isca2011} proposes variable strength error correction
codes for better performance and energy efficiency.
HI-ECC~\cite{wilkerson-isca2010} increases power efficiency for
\dhliii{high-capacity} eDRAM-based caches by integrating a strong error
correction code.

Our proposals complement existing ECC mechanisms and achieve better performance
and reliability. First, having ECC alone (regardless of ECC strength) is not
enough to guarantee correct operation with maximum latency reduction, since it
is not possible to determine the smallest value for each timing parameter
without profiling. \myprofiling can do so, enabling maximum latency reduction
while leveraging ECC support to correct failures. Second, \myshuffling enables
greater reliability in the presence of an ECC mechanism by distributing possible
errors over different ECC \dhliii{codewords}. Third, our work opens up new research
opportunities to exploit design-induced variation in combination with different
ECC schemes. For example, variable-strength ECC~\cite{alameldeen-isca2011} can
exploit awareness of design-induced variation by adjusting ECC strength based on
error probability indications/predictions from design-induced variation.

{\bf DRAM Latency Reduction with In-Memory Communication and Computation.}
Transferring data over the memory channel leads to long latency and delays other
data transfers. To reduce this latency, prior works offload bulk data
movement\sgII{~\cite{seshadri-micro2013, chang-hpca2016, lee-pact2015}} or computation
operations (e.g., \sgII{\cite{seshadri-cal2015, ahn-isca2015a, ahn-isca2015b,
patterson-ieeemicro1997, mirzadeh-asbd2015, kogge-icpp1994, kocberber-micor2013,
guo-wondp2014, gao-pact2015, farahani-hpca2015, boroumand-cal2016,
hsieh-isca2016, seshadri-arxiv2016, stone-tc1970, seshadri-micro2015,
hsieh-iccd2016, pattnaik-pact2016}}) to DRAM. These works do \dhliii{\em not}
fundamentally reduce the access \dhliii{latency} to the {\em DRAM array},
whereas our proposal \mydram does. Hence, \mydram is complementary to such
in-memory communication and computation mechanisms.

{\bf DRAM Latency Reduction Based on Memory Access Patterns.} Prior
works~\cite{shin-hpca2014, hassan-hpca2016} show that DRAM leakage affects two
DRAM timing parameters (\trcd/\tras), and recently-accessed rows have more
charge. \dhliii{This} allows \dhlii{such} rows to be \dhlii{reliably} accessed
with a lower latency than the DRAM standard. Our approach of reducing latency by
taking advantage of design-induced variation is complementary to these works.

	\section{Conclusion} 
\label{sec:conclusion}

This paper provides the first study that experimentally characterizes and
exploits the phenomenon of {\em design-induced variation} in real DRAM
chips\dhlii{.} Based on a detailed experimental analysis of 768 modern DRAM
chips from three major manufacturers, we \dhliii{find} that there is widespread
variation in the access latency required for reliable operation of DRAM cells,
depending on how close or far the cells are to the peripheral structures that
are used to access \dhlii{them}. We \dhliii{introduce} \mydram, which consists
of two novel techniques that take advantage of design-induced variation to {\em
i)} reduce DRAM latency reliably at low cost and {\em ii)} improve reliability
by making ECC more effective. \dhliii{\em \myprofiling} reduces DRAM latency by
finding the lowest latency at which to operate DRAM reliably, by dynamically
profiling certain cells that are most vulnerable to failures \dhliii{caused by
reduced-latency operation,} due to the design of the DRAM chip. \dhliii{\em
\myshuffling} improves DRAM reliability by intelligently shuffling data such
that errors induced due to \dhlii{reduced-latency} operation become correctable
by ECC. Our comprehensive experimental evaluations demonstrate that \mydram can
greatly reduce DRAM read/write latency, leading to significant system
performance improvements on a variety of workloads and system configurations,
compared to both modern DRAM and the state-of-the-art Adaptive-Latency
DRAM~\cite{lee-hpca2015}. We conclude that exploiting the design-induced latency
variation inherent in DRAM using our new techniques provides a promising,
reliable, and low-cost way of significantly reducing DRAM latency. We hope that
our comprehensive experimental characterization and analysis of design-induced
variation in \dhlii{modern DRAM chips} enables the development of other
mechanisms to improve DRAM latency and reliability.

	\section*{ACKNOWLEDGMENTS} 
\label{sec:acknowledgement}

We thank the reviewers of SIGMETRICS 2017, \dhliii{HPCA 2017}, MICRO 2016, HPCA
2016, and ISCA 2016 for their comments. \dhliv{We especially thank the
reviewers of SIGMETRICS 2017 for their constructive and insightful comments.} An
earlier version of this work was posted on arXiv~\cite{lee-arxiv2016}. We
acknowledge the generous support of Google, Intel, NVIDIA, Samsung, and VMware.
This work is supported in part by NSF grants 1212962, 1320531, and 1409723, the
Intel Science and Technology Center for Cloud Computing, and the Semiconductor
Research Corporation.

%	\newpage
	\bibliographystyle{abbrv}
	\bibliography{paper}
    \balance
	
	\newpage
	\newpage
	\appendix
    \section*{Appendix}
	%\section*{APPENDIX\vspace{0.2in}}

\section{Latency Overhead of DIVA Profiling} 
\label{sec:appendix_overhead}

In Section~\ref{sec:mech_lowlatency}, we calculate the time it takes to perform
both \myprofiling and conventional DRAM profiling (where each DRAM row is
tested)~\cite{nair-isca2013,liu-isca2012,venkatesan-hpca2006,khan-sigmetrics2014}.

DRAM profiling consists of two steps: {\em i)} writing data to the cells
\dhlii{that are} being tested, and {\em ii)} reading and verifying cell
contents. Therefore, the profiling time $t$ is calculated as:
\begin{equation}
t = \frac{\mbox{$Number Of DRAM Columns Tested$}}{\mbox{$DIMM
Bandwidth$}}\times{\mbox{$Pattern Count$}\times{2}}
\end{equation}
\noindent where we determine the fastest rate at which a column command can be
performed, and then multiply it by the number of patterns \dhlii{that are} being
tested, and by \sgII{two} because we perform a read and a write (i.e., two column
commands) to each DRAM column.

When testing a 4GB DDR3-1600 DIMM (whose DIMM bandwidth is 1600 Mbps/pin
$\times$ 64 pins $=$ 102.4~Gbps) with one test pattern, conventional DRAM
profiling mechanisms take 625~ms to test all 4GB of DRAM cells. However, since
\myprofiling needs to test \dhliii{\em only} 8MB of cells (i.e., just one row
per each 512-row subarray), it takes only 1.22~ms to complete its test.

\section{DRAM Simulation to Validate Our Hypotheses on Design-Induced Variation}
\label{sec:appendix_simulation}

We hypothesize that accessing a cell that is physically \dhliii{farther} from the
structures \dhlii{that are} required to perform the access (e.g., \dhliii{the
sense amplifiers, the wordline drivers}) takes a longer time than accessing a
cell that is closer to them. Our observations in Section~\ref{sec:profile} support this
hypothesis empirically, but they do not provide absolute proof because they are
based on observations on DRAM chips whose internal circuitry is not publicly
provided \dhlii{and thus is unknown to us}. To verify our hypothesis\dhlii{,}
\dhlii{we simulate the effects of the distance between a cell and the structures
required to perform the access in a DRAM mat by using a detailed SPICE circuit
model.} \dhlii{Our \dhliii{SPICE} simulation model and parameters are publicly
available}~\cite{safari-git}.

\begin{figure}[h]
	\centering
	\vspace{-0.10in}
	\includegraphics[width=\linewidth]{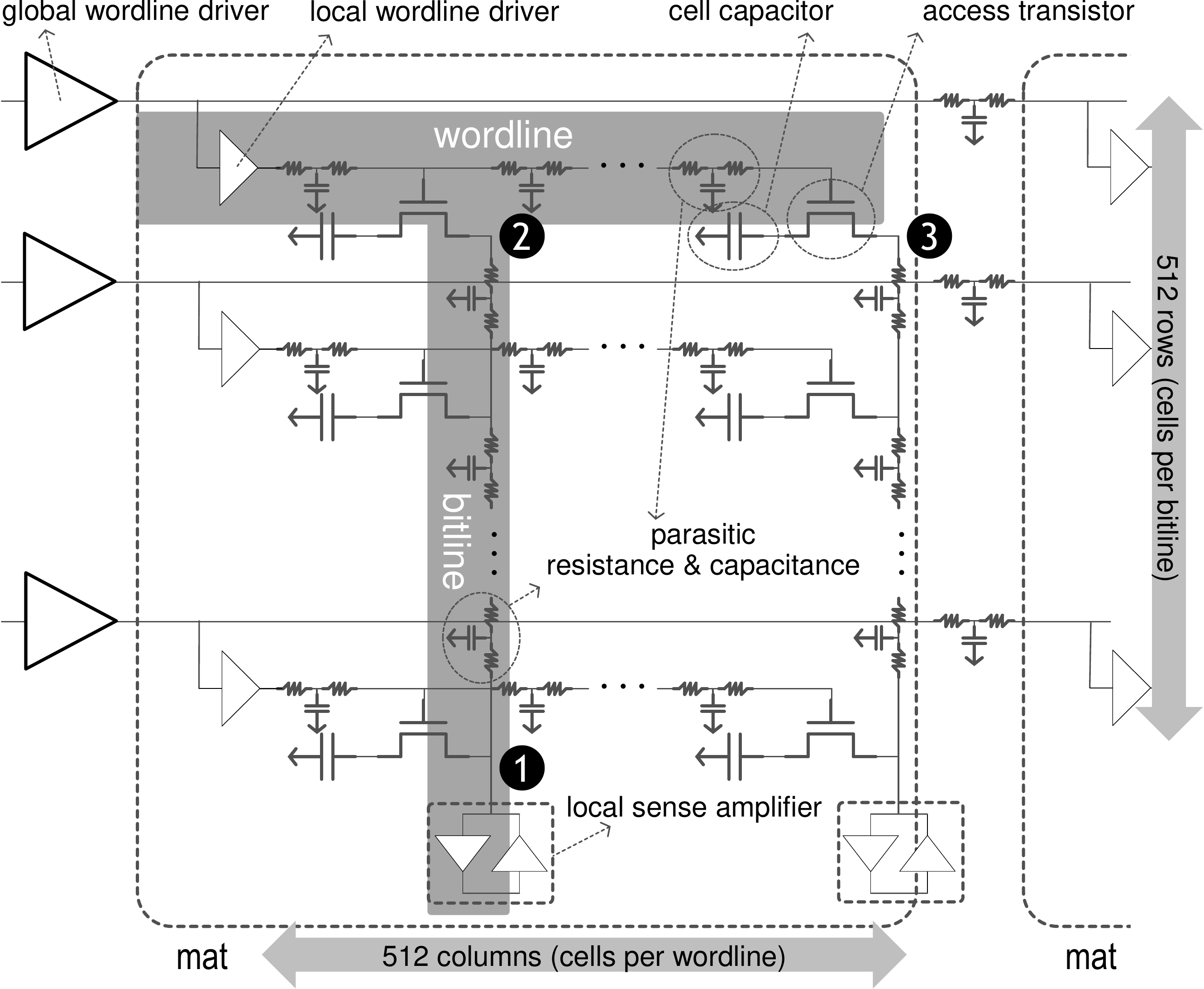}
	\vspace{-0.10in}
	\caption{Detailed Mat Model~\cite{vogelsang-micro2010, keeth-book},
	Including Parasitic Resistance and Capacitance, Used in Our Circuit
	Simulation}
	\label{fig:simmodel}
	\vspace{-0.10in}
\end{figure}

{\bf Detailed Mat Model.} We first build a DRAM mat model with a detailed wire
model, as shown in Figure~\ref{fig:simmodel}. Our mat model consists of a 512 x 512 array of
DRAM cells, which is commonly used in modern DRAM
chips~\cite{vogelsang-micro2010}. Each 512-cell column is connected to a sense
amplifier over a bitline\dhlii{,} which is plotted as \dhlii{the} vertical gray
block in Figure~\ref{fig:simmodel}. Each bitline has its own parasitic resistance and
capacitance. We expect that due to the bitline's parasitic resistance and
capacitance, accessing a cell \dhliii{\em farther} from a sense amplifier (e.g.,
cell~\ding{183}) takes a longer time than accessing a cell that is {\em closer}
to the same sense amplifier (e.g., cell~\ding{182}). Each 512-cell row is
connected to a local wordline driver over a wordline (512 local wordline drivers
in total), which is plotted as \dhlii{the} horizontal gray block in Figure~20.
Each wordline has its own parasitic resistance and capacitance. We expect that
due to the wordline's parasitic resistance and capacitance, accessing a cell
\dhliii{\em farther} from a wordline driver (e.g., cell~\ding{184}) takes a longer time
than accessing a cell that is {\em closer} to the same wordline driver (e.g.,
cell~\ding{183}). 

\begin{figure}[h]
	\vspace{0.05in}
	\centering
	\subfloat[Design-Induced Variation in Bitline] {
		\includegraphics[height=1.55in]{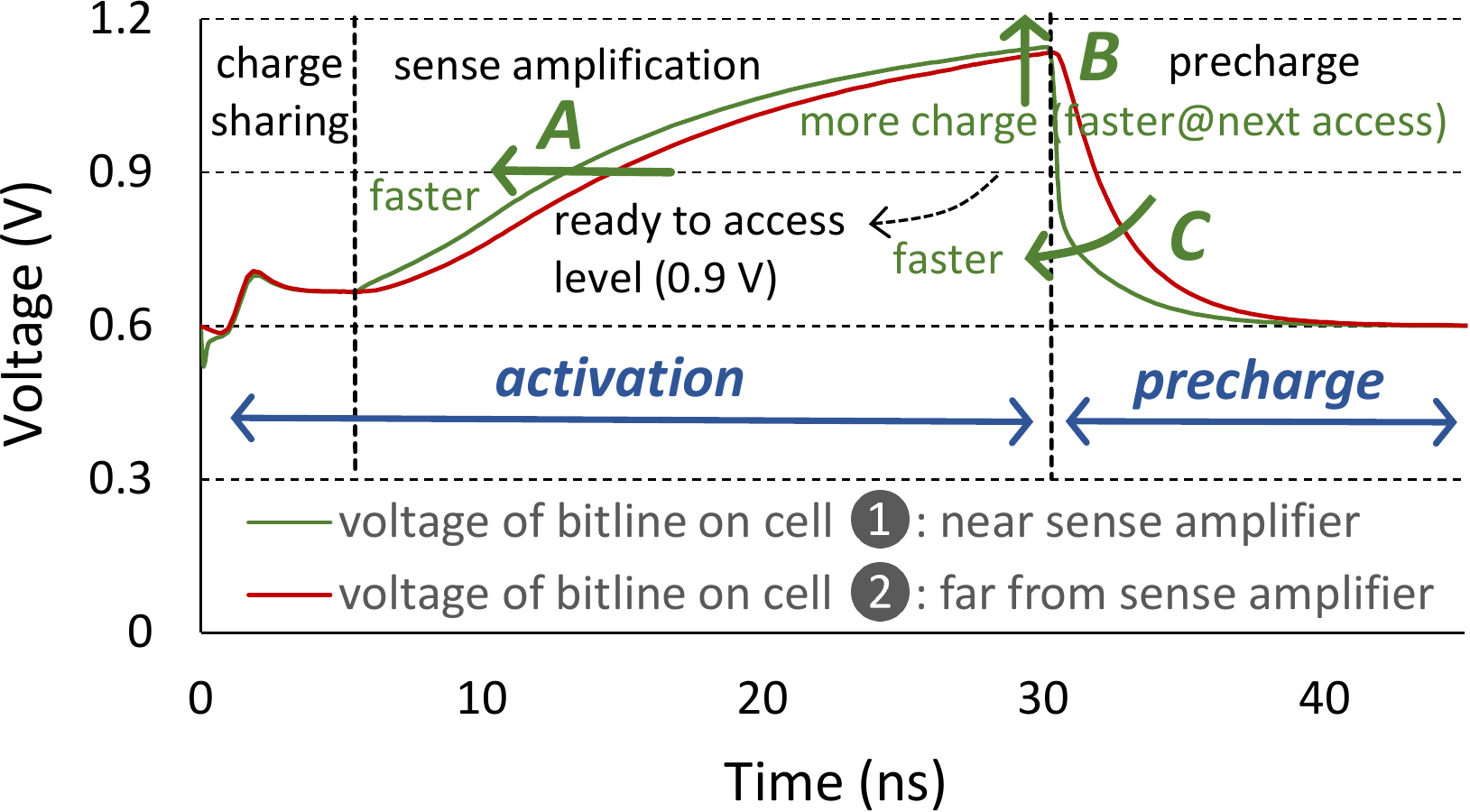}
		\label{fig:sim_bitline}
	}

	\vspace{-0.05in}
	\subfloat[Design-Induced Variation in Wordline] {
		\includegraphics[height=1.55in]{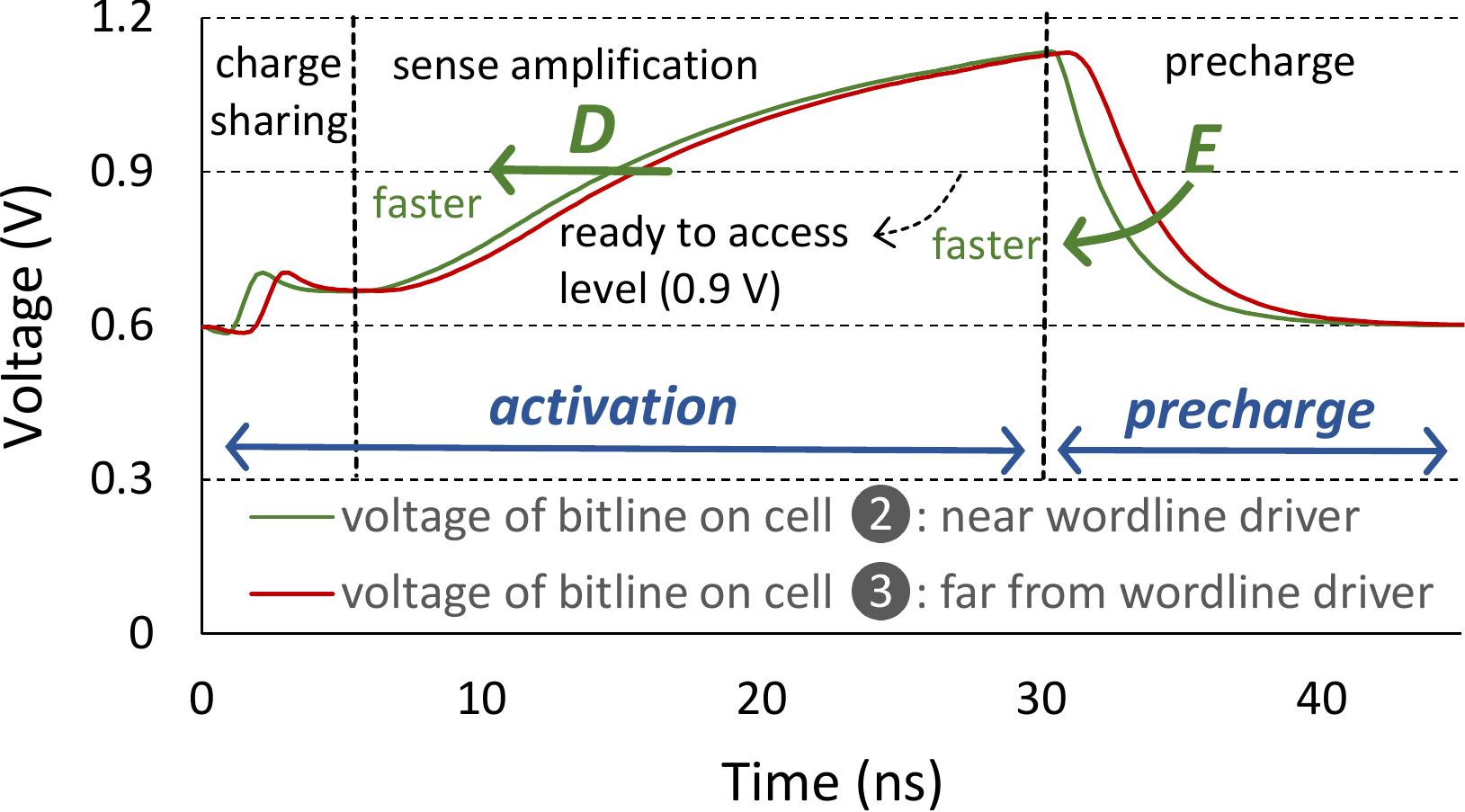}
		\label{fig:sim_wordline}
	}
	\vspace{-0.10in}
	\caption{Simulation Results Showing Access Latency Difference
	Between Cells Nearby and Far from Peripheral Structures}
	\label{fig:simulation}
	\vspace{-0.05in}
\end{figure}

{\bf Simulation Methodology.} To simulate the access latency for cells in
different locations, we use technology parameters from \dhlii{a 55 nm DRAM
model}~\cite{vogelsang-micro2010} and from \dhliii{a} 45 nm \dhliii{logic}
\dhlii{process} model~\cite{ptm, zhao-isqed2006} to construct a detailed
circuit-level SPICE simulation model. We assume that \dhlii{the} cell
capacitance is 24 fF, and the bitline capacitance is 144
fF~\cite{vogelsang-micro2010}. The cell and sense amplifier operating voltage is
1.2V, while the wordline operating voltage is 3.0V. In our evaluation, we issue
the \cmdact command at 0~ns and \cmdprech at 30~ns, which replicates the
behavior of using a {\em reduced} \tras timing parameter (the standard \tras is
35~ns~\cite{ddr3}). We plot the circuit-level SPICE simulation results in
Figure~\ref{fig:simulation}. Figure~\ref{fig:sim_bitline} shows the variation on voltage levels of the
\dhlii{bitlines} for two different cells: {\em i)} a cell that is near the sense
amplifier (cell~\ding{182} in Figure~\ref{fig:simmodel}), and {\em ii)} a cell that is far from
the sense amplifier (cell~\ding{183} in Figure~\ref{fig:simmodel}). Similarly, Figure~\ref{fig:sim_wordline} shows
the variation on voltage levels of the bitline for two different cells: {\em i)}
a cell that is near the wordline driver (cell~\ding{183} in Figure~\ref{fig:simmodel}), and {\em
ii)} a cell that is far from the wordline driver (cell~\ding{184} in Figure~\ref{fig:simmodel}).
We explain the figures and our results in detail below, but the key conclusion
is that the voltage level of the cell that is closer to the sense amplifier
(cell~\ding{182}) becomes higher (and lower) more quickly than that of the cell
that is \dhliii{farther} from the sense amplifier (cell~\ding{183}), as shown in
Figure~\ref{fig:sim_bitline}. \dhlii{The same} observation is true for the cell that is closer to
the wordline driver (cell~\ding{183}) vs.\ the cell that is \dhliii{farther}
from the wordline driver (cell~\ding{184}). Since the voltage level of a cell
that is closer to the sense amplifier or \dhlii{the} wordline driver becomes
higher (or lower) more quickly, that cell can be accessed faster. We explain
this \dhlii{phenomenon} in \dhliii{more} detail below. 

{\bf DRAM Row and Column Access.} There are three steps performed to access data
in a DRAM cell. The first step is selecting a wordline. Inside DRAM, there are
{\em i)} a global wordline (that stretches over the entire subarray) and {\em
ii)} multiple local wordlines (each of which stretches over a single mat).
Enabling a global wordline driver raises the voltage of the global wordline.
Then, the global wordline enables multiple local wordlines. Each wordline turns
on 512 access transistors, connecting one cell capacitor in each column to its
corresponding bitline. We call this step {\em charge sharing} in
\dhlii{Figure~\ref{fig:simulation}}. Figures~\ref{fig:sim_bitline} and \ref{fig:sim_wordline} \dhlii{show that charge sharing becomes}
enabled by raising a wordline between 0~ns and 5~ns. In this example, the
bitline voltage level (which is initially precharged to 0.6V, \hvdd)
\dhlii{increases} due to \dhlii{the sharing of charge from} the connected cell
(\dhliii{which we assume is} initially fully charged to 1.2V, \vdd). 

Second, after charge sharing, the sense amplifiers are enabled, starting to
detect the voltage perturbation \dhlii{caused by} the charge sharing operation
and amplifying the bitline voltage level toward 1.2V (\vdd). We call this
\dhlii{step} {\em sense amplification} \dhlii{in Figure~\ref{fig:simulation}}. Figure~\ref{fig:simulation} shows
sense amplification taking place between 5~ns to 30~ns. During sense
amplification, a sense amplifier can \dhliii{reliably} transfer the detected
data to the IO circuitry when the voltage level reaches 0.9V (half way between
\hvdd and \vdd). \dhliii{In other words, the data becomes ready to access at the
bitline voltage level of 0.9V.}

Third, after finishing sense amplification, \dhliii{in order to prepare
the subarray for an access to another row,} 
% the raised wordline voltage should be brought back down to 0V, which takes
% about 5~ns (from 25~ns to 30~ns). Then, 
the bitline voltage should be reduced to 0.6V (the initial voltage level of the
bitline, \hvdd), to allow access to cells in a different row. We call this
\dhlii{step} {\em precharge} \dhlii{in Figure~\ref{fig:simulation}}. Figures~\ref{fig:sim_bitline} and \ref{fig:sim_wordline} show the
precharge of a bitline taking place between 30~ns to 40~ns. 

For these three steps, we can understand the latency of each step by examining
the bitline voltage level. The access \dhlii{latencies} of these operations
\dhlii{are} determined based on how quickly the bitline voltage level changes.
For example, the latency of activation depends on how fast the bitline level can
\dhlii{reach} \vdd. Similarly, the latency of the precharge operation depends on
how fast the bitline level can return to \hvdd. In the next two paragraphs, we
observe the latency of accessing cells in different locations in a mat (cells
\ding{182}, \ding{183}, and \ding{184}), \dhlii{as shown in Figures~\ref{fig:simmodel} and \ref{fig:simulation}.}

{\bf Accessing Cells \dhliii{on} the Same Bitline.} We evaluate and compare two
cases for accessing cells \dhliii{on} the same bitline: {\em i)} a cell that is
near a sense amplifier (labeled cell~\ding{182}), and {\em ii)} a cell that is
far from the sense amplifier (labeled cell~\ding{183}). Note that we use the
same labels to describe the same cells in Figures~\ref{fig:simmodel} and \ref{fig:simulation}. Figure~\ref{fig:sim_bitline} shows
the voltage levels of the bitline when measured near the accessed cells
(\dhlii{cells}~\ding{182} and~\ding{183}). We make three major observations.
First, the cell that is near the sense amplifier (cell~\ding{182}) finishes {\em
sense amplification} \dhlii{\em earlier} than the cell that is far from the
sense amplifier (cell~\ding{183}), \dhlii{as} pointed to by label {\bf $A$} in
Figure~\ref{fig:sim_bitline}. This is mainly due to the {\em additional} parasitic resistance and
capacitance required for accessing cell~\ding{183}, which causes its voltage
level to rise more slowly. Second, 
% when reducing the restoration time (which \dhlii{is} represented with the
% \tras timing parameter), 
the restored voltage level of a cell (i.e., the highest voltage \dhliii{level}
of \dhlii{a bitline} in Figure~\ref{fig:sim_bitline}) that is near the sense amplifier
\dhliii{(cell~\ding{182})} is higher than the level of the cell that is far from
the sense amplifier \dhliii{(cell~\ding{183})}, \dhlii{as} pointed to by label
$B$ in Figure~\ref{fig:sim_bitline}. Therefore, when reducing the restoration time (\tras), the
cell that is far from the sense amplifier (cell~\ding{183}) holds less charge
than the cell that is near the sense amplifier (cell~\ding{182}). Due to the
smaller amount of charge in the far cell, accessing the far cell takes a longer
time than accessing the near cell. Third, precharging the bitline when accessing
the near cell takes less time than when accessing the far cell, \dhlii{as shown
by the voltage} level of cell~\ding{182} dropping much faster than that of
cell~\ding{183} during the \dhlii{\em precharge} operation (pointed to by label
$C$ in Figure~\ref{fig:sim_bitline}). Therefore, reducing \dhlii{the precharge timing parameter}
(\trp) \dhlii{might be fine for the near cell (as the bitline can still return
to full} \hvdd \dhlii{within the reduced} \trp). \dhlii{However, for the far
cell, reducing the precharge timing parameter can result} in the bitline not
fully returning to \hvdd after we access the far cell. If \dhlii{the} next
access is to a cell whose \dhlii{charge state (i.e., charged/discharged)} is
different from the cell we just accessed,
% it will take longer for that subsequent access to reach sense amplification, 
\dhliii{it will take longer for the next access to be ready,} as the bitline
voltage needs to change by a greater amount. From these detailed circuit-level
evaluations, we conclude that accessing a cell that is far from the sense
amplifier takes a longer time than accessing a cell that is near the sense
amplifier. 

{\bf Accessing Cells \dhliii{on} the Same Local Wordline.} We evaluate and
compare two cases for accessing cells \dhliii{on} the same wordline \dhlii{in
Figures~\ref{fig:simmodel} and \ref{fig:simulation}}: {\em i)} a cell that is near a local wordline driver
(labeled cell~\ding{183}), and {\em ii)} a cell that is far from the local
wordline driver (labeled cell~\ding{184}). Figure~\ref{fig:sim_wordline} shows the voltage levels
of the corresponding bitlines of the two cells when measured near the accessed
cells. The key observation from the figure is that accessing a cell that is far
from the local wordline driver takes a longer time than accessing a cell that is
near the local wordline driver. This is mainly because the wordline has a large
resistance and capacitance \dhlii{and}, thus it takes longer for the activation
signal to reach the far cell. As a result, the voltage level of the nearby cell
becomes higher than that of the far cell, after an activation operation\dhlii{,}
as pointed to by label $D$ in Figure~\ref{fig:sim_wordline}. \dhliii{Also}, precharging is faster
for the nearby cell because its voltage level \dhlii{gets} closer to \hvdd
\dhlii{(0.6 V)} much faster than that of the far cell, as pointed to by label
$E$ in Figure~\ref{fig:sim_wordline}.  Similarly, other control signals (e.g., sense amplifier
enable, equalizer enable) also \dhlii{experience} wire propagation delay
\dhlii{that is higher when accessing the far cell}. As a result, the operations
that take place when accessing a cell \dhlii{farther away} from the wordline
driver require a longer time to complete.

In summary, in our detailed circuit-level simulations, we observe that accessing
a cell that is \dhlii{farther} from the structures \dhlii{that are} required to
perform the access (e.g., \dhlii{the sense amplifiers and the wordline drivers})
takes a longer time than accessing a cell that is closer to \dhliii{such
structures}. Based on these evaluations, we conclude that cells in a DRAM mat
have different latency characteristics based on their location, which leads to a
major source of design-induced variation.

\section{Design-Induced Variation vs. Process Variation} 
\label{sec:pvt}

We observe two types of errors: {\em i)} errors caused by \dhliii{process
variation} that \sgI{is} usually randomly distributed over the entire DRAM
chip~\cite{khan-sigmetrics2014, cha-hpca2017}, and {\em ii)} errors caused by
design-induced variation that are concentrated in specific regions (as we showed
in Section~\ref{sec:mech_lowlatency}). There are cases where \dhliii{the effect of} design-induced
variation \dhliii{on latency} is greater than \dhliii{that of} \dhlii{process}
variation, and there are cases where \dhliii{the effect of} \dhlii{process}
variation is greater. \dhlii{Our mechanism, DIVA-DRAM,} enables reliable
operation in both cases. The total DRAM latency variation is the sum of
design-induced variation and \dhlii{process} variation. We provide a separate
mechanism \dhlii{to reliably handle} each type of variation: {\em i)} online
\dhlii{\myprofiling} to minimize latency \dhlii{by exploiting} design-induced
variation, and {\em ii)} \dhlii{ECC, strengthened with DIVA Shuffling, to
provide high reliability in the presence of} \dhliii{process} variation. Because
we provide ECC with improved reliability to account for the presence of
\dhliii{process} variation, we are able to safely harness the performance
improvements offered by our \dhlii{exploitation} of design-induced variation,
even when \dhliii{the effect of process} variation is higher.

\dhlii{We note that even in situations where \dhlii{process} variation changes
from DIMM to DIMM, one can still exploit design-induced variation for better
performance and reliability by embedding the DIMM-specific information
\dhliii{(i.e., the addresses of the \dhliv{slowest regions that can be used for the latency test regions in DIVA Profiling}, the
external-to-internal address mapping information)} within the DRAM module (e.g.,
inside the serial-presence-detect EEPROM in a DRAM module, as described
in~\mbox{\cite{kim-isca2012}}), and providing this information to the memory
controller.}

\newpage \onecolumn
\section{List of Tested DIMMs} 
\label{sec:appendix_summary}

We report a short summary of the properties and \dhlii{design-induced}
vulnerability of each of the 96~DIMMs (from three major DRAM vendors) presented
in this paper, separated by vendor\dhlii{,} in Tables~\ref{tbl:vendor1},
\ref{tbl:vendor2}, and \ref{tbl:vendor3}. The evaluated DIMMs are manufactured
in the period from 2010 to 2013. While newer DIMMs enable higher capacity and
bandwidth, the DRAM cell array architecture of these newer DIMMs has not changed
significantly from the architecture of the DIMMs we evaluate~\cite{keeth-book}.
Therefore, we believe that our observations on DRAM latency variation hold true
for more recently manufactured \dhliii{DRAM chips}.

\definecolor{lightgray}{gray}{0.95}

\begin{table}[h]

\setlength{\tabcolsep}{3pt}
\centering
\footnotesize

\begin{tabular}{cccccccccccc}

\toprule

\multirow{2}{*}[-2pt]{\em \footnotesize \centering Vendor} & 
\multirow{2}{*}[-2pt]{\em \footnotesize \centering Module} & 
                     {\em \footnotesize \centering Date$^{\ast}$} &
\multicolumn{2}{c}{\em \footnotesize \centering Timing$^{\dagger}$} &
\multicolumn{2}{c}{\em \footnotesize \centering Organization} &
\multicolumn{3}{c}{\em \footnotesize \centering Chip} &
\multicolumn{2}{c}{\em \footnotesize \centering Vulnerability Ratio$^{\star}$} \\

\cmidrule(lr){3-3} \cmidrule(lr){4-5} \cmidrule(lr){6-7} \cmidrule(lr){8-10} \cmidrule(lr){11-12} 

 & 
 & {\em \footnotesize (yy-ww)} 
 & {\em \footnotesize Freq~(MT/s)} 
 & {\em \footnotesize tRC (ns)} 
 & {\em \footnotesize Size (GB)} 
 & {\em \footnotesize Chips} 
 & {\em \footnotesize Size (Gb)$^{\ddagger}$} 
 & {\em \footnotesize Pins} 
 & {\em \footnotesize Die Version$^{\S}$} 
 & {\em \footnotesize tRP} 
 & {\em \footnotesize tRCD} \\

\midrule

%% HYNIX
%\cellcolor{white} 
& \module{A}{ 1}{} & 10-18 & 1333 & 49.125 & 2 	& 8 & 2 & $\times$8  & $\mathcal{A}$ 	& 9.9  & 2.3	\\
%\cellcolor{white} 
& \module{A}{ 2}{} & 10-20 & 1066 & 50.625 & 2 	& 8 & 2 & $\times$8  & $\mathcal{A}$ 	& 23.4 & 440	\\
%\cellcolor{white} 
& \module{A}{ 3}{} & 10-22 & 1066 & 50.625 & 2 	& 8 & 2 & $\times$8  & $\mathcal{A}$ 	& 29   & 16.5	\\
%\cellcolor{white} 
& \module{A}{ 4}{} & 10-23 & 1066 & 50.625 & 2 	& 8 & 2 & $\times$8  & $\mathcal{A}$ 	& 3.4  & 4.1	\\
%\cellcolor{white} 
& \module{A}{ 5}{} & 10-26 & 1333 & 49.125 & 2 	& 8 & 2 & $\times$8  & $\mathcal{B}$ 	& 5.6  & 11.2 \\
%\cellcolor{white} 
& \module{A}{ 6}{} & 10-26 & 1333 & 49.125 & 2 	& 8 & 2 & $\times$8  & $\mathcal{B}$ 	& 5.7  & 20.3 \\
%\cellcolor{white} 
& \module{A}{ 7}{} & 10-43 & 1333 & 49.125 & 1 	& 8 & 1 & $\times$8  & $\mathcal{T}$ 	& 5837 & 764	\\
%\cellcolor{white} 
& \module{A}{ 8}{} & 10-51 & 1333 & 49.125 & 2 	& 8 & 2 & $\times$8  & $\mathcal{B}$ 	& 5.6  & 290  \\
%\cellcolor{white} 
& \module{A}{ 9}{} & 11-12 & 1333 & 46.25  & 2 	& 8 & 2 & $\times$8  & $\mathcal{B}$ 	& --   & --   \\
%\cellcolor{white} 
& \module{A}{10}{} & 11-19 & 1333 & 46.25  & 2 	& 8 & 2 & $\times$8  & $\mathcal{B}$ 	& 2.4  & 2.0  \\
%\cellcolor{white} 
& \module{A}{11}{} & 11-19 & 1333 & 46.25  & 2 	& 8 & 2 & $\times$8  & $\mathcal{B}$ 	& --   & --   \\
%\cellcolor{white} 
& \module{A}{12}{} & 11-31 & 1333 & 49.125 & 2 	& 8 & 2 & $\times$8  & $\mathcal{B}$ 	& 4.3  & --	  \\
%\cellcolor{white} 
& \module{A}{13}{} & 11-42 & 1333 & 49.125 & 2 	& 8 & 2 & $\times$8  & $\mathcal{B}$ 	& 4.9  & 93.7 \\
%\cellcolor{white} 
& \module{A}{14}{} & 12-08 & 1333 & 49.125 & 2 	& 8 & 2 & $\times$8  & $\mathcal{C}$ 	& 96.7 & 28.6 \\
%\cellcolor{white} 
& \module{A}{15}{} & 12-12 & 1333 & 49.125 & 2 	& 8 & 2 & $\times$8  & $\mathcal{C}$ 	& 3.9  & 45.2 \\
%\cellcolor{white} 
& \module{A}{16}{} & 12-12 & 1333 & 49.125 & 2 	& 8 & 2 & $\times$8  & $\mathcal{C}$ 	& 103  & 373  \\
%\cellcolor{white} 
& \module{A}{17}{} & 12-20 & 1600 & 48.125 & 2 	& 8 & 2 & $\times$8  & $\mathcal{C}$ 	& 31.4 & 178  \\
%\cellcolor{white} 
& \module{A}{18}{} & 12-20 & 1600 & 48.125 & 2 	& 8 & 2 & $\times$8  & $\mathcal{C}$ 	& --   & --   \\
%\cellcolor{white} 
& \module{A}{19}{} & 12-24 & 1600 & 48.125 & 2 	& 8 & 2 & $\times$8  & $\mathcal{C}$ 	& 37.1 & 21.3 \\
%\cellcolor{white} 
& \module{A}{20}{} & 12-26 & 1600 & 48.125 & 2 	& 8 & 2 & $\times$8  & $\mathcal{C}$ 	& 26.7 & 26.9 \\
%\cellcolor{white} 
& \module{A}{21}{} & 12-32 & 1600 & 48.125 & 2 	& 8 & 2 & $\times$8  & $\mathcal{C}$ 	& 61.3 & 160  \\
%\cellcolor{white} 
& \module{A}{22}{} & 12-37 & 1600 & 48.125 & 2 	& 8 & 2 & $\times$8  & $\mathcal{C}$ 	& 9.9  & 44.3 \\
%\cellcolor{white} 
& \module{A}{23}{} & 12-37 & 1600 & 48.125 & 2 	& 8 & 2 & $\times$8  & $\mathcal{C}$ 	& 161  & 37.1 \\
%\cellcolor{white} 
& \module{A}{24}{} & 12-41 & 1600 & 48.125 & 2 	& 8 & 2 & $\times$8  & $\mathcal{C}$ 	& 54.4 & 196  \\
%\cellcolor{white} 
& \module{A}{25}{} & 12-41 & 1600 & 48.125 & 2 	& 8 & 2 & $\times$8  & $\mathcal{C}$ 	& 24.1 & 1034 \\
%\cellcolor{white} 
& \module{A}{26}{} & 12-41 & 1600 & 48.125 & 2 	& 8 & 2 & $\times$8  & $\mathcal{C}$ 	& 208  & 55.8 \\
%\cellcolor{white} 
& \module{A}{27}{} & 12-41 & 1600 & 48.125 & 2 	& 8 & 2 & $\times$8  & $\mathcal{C}$ 	& 88.3 & 20.8 \\
%\cellcolor{white} 
& \module{A}{28}{} & 12-41 & 1600 & 48.125 & 2 	& 8 & 2 & $\times$8  & $\mathcal{C}$ 	& 51.6 & 122  \\
%\cellcolor{white} 
& \module{A}{29}{} & 12-41 & 1600 & 48.125 & 2 	& 8 & 2 & $\times$8  & $\mathcal{C}$ 	& 31.8 & 100  \\
\multirow{-32}{5em}{\centering {\large\em A} \\ {\tiny \quad \\} \centering Total of \\ \centering 30 DIMMs}
%\cellcolor{white} 
& \module{A}{30}{} & 13-11 & 1600 & 48.125 & 2 	& 8 & 2 & $\times$8  & $\mathcal{C}$ 	& 478  & 1590 \\
\bottomrule
\end{tabular}

\begin{tabular}{l}

{$\,$} \vspace{-7pt}\\

\footnotesize $\ast$ We report the manufacturing date in a year-week (yy-ww)
format. For example, 15-01 means that the DIMM was \\ \hspace{0.1in}manufactured during
the first week of 2015.\\ 

{$\,$} \vspace{-7pt}\\

\footnotesize $\dagger$  We report two representative timing factors: $Freq$ (the data transfer frequency per pin) and
$tRC$ (the row access cycle time).\\

{$\,$} \vspace{-7pt}\\

\footnotesize $\ddagger$ The maximum DRAM chip size supported by our testing
platform is 2Gb.\\

{$\,$} \vspace{-7pt}\\

\footnotesize $\S$ We report the DRAM die versions that are marked on the chip
package. Since the die version changes when the DRAM design\\
\hspace{0.1in}changes, we expect and typically observe that DIMMs with the same
die version have similar design-induced variation.\\

{$\,$} \vspace{-7pt}\\

\footnotesize $\star$ We report the {\em vulnerability ratio}, which
we define in Section~\ref{sec:profile_dimms} as the ratio of the number of errors
that occur in the top \\ \hspace{0.1in}10\% most
vulnerable \dhliii{rows} and \dhliii{the top 10\%} least vulnerable rows, to show design-induced variation in timing parameters.\\
\hspace{0.1in}A larger value indicates a greater amount of design-induced variation in the DIMM.\\
\hspace{0.1in}``$-$''\xspace indicates that we did not observe design-induced variation for the timing parameter in the DIMM.\vspace{5pt}\\

\hspace{0.1in}DIMMs with the same die version usually have a similar vulnerability ratio. However, there are some cases where we observe\\ 
\hspace{0.1in}large variation in the vulnerability ratio between two DIMMs with the {\em same} die version. \dhliii{We believe this observation is a result}\\ 
\hspace{0.1in}\dhliii{of process variation, which is dominant in some cases.}\\

\end{tabular}

\caption{Sample Population of 30 DDR3 DIMMs from Vendor~A (Sorted by Manufacturing Date)} 
\label{tbl:vendor1}
\end{table}

\definecolor{lightgray}{gray}{0.95}

\begin{table}[t]

\setlength{\tabcolsep}{3pt}
\centering
\footnotesize

\begin{tabular}{cccccccccccc}

\toprule

\multirow{2}{*}[-2pt]{\em \footnotesize \centering Vendor} & 
\multirow{2}{*}[-2pt]{\em \footnotesize \centering Module} & 
                     {\em \footnotesize \centering Date$^{\ast}$} &
\multicolumn{2}{c}{\em \footnotesize \centering Timing$^{\dagger}$} &
\multicolumn{2}{c}{\em \footnotesize \centering Organization} &
\multicolumn{3}{c}{\em \footnotesize \centering Chip} &
\multicolumn{2}{c}{\em \footnotesize \centering Vulnerability Ratio$^{\star}$} \\

\cmidrule(lr){3-3} \cmidrule(lr){4-5} \cmidrule(lr){6-7} \cmidrule(lr){8-10} \cmidrule(lr){11-12} 

 & 
 & {\em \footnotesize (yy-ww)} 
 & {\em \footnotesize Freq~(MT/s)} 
 & {\em \footnotesize tRC (ns)} 
 & {\em \footnotesize Size (GB)} 
 & {\em \footnotesize Chips} 
 & {\em \footnotesize Size (Gb)$^{\ddagger}$} 
 & {\em \footnotesize Pins} 
 & {\em \footnotesize Die Version$^{\S}$} 
 & {\em \footnotesize tRP} 
 & {\em \footnotesize tRCD} \\

\midrule

%% MICRON
%\cellcolor{white} 
& \module{B}{ 1}{} & 10-09 & 1066 & 50.625 & 0.5& 4 & 1 & $\times$16 & $\mathcal{B}$ & --   & --   \\
%\cellcolor{white} 
& \module{B}{ 2}{} & 10-22 & 1066 & 50.625 & 0.5& 4 & 1 & $\times$16 & $\mathcal{B}$ & --   & --   \\ 
%\cellcolor{white} 
& \module{B}{ 3}{} & 10-23 & 1066 & 50.625 & 1  & 8 & 1 & $\times$8  & $\mathcal{F}$ & --   & --   \\
%\cellcolor{white} 
& \module{B}{ 4}{} & 10-23 & 1066 & 50.625 & 1  & 8 & 1 & $\times$8  & $\mathcal{F}$ & --   & --   \\
%\cellcolor{white} 
& \module{B}{ 5}{} & 10-23 & 1066 & 50.625 & 1  & 8 & 1 & $\times$8  & $\mathcal{F}$ & --   & --   \\
%\cellcolor{white} 
& \module{B}{ 6}{} & 11-26 & 1066 & 49.125 & 1  & 4 & 2 & $\times$16 & $\mathcal{D}$ & --   & --   \\
%\cellcolor{white} 
& \module{B}{ 7}{} & 11-35 & 1066 & 49.125 & 1  & 4 & 2 & $\times$16 & $\mathcal{D}$ & 2.1  & --   \\
%\cellcolor{white} 
& \module{B}{ 8}{} & 11-35 & 1066 & 49.125 & 1  & 4 & 2 & $\times$16 & $\mathcal{D}$ & 479  & --   \\
%\cellcolor{white} 
& \module{B}{ 9}{} & 11-35 & 1066 & 49.125 & 1  & 4 & 2 & $\times$16 & $\mathcal{D}$ & 1.9  & --   \\
%\cellcolor{white} 
& \module{B}{10}{} & 11-35 & 1066 & 49.125 & 1  & 4 & 2 & $\times$16 & $\mathcal{D}$ & 4.3  & --   \\
%\cellcolor{white} 
& \module{B}{11}{} & 12-02 & 1066 & 49.125 & 1  & 4 & 2 & $\times$16 & $\mathcal{D}$ & 161  & --   \\
%\cellcolor{white} 
& \module{B}{12}{} & 12-02 & 1066 & 49.125 & 1  & 4 & 2 & $\times$16 & $\mathcal{D}$ & 2.3  & --   \\
%\cellcolor{white} 
& \module{B}{13}{} & 12-29 & 1600 & 50.625 & 1  & 4 & 2 & $\times$16 & $\mathcal{D}$ & 16.0 & --   \\
%\cellcolor{white} 
& \module{B}{14}{} & 12-29 & 1600 & 50.625 & 1  & 4 & 2 & $\times$16 & $\mathcal{D}$ & 8.6  & --   \\
%\cellcolor{white} 
& \module{B}{15}{} & 12-26 & 1600 & 49.125 & 2  & 8 & 2 & $\times$8  & $\mathcal{M}$ & --   & --   \\
%\cellcolor{white} 
& \module{B}{16}{} & 12-26 & 1600 & 49.125 & 2  & 8 & 2 & $\times$8  & $\mathcal{M}$ & --   & --   \\
%\cellcolor{white} 
& \module{B}{17}{} & 12-41 & 1600 & 48.125 & 2  & 8 & 2 & $\times$8  & $\mathcal{K}$ & --   & --   \\
%\cellcolor{white} 
& \module{B}{18}{} & 12-41 & 1600 & 48.125 & 2  & 8 & 2 & $\times$8  & $\mathcal{K}$ & --   & --   \\
%\cellcolor{white} 
& \module{B}{19}{} & 12-41 & 1600 & 48.125 & 2  & 8 & 2 & $\times$8  & $\mathcal{K}$ & --   & --   \\
%\cellcolor{white} 
& \module{B}{20}{} & 12-41 & 1600 & 48.125 & 2  & 8 & 2 & $\times$8  & $\mathcal{K}$ & --   & --   \\
%\cellcolor{white} 
& \module{B}{21}{} & 12-41 & 1600 & 48.125 & 2  & 8 & 2 & $\times$8  & $\mathcal{K}$ & 4.3  & 11.4 \\
%\cellcolor{white} 
& \module{B}{22}{} & 12-41 & 1600 & 48.125 & 2  & 8 & 2 & $\times$8  & $\mathcal{K}$ & 472  & --   \\
%\cellcolor{white} 
& \module{B}{23}{} & 12-41 & 1600 & 48.125 & 2  & 8 & 2 & $\times$8  & $\mathcal{K}$ & 279  & --   \\
%\cellcolor{white} 
& \module{B}{24}{} & 12-41 & 1600 & 48.125 & 2  & 8 & 2 & $\times$8  & $\mathcal{K}$ & 3276 & --   \\
%\cellcolor{white} 
& \module{B}{25}{} & 13-02 & 1600 & 48.125 & 2  & 8 & 2 & $\times$8  & --            & --   & --   \\
%\cellcolor{white} 
& \module{B}{26}{} & 13-02 & 1600 & 48.125 & 2  & 8 & 2 & $\times$8  & --            & --   & --   \\
%\cellcolor{white} 
& \module{B}{27}{} & 13-33 & 1600 & 48.125 & 2  & 8 & 2 & $\times$8  & $\mathcal{K}$ & --   & --   \\
%\cellcolor{white} 
& \module{B}{28}{} & 13-33 & 1600 & 48.125 & 2  & 8 & 2 & $\times$8  & $\mathcal{K}$ & 78.3 & 8.2  \\
%\cellcolor{white} 
& \module{B}{29}{} & 13-33 & 1600 & 48.125 & 2  & 8 & 2 & $\times$8  & $\mathcal{K}$ & 23.4 & 5.8  \\
\multirow{-32}{5em}{\centering {\large\em B} \\ {\tiny \quad \\} \centering Total of \\ \centering 30 DIMMs}
%\cellcolor{white} 
& \module{B}{30}{} & 14-09 & 1600 & 48.125 & 2 	& 8 & 2 & $\times$8  & $\mathcal{K}$ & --   & --   \\
\bottomrule
\end{tabular}

\begin{tabular}{l}

{$\,$} \vspace{-7pt}\\

\footnotesize $\ast$ We report the manufacturing date in a year-week (yy-ww)
format. For example, 15-01 means that the DIMM was \\ \hspace{0.1in}manufactured during
the first week of 2015.\\ 

{$\,$} \vspace{-7pt}\\

\footnotesize $\dagger$  We report two representative timing factors: $Freq$ (the data transfer frequency per pin) and
$tRC$ (the row access cycle time).\\

{$\,$} \vspace{-7pt}\\

\footnotesize $\ddagger$ The maximum DRAM chip size supported by our testing
platform is 2Gb.\\

{$\,$} \vspace{-7pt}\\

\footnotesize $\S$ We report the DRAM die versions that are marked on the chip
package. Since the die version changes when the DRAM design\\
\hspace{0.1in}changes, we expect and typically observe that DIMMs with the same
die version have similar design-induced variation.\\

{$\,$} \vspace{-7pt}\\

\footnotesize $\star$ We report the {\em vulnerability ratio}, which
we define in Section~\ref{sec:profile_dimms} as the ratio of the number of errors
that occur in the top \\ \hspace{0.1in}10\% most
vulnerable \dhliii{rows} and \dhliii{the top 10\%} least vulnerable rows, to show design-induced variation in timing parameters.\\
\hspace{0.1in}A larger value indicates a greater amount of design-induced variation in the DIMM.\\
\hspace{0.1in}``$-$''\xspace indicates that we did not observe design-induced variation for the timing parameter in the DIMM.\vspace{5pt}\\

\hspace{0.1in}DIMMs with the same die version usually have a similar vulnerability ratio. However, there are some cases where we observe\\ 
\hspace{0.1in}large variation in the vulnerability ratio between two DIMMs with the {\em same} die version. \dhliii{We believe this observation is a result}\\ 
\hspace{0.1in}\dhliii{of process variation, which is dominant in some cases.}\\

\end{tabular}

\caption{Sample Population of 30 DDR3 DIMMs from Vendor~B (Sorted by Manufacturing Date)} 
\label{tbl:vendor2}
\end{table}

\newpage

\definecolor{lightgray}{gray}{0.95}

\begin{table}[t]

\setlength{\tabcolsep}{3pt}
\centering
\footnotesize

\begin{tabular}{cccccccccccc}

\toprule

\multirow{2}{*}[-2pt]{\em \footnotesize \centering Vendor} & 
\multirow{2}{*}[-2pt]{\em \footnotesize \centering Module} & 
                     {\em \footnotesize \centering Date$^{\ast}$} &
\multicolumn{2}{c}{\em \footnotesize \centering Timing$^{\dagger}$} &
\multicolumn{2}{c}{\em \footnotesize \centering Organization} &
\multicolumn{3}{c}{\em \footnotesize \centering Chip} &
\multicolumn{2}{c}{\em \footnotesize \centering Vulnerability Ratio$^{\star}$} \\

\cmidrule(lr){3-3} \cmidrule(lr){4-5} \cmidrule(lr){6-7} \cmidrule(lr){8-10} \cmidrule(lr){11-12} 

 & 
 & {\em \footnotesize (yy-ww)} 
 & {\em \footnotesize Freq~(MT/s)} 
 & {\em \footnotesize tRC (ns)} 
 & {\em \footnotesize Size (GB)} 
 & {\em \footnotesize Chips} 
 & {\em \footnotesize Size (Gb)$^{\ddagger}$} 
 & {\em \footnotesize Pins} 
 & {\em \footnotesize Die Version$^{\S}$} 
 & {\em \footnotesize tRP} 
 & {\em \footnotesize tRCD} \\

\midrule

%% SAMSUNG
%\cellcolor{white} 
& \module{C}{ 1}{} & 08-49 & 1066 & 50.625 & 1 	& 8 & 1 & $\times$8  & $\mathcal{D}$ 	& --   & --   \\
%\cellcolor{white} 
& \module{C}{ 2}{} & 09-49 & 1066 & 50.625 & 1 	& 8 & 1 & $\times$8  & $\mathcal{E}$ 	& --   & --   \\
%\cellcolor{white} 
& \module{C}{ 3}{} & 10-19 & 1066 & 50.625 & 1 	& 8 & 1 & $\times$8  & $\mathcal{F}$ 	& --   & --   \\
%\cellcolor{white} 
& \module{C}{ 4}{} & 11-16 & 1066 & 50.625 & 1 	& 8 & 1 & $\times$8  & $\mathcal{F}$ 	& --   & --   \\
%\cellcolor{white} 
& \module{C}{ 5}{} & 11-19 & 1066 & 50.625 & 1 	& 8 & 1 & $\times$8  & $\mathcal{F}$ 	& --   & --   \\
%\cellcolor{white} 
& \module{C}{ 6}{} & 11-25 & 1333 & 49.125 & 2 	& 8 & 2 & $\times$8  & $\mathcal{C}$ 	& --   & --   \\
%\cellcolor{white} 
& \module{C}{ 7}{} & 11-37 & 1333 & 49.125 & 2 	& 8 & 2 & $\times$8  & $\mathcal{D}$ 	& --   & 2.6  \\
%\cellcolor{white} 
& \module{C}{ 8}{} & 11-46 & 1333 & 49.125 & 2 	& 8 & 2 & $\times$8  & $\mathcal{D}$ 	& --   & 32.9 \\
%\cellcolor{white} 
& \module{C}{ 9}{} & 11-46 & 1333 & 49.125 & 2 	& 8 & 2 & $\times$8  & $\mathcal{D}$ 	& --   & 42.3 \\
%\cellcolor{white} 
& \module{C}{10}{} & 11-49 & 1333 & 49.125 & 2 	& 8 & 2 & $\times$8  & $\mathcal{C}$ 	& --   & --   \\
%\cellcolor{white} 
& \module{C}{11}{} & 12-10 & 1866 & 47.125 & 2 	& 8 & 2 & $\times$8  & $\mathcal{D}$ 	& --   & 104  \\
%\cellcolor{white} 
& \module{C}{12}{} & 12-10 & 1866 & 47.125 & 2 	& 8 & 2 & $\times$8  & $\mathcal{D}$ 	& --   & 117  \\
%\cellcolor{white} 
& \module{C}{13}{} & 12-10 & 1866 & 47.125 & 2 	& 8 & 2 & $\times$8  & $\mathcal{D}$ 	& --   & 291  \\
%\cellcolor{white} 
& \module{C}{14}{} & 12-10 & 1866 & 47.125 & 2 	& 8 & 2 & $\times$8  & $\mathcal{D}$ 	& --   & --   \\
%\cellcolor{white} 
& \module{C}{15}{} & 12-10 & 1866 & 47.125 & 2 	& 8 & 2 & $\times$8  & $\mathcal{D}$ 	& --   & 97.0 \\
%\cellcolor{white} 
& \module{C}{16}{} & 12-10 & 1866 & 47.125 & 2 	& 8 & 2 & $\times$8  & $\mathcal{D}$ 	& --   & 493  \\
%\cellcolor{white} 
& \module{C}{17}{} & 12-10 & 1866 & 47.125 & 2 	& 8 & 2 & $\times$8  & $\mathcal{D}$ 	& --   & 61.8 \\
%\cellcolor{white} 
& \module{C}{18}{} & 12-25 & 1600 & 48.125 & 2 	& 8 & 2 & $\times$8  & $\mathcal{E}$ 	& 2.2  & 3.3  \\
%\cellcolor{white} 
& \module{C}{19}{} & 12-28 & 1600 & 48.125 & 2 	& 8 & 2 & $\times$8  & $\mathcal{E}$ 	& 473  & 3.1  \\
%\cellcolor{white} 
& \module{C}{20}{} & 12-28 & 1600 & 48.125 & 2 	& 8 & 2 & $\times$8  & $\mathcal{E}$ 	& 5.4  & 2.7  \\
%\cellcolor{white} 
& \module{C}{21}{} & 12-28 & 1600 & 48.125 & 2 	& 8 & 2 & $\times$8  & $\mathcal{E}$ 	& 3.5  & 3.0  \\
%\cellcolor{white} 
& \module{C}{22}{} & 12-28 & 1600 & 48.125 & 2 	& 8 & 2 & $\times$8  & $\mathcal{E}$ 	& 545  & 3.0  \\
%\cellcolor{white} 
& \module{C}{23}{} & 12-28 & 1600 & 48.125 & 2 	& 8 & 2 & $\times$8  & $\mathcal{E}$ 	& 2.7  & 3.0  \\
%\cellcolor{white} 
& \module{C}{24}{} & 12-28 & 1600 & 48.125 & 2 	& 8 & 2 & $\times$8  & $\mathcal{E}$ 	& 27.2 & 2.9  \\
%\cellcolor{white} 
& \module{C}{25}{} & 12-28 & 1600 & 48.125 & 2 	& 8 & 2 & $\times$8  & $\mathcal{E}$ 	& --   & 3.3  \\
%\cellcolor{white} 
& \module{C}{26}{} & 12-28 & 1600 & 48.125 & 2 	& 8 & 2 & $\times$8  & $\mathcal{E}$ 	& 54.2 & 19.1 \\
%\cellcolor{white} 
& \module{C}{27}{} & 12-28 & 1600 & 48.125 & 2 	& 8 & 2 & $\times$8  & $\mathcal{E}$ 	& --   & 3.1  \\
%\cellcolor{white} 
& \module{C}{28}{} & 12-31 & 1600 & 48.125 & 2 	& 8 & 2 & $\times$8  & $\mathcal{E}$ 	& 29.0 & 5.4  \\
%\cellcolor{white} 
& \module{C}{29}{} & 12-31 & 1600 & 48.125 & 2 	& 8 & 2 & $\times$8  & $\mathcal{E}$ 	& 120  & 6.7  \\
%\cellcolor{white} 
& \module{C}{30}{} & 12-31 & 1600 & 48.125 & 2 	& 8 & 2 & $\times$8  & $\mathcal{E}$ 	& 196  & 3.2  \\
%\cellcolor{white} 
& \module{C}{31}{} & 12-31 & 1600 & 48.125 & 2 	& 8 & 2 & $\times$8  & $\mathcal{E}$ 	& 599  & 8.5  \\
%\cellcolor{white} 
& \module{C}{32}{} & 12-31 & 1600 & 48.125 & 2 	& 8 & 2 & $\times$8  & $\mathcal{E}$ 	& 51.6 & --   \\
%\cellcolor{white} 
& \module{C}{33}{} & 13-19 & 1600 & 48.125 & 2 	& 8 & 2 & $\times$8  & $\mathcal{E}$ 	& --   & 2.5  \\
%\cellcolor{white} 
& \module{C}{34}{} & 13-19 & 1600 & 48.125 & 2 	& 8 & 2 & $\times$8  & $\mathcal{E}$ 	& --   & 1.6  \\
%\cellcolor{white} 
& \module{C}{35}{} & 13-19 & 1600 & 48.125 & 2 	& 8 & 2 & $\times$8  & $\mathcal{E}$ 	& --   & 2.6  \\
\multirow{-38}{5em}{\centering {\large\em C} \\ {\tiny \quad \\} \centering Total of \\ \centering 36 DIMMs}
%\cellcolor{white} 
& \module{C}{36}{} & 13-19 & 1600 & 48.125 & 2 	& 8 & 2 & $\times$8  & $\mathcal{E}$ 	& --   & 1.9  \\
\bottomrule
\end{tabular}

\begin{tabular}{l}

{$\,$} \vspace{-7pt}\\

\footnotesize $\ast$ We report the manufacturing date in a year-week (yy-ww)
format. For example, 15-01 means that the DIMM was \\ \hspace{0.1in}manufactured during
the first week of 2015.\\ 

{$\,$} \vspace{-7pt}\\

\footnotesize $\dagger$  We report two representative timing factors: $Freq$ (the data transfer frequency per pin) and
$tRC$ (the row access cycle time).\\

{$\,$} \vspace{-7pt}\\

\footnotesize $\ddagger$ The maximum DRAM chip size supported by our testing
platform is 2Gb.\\

{$\,$} \vspace{-7pt}\\

\footnotesize $\S$ We report the DRAM die versions that are marked on the chip
package. Since the die version changes when the DRAM design\\
\hspace{0.1in}changes, we expect and typically observe that DIMMs with the same
die version have similar design-induced variation.\\

{$\,$} \vspace{-7pt}\\

\footnotesize $\star$ We report the {\em vulnerability ratio}, which
we define in Section~\ref{sec:profile_dimms} as the ratio of the number of errors
that occur in the top \\ \hspace{0.1in}10\% most
vulnerable \dhliii{rows} and \dhliii{the top 10\%} least vulnerable rows, to show design-induced variation in timing parameters.\\
\hspace{0.1in}A larger value indicates a greater amount of design-induced variation in the DIMM.\\
\hspace{0.1in}``$-$''\xspace indicates that we did not observe design-induced variation for the timing parameter in the DIMM.\vspace{5pt}\\

\hspace{0.1in}DIMMs with the same die version usually have a similar vulnerability ratio. However, there are some cases where we observe\\ 
\hspace{0.1in}large variation in the vulnerability ratio between two DIMMs with the {\em same} die version. \dhliii{We believe this observation is a result}\\ 
\hspace{0.1in}\dhliii{of process variation, which is dominant in some cases.}\\

\end{tabular}

\caption{Sample Population of 36 DDR3 DIMMs from Vendor~C (Sorted by Manufacturing Date)} 
\label{tbl:vendor3}

\end{table}

\end{document}